\documentclass[aps,twocolumn,prb,floatfix,superscriptaddress,amsmath,amssymb]{revtex4-2} 
\usepackage{dcolumn}
\usepackage{amsmath}
\usepackage{mathrsfs}
\usepackage{txfonts}
\usepackage{bm}
\usepackage[T1]{fontenc}
\usepackage{xspace}
\usepackage{comment}
\usepackage{braket}
\usepackage{bbold}
\usepackage{mathtools}
\newcommand{\red}[1]{#1}

\usepackage{graphicx}
\usepackage{hyperref}
\usepackage{color}
\usepackage{xcolor}
\usepackage[version=3]{mhchem}
\hypersetup{
        colorlinks=true,
        citecolor=blue,
        urlcolor=blue,
        linkcolor=blue
}

\begin{document}
\let\emph\textit

\title{
    \texorpdfstring{Schwinger boson perturbation theory for spin-$S$ Kitaev-Heisenberg magnets:
    phase diagram and dynamical response near Kitaev spin liquids}{Schwinger boson perturbation theory for spin-S Kitaev-Heisenberg magnets: phase diagram and dynamical response near Kitaev spin liquids}
}
\author{Daiki Sasamoto}
\email[sasamoto.daiki.r6@dc.tohoku.ac.jp]{}
\author{Joji Nasu}
\affiliation{
  Department of Physics, Graduate School of Science, Tohoku University, Sendai, Miyagi 980-8578, Japan
}

\date{\today}
\begin{abstract}
We develop a Schwinger boson perturbative framework for the spin-$S$
Kitaev-Heisenberg model.  The spin-liquid saddle point of the pure Kitaev
model is used as the unperturbed state, and magnetic instabilities and
dynamical spin correlations are evaluated around this saddle point.
\red{We decompose the Hamiltonian exactly into two parts by exploiting the
Klein duality intrinsic to the Kitaev-Heisenberg model.  We perform the
random-phase approximation by taking the part invariant under the duality
transformation as the unperturbed term and treating the remaining term, which
changes sign under it, as the perturbation.}  We determine the phase boundaries separating the quantum spin-liquid
regimes from the adjacent magnetically ordered phases for $S=1/2$, $1$,
$3/2$, and $2$.  \red{The spin-liquid regions shrink rapidly with increasing
$S$ and become very narrow at $S=2$.}  We also compute the spin dynamics of the spin-$S$
Kitaev-Heisenberg model, focusing on $S=1$, and find that the dressed
dynamical spin structure factor retains a broad two-spinon continuum at finite
energies, while the low-energy spectral weight softens at the ordering wave
vectors of the adjacent magnetic phases.  Our framework thus enables
thermodynamic-limit calculations of spin dynamics near magnetic
instabilities.
\end{abstract}
\maketitle


\section{Introduction}
\label{sec:introduction}

Quantum spin liquids (QSLs) are quantum-disordered states of magnets in which
magnetic moments avoid conventional long-range order even at zero temperature
because of strong quantum fluctuations.  Since Anderson's proposal of the
resonating-valence-bond state, QSLs have attracted continuing interest as
platforms for long-range entanglement, emergent gauge fields, and fractionalized
elementary excitations~\cite{Anderson-1973,Kalmeyer-Laughlin-1987,
Wen-1989,Wen-2002,Balents-2010,Savary-Balents-2016,Zhou-Kanoda-2017,
Broholm-Cava-2020,Knolle-Moessner-2019}.
One of the central difficulties in this field is that such phases are not
characterized by a local order parameter.  It is therefore essential to develop
microscopic models and theoretical methods that can identify not only the
absence of magnetic order but also the nature of the elementary excitations and
their dynamical signatures.

The Kitaev model on the honeycomb lattice provides a rare and highly influential
example of a QSL in more than one spatial dimension~\cite{Kitaev-2006}.  In the
spin-$1/2$ case, the bond-dependent Ising interactions make the model exactly
solvable, and the spin degrees of freedom fractionalize into itinerant Majorana
fermions coupled to static ${\mathbb Z}_2$ gauge fluxes.  This exact solution
has made the Kitaev model a paradigmatic setting in which fractionalization,
topological character, and spin dynamics can be discussed on firm theoretical
grounds~\cite{Knolle-2014,Knolle-2015,Hermanns-Kimchi-2018,Motome-Nasu-2020,Nasu-Motome-2021}.
The model has also
become central to the study of spin-orbit-coupled Mott insulators, where
bond-directional exchange interactions may arise from the combined effect of
strong spin-orbit coupling and electronic correlations~\cite{Jackeli-Khaliullin-2009,Foyevtsova-2013,Katukuri-2014,Yamaji-2014,Chun-2015,Yadav-2016,Rau-Lee-Kee-2014,Rau-Lee-Kee-2016,Winter-2017,Takagi-2019,Motome-2020,Trebst-Hickey-2022,Rousochatzakis-Perkins-Luo-Kee-2024}.

In real materials, however, the pristine Kitaev model is not realized by itself.
\red{Additional interactions, such as Heisenberg exchange,
are generically present and can destabilize the Kitaev QSL}
ground state.  The Kitaev-Heisenberg model has therefore served
as a \mbox{useful setting} for studying the stability of Kitaev QSLs against
conventional magnetic instabilities~\cite{Chaloupka-2010,Jiang-2011,Reuther-2011,Schaffer-2012,Price-Perkins-2012,Chaloupka-2013,Rau-Lee-Kee-2014,Iregui-2014,Sela-2014,Chaloupka-Khaliullin-2015,Winter-2016,Gotfryd-2017}.
This problem is especially important because the putative Kitaev materials
usually display magnetic ordering at low temperatures, while retaining strong
signatures of Kitaev physics in thermodynamic and dynamical quantities.
Understanding how the QSL is destabilized, and how its fractionalized excitations
evolve near magnetic instabilities, remains a key issue for connecting theory
with experiments~\cite{Singh-Gegenwart-2010,Singh-Manni-2012,Plumb-2014,
Kubota-2015,Sinn-2016,Banerjee-2016,Banerjee-2017,Knolle-2014,
Yoshitake-Nasu-Motome-2016,Yoshitake-Nasu-2017}.

The extension of Kitaev physics beyond spin-$1/2$ moments has recently opened a
second important direction.  Higher-spin Kitaev models are motivated both by
theoretical questions about the robustness of fractionalization and by candidate
materials with larger local moments~\cite{Xu-2018,Stavropoulos-2019,
Lee-2020,Stavropoulos-2021,Xu-2020}.
Numerical and analytical studies of
spin-$1$ and higher-spin Kitaev models have revealed QSL behavior and
characteristic thermodynamic properties, while also indicating a strong
dependence on the spin length~\cite{Baskaran-Mandal-2007,Baskaran-Sen-Shankar-2008,
Koga-2018,Suzuki-2018,Oitmaa-2018,Minakawa-2019,Koga-2020,
Khait-Stavropoulos-2021,Lee-Kawashima-Kim-2020,Hickey-2020,Zhu-2020,
Bradley-2022,Chen-2022,Jin-2022,Ralko-Merino-2024,Sasamoto-Nasu-2025,
Sasamoto-Ralko-Merino-Nasu-2026}.
For the spin-$S$ Kitaev-Heisenberg model, studies based on
pseudofermion functional renormalization group (PFFRG) and the high-order
coupled cluster method (CCM) have provided systematic phase diagrams as a
function of $S$, showing that the QSL regions shrink rapidly with increasing
spin length~\cite{Dong-Sheng-2020,Fukui-Kato-Nasu-Motome-2022,Georgiou-2024}.
These results motivate a complementary approach that starts from the Kitaev
spin-liquid state itself and treats magnetic instabilities and dynamical
responses within a single framework.

Schwinger boson approaches provide a natural language \red{for treating
spin-$S$ physics, because they represent} quantum-disordered magnets in terms of bosonic spin-carrying
quasiparticles and can be formulated for general spin length.  Once a
spin-liquid saddle point is obtained, the same quasiparticle description also
provides access to momentum- and frequency-dependent dynamical spin
correlations, not only to static instabilities.  Connections between spin-$1/2$ and higher-spin Kitaev physics have also been
examined through bilayer and multilayer Kitaev models, in which spin-$1/2$
Kitaev layers are coupled by an interlayer Heisenberg
interaction~\cite{Tomishige-Nasu-Koga-2018,Seifert-2018,Tomishige-Nasu-Koga-2019,Merino-Ralko-2025}.
In the bilayer case, an antiferromagnetic interlayer coupling drives a
transition from the Kitaev spin liquid to an interlayer dimer
phase~\cite{Tomishige-Nasu-Koga-2018,Seifert-2018}, whereas a ferromagnetic
coupling adiabatically connects the spin-$1/2$ Kitaev spin liquid to a
spin-$1$ Kitaev spin liquid~\cite{Tomishige-Nasu-Koga-2019}, while in multilayer
systems, the nature of the spin liquid has been found to depend on the parity
of the number of layers~\cite{Merino-Ralko-2025}.
For the higher-spin Kitaev model itself, an analysis based on the
Majorana fermion representation has established an even--odd effect in which
the statistics of the $\mathbb{Z}_2$ gauge charge depends on the parity of $2S$,
being fermionic for half-integer spins and bosonic for integer
spins~\cite{Ma-2023}.  Motivated by this, Schwinger boson mean-field theory
(SBMFT) studies of integer-spin Kitaev models have been carried out, in which
the spin-liquid state is described in terms of bosonic elementary excitations
and the dynamical spin structure factor has been computed within
SBMFT~\cite{Ralko-Merino-2024,Sasamoto-Nasu-2025,
Sasamoto-Ralko-Merino-Nasu-2026}.
This bosonic perspective is complementary to
Majorana-based descriptions of the spin-$1/2$ Kitaev model and is well suited
for exploring dynamical spin correlations in integer-spin systems.  It also
motivates using an SBMFT spin-liquid saddle point as the unperturbed background
for response calculations near the Kitaev limit.

In this paper, we develop a Schwinger boson perturbative framework for the
spin-$S$ Kitaev-Heisenberg model.  Starting from an SBMFT spin-liquid saddle
point of the pure Kitaev model, we treat the remaining interaction
through the random-phase approximation (RPA).
\red{The Klein duality, which is intrinsic to this model, allows us to separate
the Kitaev-Heisenberg Hamiltonian and to regard the Klein-odd part as the
perturbation in the RPA calculation.}
This construction allows both static magnetic instabilities and dynamical spin
correlations around the pure-Kitaev spin-liquid saddle to be evaluated directly
in the thermodynamic limit.  \red{Using the present RPA framework, we first
determine the QSL boundaries for $S=1/2$, $1$, $3/2$, and $2$ by locating the
instability of the QSL state in the static RPA susceptibility.}  We combine
these boundaries with auxiliary linear spin-wave boundaries between the
conventional ordered phases to obtain a general-$S$ phase diagram.  \red{The
QSL regions shrink rapidly with increasing $S$ and become very narrow at
$S=2$.}
Second, to clarify the spin dynamics of the spin-$S$ Kitaev-Heisenberg
model, we focus on the representative $S=1$ case and compute the RPA-dressed
dynamical spin structure factor near the QSL phase boundaries.  The spectra retain a broad finite-energy
two-spinon continuum, while the low-energy spectral weight softens at the same
ordering wave vectors that become singular in the static susceptibility.  Thus the
present framework connects the general-$S$ phase-boundary construction with the
corresponding dynamical signatures of the Kitaev spin liquid near magnetic
instabilities.

The remainder of this paper is organized as follows.  Section~\ref{sec:model}
introduces the model and the constraints imposed by Klein duality.
Section~\ref{sec:method} presents the SBMFT and perturbative RPA formulations,
together with the relation between the RPA susceptibility and the dynamical
spin structure factor.  Section~\ref{sec:results}
reports the resulting phase boundaries of the spin-$S$ Kitaev-Heisenberg
model and the dynamical spin structure factor near the spin-$1$ Kitaev
spin-liquid regime.  Section~\ref{sec:discussion} compares the phase boundaries
with previous studies and discusses the role of the Hamiltonian decomposition
constrained by Klein duality.  Section~\ref{sec:summary} summarizes the main
conclusions.  Technical details of the RPA construction and the linear
spin-wave calculation are given in Appendices~\ref{app:rpa_calculation} and
\ref{app:linear_spin_wave}, respectively.

\section{Model}
\label{sec:model}

\begin{figure}[t]
\centering
\includegraphics[width=\columnwidth]{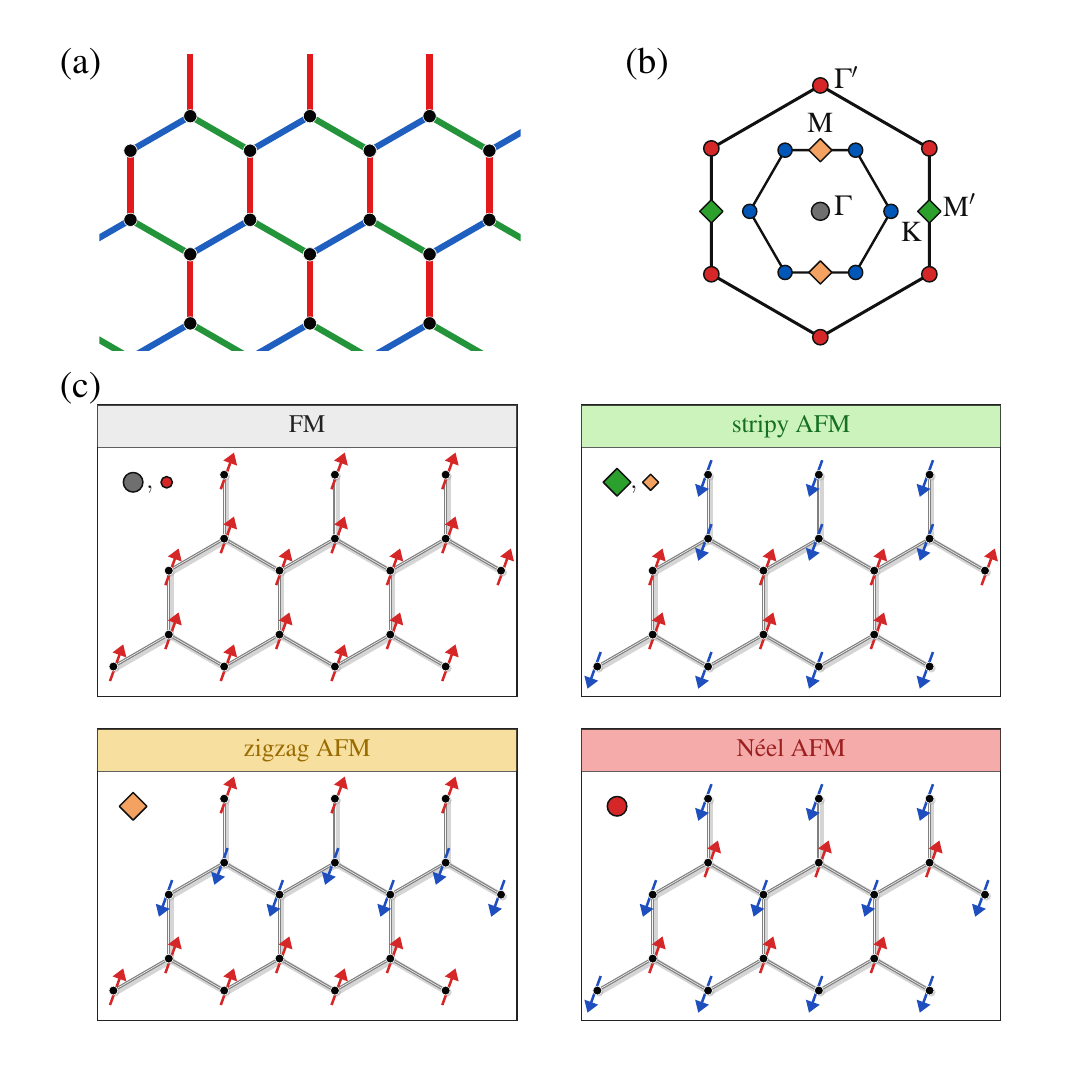}
\caption{
(a) Honeycomb lattice on which the spin-$S$ Kitaev-Heisenberg Hamiltonian is
defined.  Blue, green, and red links denote the $x$, $y$, and $z$ Kitaev bonds,
respectively.
(b) First and extended Brillouin zones.  The colored symbols mark the
high-symmetry wave vectors, $\Gamma$, $\mathrm{K}$, $\mathrm{M}$,
$\Gamma^{\prime}$, and $\mathrm{M}^{\prime}$.
(c) Real-space spin configurations of the FM, stripy AFM, zigzag AFM, and
N\'eel AFM states.  The symbols shown in each panel indicate the corresponding
ordering wave vectors, using the same symbols as in (b).  Larger and smaller
symbols denote dominant and subdominant susceptibility channels, respectively.
}
\label{fig:Kitaev_Heisenberg}
\end{figure}

We consider the spin-$S$ Kitaev-Heisenberg model on the honeycomb lattice,
defined by
\begin{align}
\mathcal{H}
=
\sum_{\gamma=x,y,z}\sum_{\langle i,j\rangle_{\gamma}}
\left[
K S_i^{\gamma}S_j^{\gamma}
+J \bm{\mathit{S}}_i\cdot\bm{\mathit{S}}_j
\right],
\label{eq:kitaev_heisenberg_hamiltonian}
\end{align}
where $\langle i,j\rangle_{\gamma}$ denotes a nearest-neighbor bond of type
$\gamma=x,y,z$, and
$\bm{\mathit{S}}_i=(S_i^x,S_i^y,S_i^z)$ is a spin-$S$ operator.  The first and
second terms in Eq.~\eqref{eq:kitaev_heisenberg_hamiltonian} are the Kitaev and
Heisenberg interactions, respectively.  Positive and negative values of $K$
correspond to antiferromagnetic and ferromagnetic Kitaev interactions, while
positive and negative values of $J$ correspond to antiferromagnetic and
ferromagnetic Heisenberg interactions.  We parametrize the couplings as
\begin{align}
K=\sin\phi,\qquad
J=\cos\phi,
\label{eq:kitaev_heisenberg_parametrization}
\end{align}
with $0\leq\phi<2\pi$ and $\sqrt{K^2+J^2}=1$, following the standard angular
parametrization of the Kitaev-Heisenberg model~\cite{Chaloupka-2010,
Chaloupka-2013,Georgiou-2024}.  In this convention, the antiferromagnetic and
ferromagnetic Kitaev points are located at $\phi=\pi/2$ and $3\pi/2$,
respectively.  The antiferromagnetic and ferromagnetic Heisenberg points are at
$\phi=0$ and $\pi$, respectively.  Spin-$S$ generalizations of the
Kitaev-Heisenberg problem have recently been studied by complementary numerical
methods, including PFFRG and CCM~\cite{Fukui-Kato-Nasu-Motome-2022,Georgiou-2024}.

An important constraint on the phase diagram is provided by the four-sublattice
Klein duality of the Kitaev-Heisenberg model~\cite{Khaliullin-2005,
Jackeli-Khaliullin-2009,Chaloupka-2010,Chaloupka-2013,
Kimchi-Vishwanath-2014,Chaloupka-Khaliullin-2015,Winter-2017,
Trebst-Hickey-2022,Rousochatzakis-Perkins-Luo-Kee-2024,Georgiou-2024}.  Under
this transformation, the honeycomb lattice is decomposed into four sublattices,
and the spins on three of them are rotated by $\pi$ about mutually orthogonal
spin axes.  The transformation preserves the SU($2$) spin algebra and maps the
Hamiltonian in Eq.~\eqref{eq:kitaev_heisenberg_hamiltonian} to the same form
with transformed couplings
\begin{align}
\widetilde{J}=-J,\qquad
\widetilde{K}=K+2J.
\label{eq:klein_duality_couplings}
\end{align}
After normalizing the energy scale back to
$\sqrt{\widetilde{K}^{2}+\widetilde{J}^{2}}=1$, this gives the angular map
\begin{align}
\tan\widetilde{\phi}= -\left(\tan\phi+2\right),
\label{eq:klein_duality_angle}
\end{align}
with the branch of $\widetilde{\phi}$ chosen in $[0,2\pi)$.  The two pure
Kitaev points are self-dual.  In addition, the points satisfying $K=-2J$ are
mapped to pure Heisenberg models in the rotated frame.  These are the hidden
SU($2$) points at $\phi=\pi-\arctan 2$ and $2\pi-\arctan 2$.

The same duality imposes a corresponding relation between phase boundaries.
Using the ratio
\begin{align}
j=\frac{J}{K}=\cot\phi,
\end{align}
Eq.~\eqref{eq:klein_duality_couplings} gives
\begin{align}
\widetilde{j}
=
\frac{\widetilde{J}}{\widetilde{K}}
=
-\frac{j}{1+2j}.
\label{eq:klein_duality_ratio}
\end{align}
Thus, if a critical point adjacent to a Kitaev spin liquid occurs at $j=j_c$,
the dual critical point must occur at
\begin{align}
j_c^{\mathrm{dual}}
=
-\frac{j_c}{1+2j_c}.
\label{eq:klein_duality_transition_relation}
\end{align}
This relation connects the N\'eel and zigzag boundaries around the
antiferromagnetic Kitaev point, and the FM and stripy AFM boundaries around the
ferromagnetic Kitaev point.  We use this duality relation below to organize
the RPA vertex in Klein-even and Klein-odd channels, namely parts that are even
or odd under the Klein transformation, and to check the paired instability lines
obtained from the static susceptibility.

\section{Method}
\label{sec:method}

\subsection{Schwinger boson mean-field theory}
\label{sec:schwinger_boson_mean_field_theory}

We begin by introducing the Schwinger boson representation and the associated
Schwinger boson mean-field theory (SBMFT)~\cite{Read-Sachdev-1991,
Arovas-Auerbach-1998,Sachdev-Read-1991,Sachdev-1992,Sasamoto-Nasu-2025}.
This subsection fixes the notation for the bosonic Hilbert-space constraint,
the bond-operator mean-field decoupling, and the resulting bosonic
Bogoliubov-de Gennes Hamiltonian used in the response calculation below.
The basic idea is to rewrite a spin of length $S$ in terms of two bosonic
spinon flavors, $b_{i\uparrow}$ and $b_{i\downarrow}$, \red{which obey the
canonical commutation relations
$[b_{i\mu},b_{j\nu}^{\dagger}]=\delta_{ij}\delta_{\mu\nu}$ and
$[b_{i\mu},b_{j\nu}]=[b_{i\mu}^{\dagger},b_{j\nu}^{\dagger}]=0$.}
In terms of these bosons, the spin operator is represented as
\begin{align}
S_i^\alpha
=
\frac{1}{2}
\sum_{\mu,\nu=\uparrow,\downarrow}
b_{i\mu}^{\dagger}
\sigma^\alpha_{\mu\nu}
b_{i\nu}.
\label{eq:schwinger_boson_representation}
\end{align}
Here $\sigma^\alpha$ $(\alpha=x,y,z)$ is a Pauli matrix, while $\mu,\nu$ label
\red{the two boson spin flavors, $\uparrow$ and $\downarrow$}.  We also define
the boson-number operator at site $i$ by
\begin{align}
n_i
\equiv
\sum_{\mu=\uparrow,\downarrow}
b_{i\mu}^{\dagger}b_{i\mu}.
\label{eq:schwinger_boson_number_operator}
\end{align}
The representation in Eq.~\eqref{eq:schwinger_boson_representation} enlarges
the local Hilbert space, because the boson occupation number is not fixed a
priori.  To recover the physical spin-$S$ Hilbert space, one must impose the
local constraint
\begin{align}
n_i=2S.
\label{eq:schwinger_boson_local_constraint}
\end{align}
The fixed-occupancy subspace specified by Eq.~\eqref{eq:schwinger_boson_local_constraint}
contains $2S+1$ states and realizes the spin-$S$ irreducible representation.
Within this subspace, Eq.~\eqref{eq:schwinger_boson_representation} reproduces
the spin algebra and $\bm{S}_i^2=S(S+1)$.  Thus the Schwinger boson
representation is exact only when the local constraint is enforced at every
site.
In SBMFT, the local constraint in Eq.~\eqref{eq:schwinger_boson_local_constraint}
is imposed at the saddle-point level through Lagrange multipliers.  We
therefore add
\begin{align}
\mathcal{H}_{\lambda}
=
\sum_i
\lambda_i
\left(
n_i-2S
\right)
\label{eq:schwinger_boson_constraint_term}
\end{align}
to the bosonic Hamiltonian.  For a translationally invariant Ansatz, $\lambda_i$
is taken to be periodic in the mean-field unit cell and may depend on the
sublattice.  \red{The spin length fixes the right-hand side of the local
constraint in Eq.~\eqref{eq:schwinger_boson_local_constraint}.}  Although the
operator identities below are independent of $S$, the self-consistent saddle
point changes with the required boson density $2S$.

The mean-field decoupling is performed in terms of bond operators.  For the
Heisenberg interaction, conventional SBMFT uses the SU($2$)-invariant hopping and
singlet-pairing channels.  The Kitaev interaction, however, selects a spin
component depending on the bond direction.  We therefore use a bond-channel
formulation that supplements the SU($2$)-invariant channels by SU($2$)-breaking
anisotropic channels~\cite{Kargarian-Langari-Fiete-2012,Kos-Punk-2017,
Samajdar-Scheurer-2019,Mondal-2021,Schneider-2022,Messio-Cepas-2010,
Mondal-2017,Rossi-Motruk-2023,Ralko-Merino-2024,Sasamoto-Nasu-2026,
Sasamoto-Nasu-2025}.  Related mixed singlet/triplet Schwinger boson
descriptions have also been applied to the integer-spin
Heisenberg-Kitaev model~\cite{Ralko-Merino-2024,Sasamoto-Nasu-2025}.  On a directed
nearest-neighbor bond $(i,j)$, we define
\begin{subequations}
\label{eq:schwinger_boson_bond_operators}
\begin{align}
\mathcal{A}_{ij}
&=
\frac{i}{2}
\sum_{\mu,\nu=\uparrow,\downarrow}
b_{i\mu}\sigma^y_{\mu\nu}b_{j\nu},
\label{eq:schwinger_boson_bond_operator_A}
\\
\mathcal{B}_{ij}
&=
\frac{1}{2}
\sum_{\mu,\nu=\uparrow,\downarrow}
b_{i\mu}^{\dagger}\sigma^0_{\mu\nu}b_{j\nu},
\label{eq:schwinger_boson_bond_operator_B}
\\
\mathcal{C}_{ij}^{\alpha}
&=
\frac{1}{2}
\sum_{\mu,\nu=\uparrow,\downarrow}
b_{i\mu}^{\dagger}\sigma^\alpha_{\mu\nu}b_{j\nu},
\label{eq:schwinger_boson_bond_operator_C}
\\
\mathcal{D}_{ij}^{\alpha}
&=
\frac{i}{2}
\sum_{\mu,\nu=\uparrow,\downarrow}
b_{i\mu}
\left(\sigma^y\sigma^\alpha\right)_{\mu\nu}
b_{j\nu},
\label{eq:schwinger_boson_bond_operator_D}
\end{align}
\end{subequations}
where $\sigma^0$ denotes the identity matrix for the spin-flavor indices.  The operators
$\mathcal{B}_{ij}$ and $\mathcal{C}_{ij}^{\alpha}$ are hopping-type channels
of the form $b_i^\dagger b_j$, while $\mathcal{A}_{ij}$ and
$\mathcal{D}_{ij}^{\alpha}$ are pairing-type channels of the form $b_i b_j$.
The pair $\mathcal{A}_{ij}$ and $\mathcal{B}_{ij}$ is invariant under
global spin rotations and corresponds to the usual Schwinger boson
singlet-pairing and hopping amplitudes.  The anisotropic pair,
$\mathcal{C}_{ij}^{\alpha}$ and $\mathcal{D}_{ij}^{\alpha}$, carries an
explicit spin-component index $\alpha=x,y,z$.  On a Kitaev $\gamma$ bond,
the matching component $\alpha=\gamma$ encodes the bond-spin locking of the Kitaev
interaction.

It is useful to collect these channels into
\begin{align}
\bm{\mathcal{Q}}_{ij}
=
\left(
\mathcal{A}_{ij},
\mathcal{B}_{ij},
\mathcal{C}_{ij}^{x},
\mathcal{C}_{ij}^{y},
\mathcal{C}_{ij}^{z},
\mathcal{D}_{ij}^{x},
\mathcal{D}_{ij}^{y},
\mathcal{D}_{ij}^{z}
\right)^{T}.
\label{eq:bond_operator_vector}
\end{align}
The set in Eq.~\eqref{eq:bond_operator_vector} is complete for rewriting
nearest-neighbor bilinear spin interactions in terms of Schwinger boson bond
channels.  More generally, a two-spin operator can be represented as
\begin{align}
S_i^\alpha S_j^{\alpha^{\prime}}
=
\sum_{p,q}
A_{pq}^{\alpha\alpha^{\prime}}
\mathopen{:}
\mathcal{Q}_{ij}^{p\dagger}
\mathcal{Q}_{ij}^{q}
\mathclose{:},
\label{eq:general_bond_operator_identity}
\end{align}
with coefficients $A_{pq}^{\alpha\alpha^{\prime}}$ fixed by the spin components and by
the chosen algebraically equivalent representation.  The present Kitaev problem
only requires the diagonal Ising component.  On a $\gamma$ bond it is expressed
as
\begin{align}
S_i^\gamma S_j^\gamma
=
\frac{1}{2}
\left[
\mathopen{:}\mathcal{B}_{ij}^{\dagger}\mathcal{B}_{ij}\mathclose{:}
-\mathcal{D}_{ij}^{\gamma\dagger}\mathcal{D}_{ij}^{\gamma}
+\mathopen{:}\mathcal{C}_{ij}^{\gamma\dagger}\mathcal{C}_{ij}^{\gamma}\mathclose{:}
-\mathcal{A}_{ij}^{\dagger}\mathcal{A}_{ij}
\right],
\label{eq:kitaev_bond_operator_identity}
\end{align}
where $\mathopen{:}O_{1}O_{2}\mathclose{:}$ denotes normal ordering.  In the
present bosonic representation this means that the boson creation operators are
moved to the left of
annihilation operators before the mean-field decoupling, so that no additional
constant from bosonic commutation relations is included.  By introducing the
SU($2$)-breaking bond operators in this way, anisotropic interactions such as
the Kitaev interaction can also be expressed as bilinear forms of bond
operators, as in Eq.~\eqref{eq:kitaev_bond_operator_identity}, so that the
mean-field prescription of SBMFT described below can be applied directly.

Substituting Eq.~\eqref{eq:kitaev_bond_operator_identity} into the pure Kitaev
Hamiltonian gives a quartic bosonic Hamiltonian, supplemented by
Eq.~\eqref{eq:schwinger_boson_constraint_term}.  SBMFT replaces this interacting
boson problem by a quadratic one by allowing the bond operators to acquire
static expectation values.  For each channel we write
\begin{align}
\mathcal{Q}_{ij}= \langle \mathcal{Q}_{ij}\rangle+\delta \mathcal{Q}_{ij},
\qquad
\mathcal{Q}_{ij}\in
\{\mathcal{A}_{ij},\mathcal{B}_{ij},
\mathcal{C}_{ij}^{\gamma},\mathcal{D}_{ij}^{\gamma}\},
\label{eq:bond_field_decomposition}
\end{align}
and neglect terms quadratic in the fluctuations.  Equivalently, a product of
two bond operators is decoupled as
\begin{align}
\mathcal{Q}_{ij}^{p\dagger}\mathcal{Q}_{ij}^{q}
\rightarrow
\langle \mathcal{Q}_{ij}^{p\dagger}\rangle
\mathcal{Q}_{ij}^{q}
+
\mathcal{Q}_{ij}^{p\dagger}
\langle \mathcal{Q}_{ij}^{q}\rangle
-
\langle \mathcal{Q}_{ij}^{p\dagger}\rangle
\langle \mathcal{Q}_{ij}^{q}\rangle .
\label{eq:bond_operator_mean_field_decoupling}
\end{align}
The complex numbers $\langle\mathcal{Q}_{ij}^{p}\rangle$ are the mean-field
parameters.  They are assigned to the oriented bonds in a chosen mean-field unit
cell and are allowed, in general, to be complex.  Gauge-inequivalent patterns of
these bond amplitudes correspond to distinct projective mean-field sectors.

The saddle point is obtained by solving the quadratic Hamiltonian
self-consistently.  If $\langle\cdots\rangle_{\mathrm{MF}}$ denotes the
expectation value in the bosonic mean-field ground state, the self-consistency
conditions are
\begin{align}
\langle \mathcal{Q}_{ij}^{p}\rangle
&=
\langle \mathcal{Q}_{ij}^{p}\rangle_{\mathrm{MF}},
&
\langle n_i\rangle_{\mathrm{MF}}
&=
2S,
\label{eq:schwinger_boson_self_consistency}
\end{align}
for all independent bond fields and sublattices in the mean-field unit cell.
In practice one starts from a chosen mean-field Ansatz, diagonalizes the resulting
quadratic Hamiltonian, recomputes the right-hand side of
Eq.~\eqref{eq:schwinger_boson_self_consistency}, and iterates until convergence.
This procedure determines both the bond amplitudes and the Lagrange multipliers.
The role of $S$ is transparent in Eq.~\eqref{eq:schwinger_boson_self_consistency}, because
changing $S$ changes the target boson density and therefore the values taken
by the saddle-point parameters, while the bond-operator algebra and the structure of the quadratic
Hamiltonian remain the same.

We denote the number of sublattices in this mean-field unit cell by $M$ and the
total number of sites by $N$, and hence the number of Bravais unit cells is $N/M$.
After Fourier transformation, we introduce the Nambu spinor
\begin{align}
\Psi_{\bm{k}}^{\dagger}
=
\Bigl(
&
b_{\bm{k},1,\uparrow}^{\dagger}\cdots
b_{\bm{k},M,\uparrow}^{\dagger},
b_{\bm{k},1,\downarrow}^{\dagger}\cdots
b_{\bm{k},M,\downarrow}^{\dagger},
\notag\\
&\quad
b_{-\bm{k},1,\uparrow}\cdots
b_{-\bm{k},M,\uparrow},
b_{-\bm{k},1,\downarrow}\cdots
b_{-\bm{k},M,\downarrow}
\Bigr),
\label{eq:schwinger_boson_nambu_spinor}
\end{align}
whose ordering follows the spin flavor, sublattice, and particle-hole
structure.  The resulting quadratic Hamiltonian is written in the bosonic
Bogoliubov-de Gennes form
\begin{align}
\mathcal{H}_{\mathrm{MF}}
=
\frac{1}{2}\sum_{\bm{k}}
\Psi_{\bm{k}}^{\dagger}
\mathcal{M}_{\bm{k}}
\Psi_{\bm{k}}
 + E_{\mathrm{MF}},
\label{eq:schwinger_boson_bdg}
\end{align}
where $\mathcal{M}_{\bm{k}}$ is a $4M\times4M$ matrix.  The Nambu spinor
therefore has $2\times 2\times M$ components, corresponding to two spin
flavors, particle-hole doubling, and $M$ sublattices.
Because the Hamiltonian is bosonic, $\mathcal{M}_{\bm{k}}$ is
diagonalized by a paraunitary, rather than unitary, transformation that preserves
the canonical bosonic commutation relations in Nambu
space~\cite{Colpa-1978}.  The positive
eigenvalues are the spinon excitation energies.  A stable spin-liquid saddle
has positive excitation energies throughout the Brillouin zone.  In ordinary
SBMFT, the softening of a bosonic mode signals Bose condensation and the onset
of magnetic order.  The same paraunitary transformation also gives the
mean-field Green's function used below in the response calculation.

In the response calculation described next, we separate the Hamiltonian
based on Klein duality into a pure-Kitaev part with coupling $K+J$ and a
remaining Klein-odd interaction.
The saddle is solved at the target value of $S$.  Changing $S$ therefore only
requires solving the same set of mean-field equations with a different
boson-number constraint.  This remaining interaction is treated
perturbatively in the RPA calculation, as described in
Sec.~\ref{sec:perturbative_rpa_response}.
This separation lets us ask how a Kitaev spin-liquid saddle becomes
unstable when the non-Kitaev part of the Kitaev-Heisenberg interaction is
introduced in the response channel, rather than allowing the saddle point
itself to immediately deform into a magnetically ordered mean-field state.

\subsection{Perturbative RPA formulation}
\label{sec:perturbative_rpa_response}

\begin{figure}[t]
\centering
\includegraphics[width=\columnwidth]{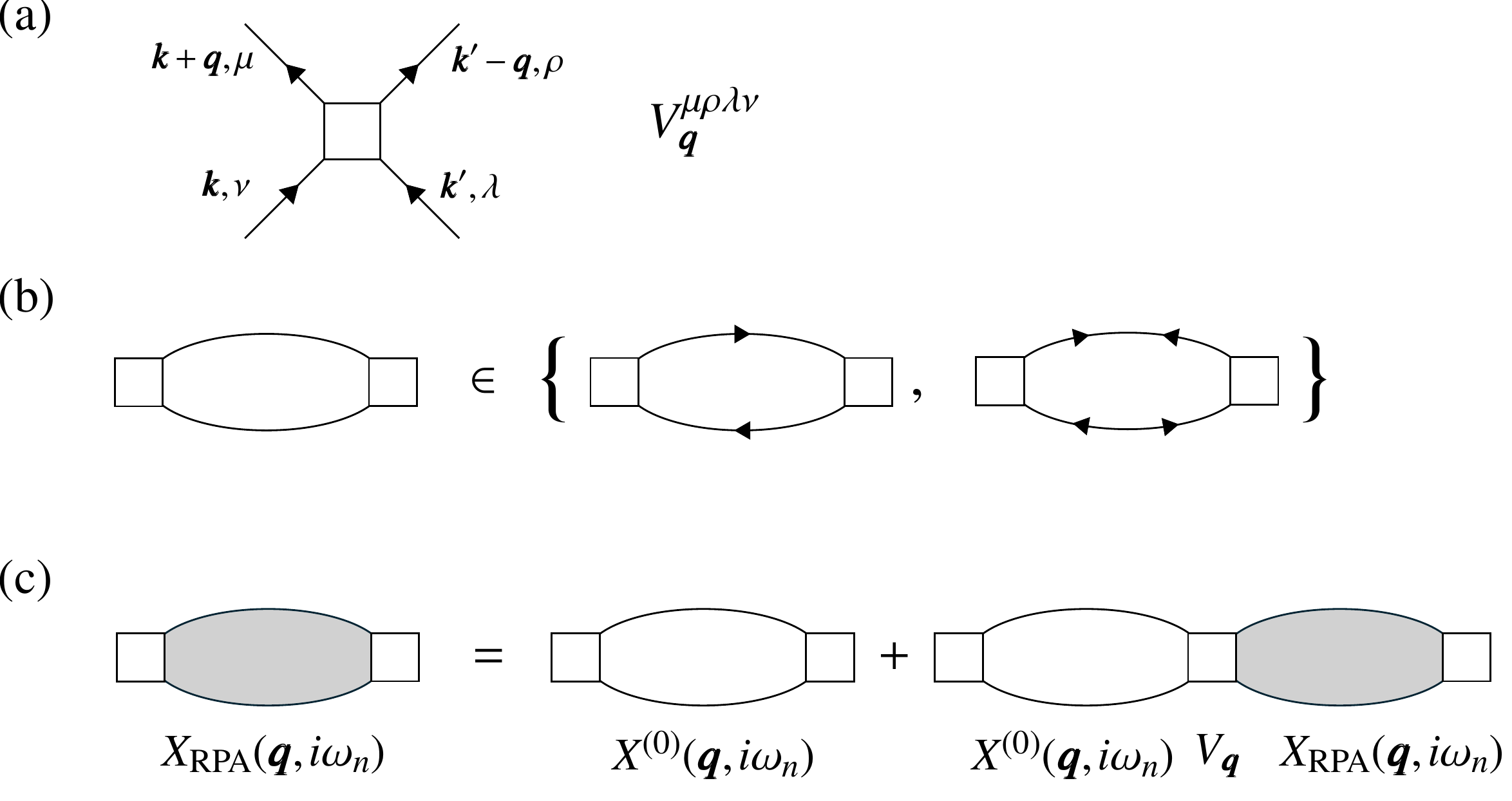}
\caption{
(a) Four-boson interaction vertex $V_{\bm q}$ obtained from the
perturbing Hamiltonian in Eq.~\eqref{eq:rpa_perturbation_hamiltonian}.
(b) Coefficient-free bare bubble
$X^{(0)}(\bm{q},i\omega_n)$ obtained from normal and anomalous Schwinger boson
Green's functions of the Kitaev SBMFT Hamiltonian.
(c) Dyson equation for the RPA-dressed internal bubble
$X_{\mathrm{RPA}}(\bm{q},i\omega_n)$.  The physical susceptibility is obtained
after attaching the external spin vertices.
}
\label{fig:RPA_response}
\end{figure}

We next describe the response calculation used to detect magnetic
instabilities of the Kitaev spin liquid.  Because our focus is on the regime
proximate to the Kitaev spin liquid, the pure-Kitaev SBMFT BdG Hamiltonian
$\mathcal{H}_{\mathrm{MF}}$ in Eq.~\eqref{eq:schwinger_boson_bdg} is taken as
the unperturbed Hamiltonian.  Near the Kitaev point, it is useful to
separate the microscopic Hamiltonian into Klein-even and Klein-odd parts.
Equation~\eqref{eq:kitaev_heisenberg_hamiltonian} can be written exactly as
\begin{align}
\mathcal H
&=
\sum_{\gamma=x,y,z}
\sum_{\langle i,j\rangle_{\gamma}}
\left[
(K+J)S_i^\gamma S_j^\gamma
+J\left(
\bm{\mathit{S}}_i\cdot\bm{\mathit{S}}_j
-S_i^\gamma S_j^\gamma
\right)
\right].
\label{eq:klein_even_odd_decomposition}
\end{align}
\red{The first term in the square brackets has the form of a pure Kitaev
Hamiltonian and defines the SBMFT saddle point.  This Klein-even part is
invariant under the Klein transformation.  The remaining term is}
\begin{align}
\mathcal{H}_{\mathrm{pert}}
&=
\sum_{\gamma=x,y,z}
\sum_{\langle i,j\rangle_{\gamma}}
J\left[
\bm{\mathit{S}}_i\cdot\bm{\mathit{S}}_j
-S_i^\gamma S_j^\gamma
\right].
\label{eq:rpa_perturbation_hamiltonian}
\end{align}
\red{This term changes sign under the Klein transformation,
$\mathcal{H}_{\mathrm{pert}}\to-\mathcal{H}_{\mathrm{pert}}$, and we regard
this Klein-odd part as the perturbation entering the RPA calculation.}
The Fourier transform of
Eq.~\eqref{eq:rpa_perturbation_hamiltonian} gives the four-boson RPA vertex
$V_{\bm q}$ shown in Fig.~\ref{fig:RPA_response}(a), with its explicit matrix
form summarized in Appendix~\ref{app:rpa_calculation}.  Accordingly, all expectation values and
imaginary-time evolution in this subsection are evaluated with respect to
$\mathcal{H}_{\mathrm{MF}}$.  Following the Matsubara
formulation of spin correlations used in Schwinger boson
theory~\cite{Arovas-Auerbach-1998,Sasamoto-Nasu-2025} and recent RPA-type
treatments of dynamical responses in parton theories, including the
generalized bilinear response formulation~\cite{Rao-Moessner-Knolle-2025,
Sasamoto-Nasu-2026,Willsher-Knolle-2025}, we first
introduce the imaginary-time Heisenberg representation with respect to
$\mathcal{H}_{\mathrm{MF}}$.
\begin{align}
b_{i\mu}(\tau)
&=
e^{\tau\mathcal{H}_{\mathrm{MF}}}
b_{i\mu}
e^{-\tau\mathcal{H}_{\mathrm{MF}}},
&
\bar{b}_{i\mu}(\tau)
&=
e^{\tau\mathcal{H}_{\mathrm{MF}}}
b_{i\mu}^{\dagger}
e^{-\tau\mathcal{H}_{\mathrm{MF}}}.
\label{eq:imaginary_time_bosons}
\end{align}
Here $0\leq\tau<\beta$, with $\beta=1/T$ the inverse temperature.
\red{The operator $\bar b_{i\mu}(\tau)$ denotes the imaginary-time-evolved
creation operator and is not identical to the Hermitian conjugate of
$b_{i\mu}(\tau)$ at the same imaginary time.}  The spin
operator in this representation is
\begin{align}
S_i^\alpha(\tau)
=
\frac{1}{2}
\sum_{\mu,\nu=\uparrow,\downarrow}
\bar{b}_{i\mu}(\tau)
\sigma^\alpha_{\mu\nu}
b_{i\nu}(\tau).
\label{eq:imaginary_time_spin_operator}
\end{align}
Here $\alpha=x,y,z$ labels the spin component of the diagonal susceptibility,
while $\gamma$ is reserved for the Kitaev bond direction.  We keep this
spin-component label distinct from $\gamma$ to avoid identifying a spin component
with a bond label.

The diagonal bare Matsubara susceptibility for spin component $\alpha$
is then defined as the spin correlation function evaluated with respect to
$\mathcal{H}_{\mathrm{MF}}$,
\begin{align}
\left[\chi^{(0)\alpha\alpha}(\bm{q},i\omega_n)\right]_{ab}
=&
\frac{M}{N}
\sum_{\bm{R},\bm{R}^{\prime}}
e^{-i\bm{q}\cdot
\left(\bm{r}_{\bm{R}a}-\bm{r}_{\bm{R}^{\prime}b}\right)}
\nonumber\\
&\times
\int_{0}^{\beta}d\tau\,e^{i\omega_n\tau}
\left\langle
\mathcal{T}_{\tau}
S_{\bm{R}a}^{\alpha}(\tau)
S_{\bm{R}^{\prime}b}^{\alpha}(0)
\right\rangle,
\label{eq:bare_matsubara_susceptibility}
\end{align}
where $a,b=1,\ldots,M$ denote sublattices in the mean-field unit cell,
$\bm{r}_{\bm{R}a}$ is the position of the site on sublattice $a$ in the unit
cell specified by the Bravais lattice vector $\bm{R}$, and
$\omega_n=2\pi n/\beta$ is a bosonic Matsubara frequency with integer
$n$.  The saddle point has
no static magnetic moment, and hence the disconnected one-point contribution is
absent.  Since $S_i^\alpha(\tau)$ is bilinear in Schwinger bosons and
$\mathcal{H}_{\mathrm{MF}}$ is quadratic,
Eq.~\eqref{eq:bare_matsubara_susceptibility} is evaluated as products of
bosonic two-point correlation functions.
Both normal and anomalous two-point correlation functions contribute, because
the bosonic BdG Hamiltonian contains pairing terms.

We use the following time-split two-point functions
\begin{subequations}
\label{eq:time_split_bond_contractions}
\begin{align}
\mathscr{A}_{ij}(\tau)
&=
\frac{i}{2}
\sum_{\mu,\nu=\uparrow,\downarrow}
\sigma^y_{\mu\nu}
\left\langle
\mathcal{T}_{\tau}
b_{i\mu}(\tau)b_{j\nu}(0)
\right\rangle,
\label{eq:time_split_bond_contraction_A}
\\
\mathscr{B}_{ij}(\tau)
&=
\frac{1}{2}
\sum_{\mu=\uparrow,\downarrow}
\left\langle
\mathcal{T}_{\tau}
\bar{b}_{i\mu}(\tau)b_{j\mu}(0)
\right\rangle,
\label{eq:time_split_bond_contraction_B}
\\
\mathscr{C}_{ij}^{\alpha}(\tau)
&=
\frac{1}{2}
\sum_{\mu,\nu=\uparrow,\downarrow}
\sigma^\alpha_{\mu\nu}
\left\langle
\mathcal{T}_{\tau}
\bar{b}_{i\mu}(\tau)b_{j\nu}(0)
\right\rangle,
\label{eq:time_split_bond_contraction_C}
\\
\mathscr{D}_{ij}^{\alpha}(\tau)
&=
\frac{i}{2}
\sum_{\mu,\nu=\uparrow,\downarrow}
\left(\sigma^y\sigma^\alpha\right)_{\mu\nu}
\left\langle
\mathcal{T}_{\tau}
b_{i\mu}(\tau)b_{j\nu}(0)
\right\rangle.
\label{eq:time_split_bond_contraction_D}
\end{align}
\end{subequations}
The four correlators in
Eqs.~\eqref{eq:time_split_bond_contraction_A}--\eqref{eq:time_split_bond_contraction_D}
are the time-split counterparts of the bond operators in
Eqs.~\eqref{eq:schwinger_boson_bond_operator_A}--\eqref{eq:schwinger_boson_bond_operator_D},
respectively.  For the anisotropic correlators, the superscript $\alpha$
denotes the spin component selected in the susceptibility, not a bond label.
The quantities with an overbar,
$\overline{\mathscr{A}}_{ij}$,
$\overline{\mathscr{B}}_{ij}$,
$\overline{\mathscr{C}}_{ij}^{\alpha}$, and
$\overline{\mathscr{D}}_{ij}^{\alpha}$ are defined analogously as the
time-split two-point correlation functions of the Hermitian-conjugate channels
$\mathcal{A}_{ij}^{\dagger}$,
$\mathcal{B}_{ij}^{\dagger}$,
$\mathcal{C}_{ij}^{\alpha\dagger}$, and
$\mathcal{D}_{ij}^{\alpha\dagger}$, respectively.  Thus the diagonal
spin component entering the bare susceptibility is evaluated as
\begin{align}
\chi_{ij}^{\alpha\alpha,(0)}(\tau)
=
\frac{1}{2}
\Bigl[
\overline{\mathscr{B}}_{ij}(\tau)\mathscr{B}_{ij}(\tau)
-
\overline{\mathscr{D}}_{ij}^{\alpha}(\tau)\mathscr{D}_{ij}^{\alpha}(\tau)
\nonumber\\
\qquad
+
\overline{\mathscr{C}}_{ij}^{\alpha}(\tau)\mathscr{C}_{ij}^{\alpha}(\tau)
-
\overline{\mathscr{A}}_{ij}(\tau)\mathscr{A}_{ij}(\tau)
\Bigr].
\label{eq:bond_channel_bare_correlator}
\end{align}
This equation is the explicit form of the statement that the susceptibility is
computed in the bond-operator basis~\cite{Sasamoto-Nasu-2025}.  The momentum-
and frequency-dependent
$\left[\chi^{(0)\alpha\alpha}(\bm{q},i\omega_n)\right]_{ab}$ in
Eq.~\eqref{eq:bare_matsubara_susceptibility} is obtained by Fourier
transforming these imaginary-time correlators and integrating over $\tau$.

\red{For the RPA procedure, we introduce the coefficient-free bubble}
$X^{(0)}_{rs}=\langle
\mathcal{T}_\tau\mathcal{X}_r(\tau)\mathcal{X}_s(0)\rangle_c$, where
$\mathcal{X}_r$ denotes an elementary bosonic bilinear such as
\red{$\bar b b$}, $bb$, or the corresponding conjugate channel, and $r$
collects the associated internal labels.  The physical bare susceptibility is then
the external-vertex contraction
\begin{align*}
\left[\chi^{(0)\alpha\alpha}(\bm{q},i\omega_n)\right]_{ab}
&=
\sum_{r,s}
\Gamma_{r}^{(a\alpha)}
\left[X^{(0)}(\bm{q},i\omega_n)\right]_{rs}
\Gamma_{s}^{(b\alpha)}.
\end{align*}
Appendix~\ref{app:rpa_calculation} gives the explicit channel
definitions and normalization.  In the numerical implementation, the resulting
coefficient-free bubble is arranged in the 24-component internal-channel basis
$(\alpha,a)$.  The perturbation in
Eq.~\eqref{eq:rpa_perturbation_hamiltonian} defines the internal RPA vertex
$V(\bm q)$, and the Dyson equation reads
\begin{align}
X_{\mathrm{RPA}}(\bm{q},i\omega_n)
&=
X^{(0)}(\bm{q},i\omega_n)
-
X^{(0)}(\bm{q},i\omega_n)
V(\bm{q})
X_{\mathrm{RPA}}(\bm{q},i\omega_n),
\label{eq:rpa_dyson}
\end{align}
or, with the sign convention used here,
\begin{align}
X_{\mathrm{RPA}}(\bm{q},i\omega_n)
&=
\left[
\bm{1}_{24}
+
X^{(0)}(\bm{q},i\omega_n)V(\bm{q})
\right]^{-1}
X^{(0)}(\bm{q},i\omega_n).
\label{eq:rpa_solution}
\end{align}
Here $\bm{1}_{24}$ denotes the $24\times24$ identity
matrix.  \red{The physical RPA susceptibility is obtained within this
procedure by replacing $X^{(0)}$ with $X_{\mathrm{RPA}}$ in the
external-vertex contraction above.}  The explicit normalization of
$X^{(0)}$, $V$, and the final projection back to $\chi_{\mathrm{RPA}}$
is given in Appendix~\ref{app:rpa_calculation}.
\red{The real-frequency RPA susceptibility is obtained by the analytic continuation}
$i\omega_n\to\omega+i\delta$, where the positive infinitesimal is replaced by a
\red{finite value of $\delta$} in the numerical calculation.  The
24-component implementation of the perturbation in
Eq.~\eqref{eq:rpa_perturbation_hamiltonian} is detailed in
Appendix~\ref{app:rpa_calculation}.

For the static instability analysis, we monitor the static RPA susceptibility
\begin{align}
\red{\chi_{\mathrm{RPA}}(\bm{q})}
&=
\operatorname{Re}
\sum_{\alpha=x,y,z}
\sum_{a,b}
\left[
\chi_{\mathrm{RPA}}(\bm{q},\omega=0)
\right]_{a\alpha,b\alpha}.
\label{eq:static_sublattice_response}
\end{align}
\red{An instability is signaled by a strong enhancement of this quantity,
equivalently when
$\det[\bm{1}_{24}+X^{(0)}(\bm{q},0)V(\bm{q})]$ vanishes.}

\subsection{Dynamical spin structure factor}
\label{sec:dynamical_spin_structure_factor}

The quantity directly comparable with inelastic neutron-scattering spectra is
the dynamical spin structure factor.  We use a diagonal component-resolved
notation in which the spin index is written as a superscript and sublattice
indices as subscripts.  With the same sublattice convention as in
Eq.~\eqref{eq:bare_matsubara_susceptibility}, the sublattice-resolved structure
factor is defined by
\begin{align}
\left[
S^{\alpha\alpha}(\bm{q},\omega)
\right]_{ab}
=&
\frac{M}{N}
\sum_{\bm{R},\bm{R}^{\prime}}
e^{-i\bm{q}\cdot
\left(\bm{r}_{\bm{R}a}-\bm{r}_{\bm{R}^{\prime}b}\right)}
\nonumber\\
&\times
\int_{-\infty}^{\infty}\frac{dt}{2\pi}\,e^{i\omega t}
\left\langle
S_{\bm{R}a}^{\alpha}(t)
S_{\bm{R}^{\prime}b}^{\alpha}(0)
\right\rangle ,
\label{eq:dynamical_spin_structure_factor_definition}
\end{align}
where $S_i^\alpha(t)=e^{i\mathcal{H}_{\mathrm{MF}}t}S_i^\alpha
e^{-i\mathcal{H}_{\mathrm{MF}}t}$ is the real-time Heisenberg representation
with respect to the unperturbed SBMFT Hamiltonian $\mathcal{H}_{\mathrm{MF}}$.
\red{We keep the sublattice labels explicit because the RPA susceptibility matrix is}
computed as a sublattice matrix.  After the analytic continuation
$i\omega_n\to\omega+i\delta$, the fluctuation-dissipation theorem gives
\begin{align}
\left[
S^{\alpha\alpha}(\bm{q},\omega)
\right]_{ab}
=
\frac{1}{\pi}
\frac{1}{1-e^{-\beta\omega}}
\operatorname{Im}
\left[
\chi_{\mathrm{RPA}}(\bm{q},\omega+i\delta)
\right]_{a\alpha,b\alpha}.
\label{eq:dssf_chi_relation}
\end{align}
\red{In the numerical calculation, $\delta$ is kept finite as the
analytic-continuation parameter.}
Since the starting SBMFT state
has no static magnetic moment, no elastic Bragg contribution is present in
Eq.~\eqref{eq:dynamical_spin_structure_factor_definition}.

\section{Results}
\label{sec:results}

Throughout this section, we focus exclusively on the ground state and evaluate
all quantities in the zero-temperature limit $T\to0$, or equivalently
$\beta\to\infty$.

\subsection{Mean-field Ansatz}
\label{sec:mean_field_ansatz}

\begin{figure}[t]
\centering
\includegraphics[width=\columnwidth]{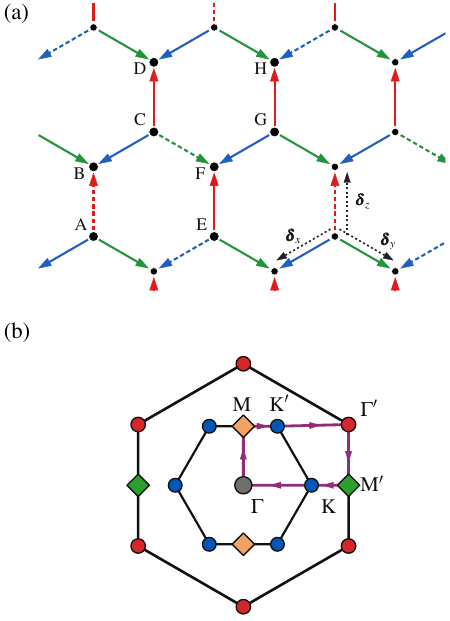}
\caption{
\red{(a) Real-space mean-field pattern of the $\pi/2$-flux Ansatz in the eight-site
unit cell.  Blue, green, and red arrows denote the directed $x$,
$y$, and $z$ bonds, respectively.  Each arrow fixes the ordered pair
$(i,j)$, from site $i$ to site $j$, used for
all bond operators and for $\eta_{ij}^{\gamma}$.  Solid directed bonds have
$\eta_{ij}^{\gamma}=1$, whereas dashed directed bonds have
$\eta_{ij}^{\gamma}=i$.}  The offset black dotted arrows indicate
the directed bond vectors $\bm{\delta}_{x}$, $\bm{\delta}_{y}$, and
$\bm{\delta}_{z}$ defined in Eq.~\eqref{eq:appendix_rpa_directed_bonds}.
The sublattices are labeled
$\mathrm{A}$--$\mathrm{H}$ in the order used in the mean-field Hamiltonian.
(b) First and extended Brillouin zones with the high-symmetry symbols defined
in Fig.~\ref{fig:Kitaev_Heisenberg}(b).  Purple arrows show the path
$\mathrm{K}\!\rightarrow\!\Gamma\!\rightarrow\!\mathrm{M}
\!\rightarrow\!\mathrm{K}^{\prime}\!\rightarrow\!\Gamma^{\prime}
\!\rightarrow\!\mathrm{M}^{\prime}\!\rightarrow\!\mathrm{K}$ used for the
dynamical spin structure factor.
}
\label{fig:mean_field_ansatz}
\end{figure}

For the numerical calculations, we choose the $\pi/2$-flux SBMFT saddle point
of the pure Kitaev model~\cite{Ralko-Merino-2024,Sasamoto-Nasu-2025,
Sasamoto-Ralko-Merino-Nasu-2026}.
\red{In this Ansatz, the only nonzero intersite mean-field channel is the
anisotropic pairing channel on each directed $\gamma$ bond.  In all bond
operators, $\mathcal{A}_{ij}$, $\mathcal{B}_{ij}$,
$\mathcal{C}_{ij}^{\gamma}$, and $\mathcal{D}_{ij}^{\gamma}$, the ordered pair
$(i,j)$ follows the arrow direction in Fig.~\ref{fig:mean_field_ansatz}(a),
from site $i$ to site $j$.  The intersite
$\mathcal{A}_{ij}$, $\mathcal{B}_{ij}$, and
$\mathcal{C}_{ij}^{\gamma}$ fields vanish in the self-consistent solution, and the remaining field is
parametrized as
$\langle\mathcal{D}_{ij}^{\gamma}\rangle
=\bar{D}\eta_{ij}^{\gamma}$, with
$|\eta_{ij}^{\gamma}|=1$.  Here $\bar{D}$ is a real, bond-independent amplitude determined
self-consistently together with the Lagrange multipliers for each $S$, whereas
$\eta_{ij}^{\gamma}$ is a fixed phase factor specifying the $\pi/2$-flux
mean-field pattern.  Thus all nonzero intersite
$\langle\mathcal{D}_{ij}^{\gamma}\rangle$ fields have the same magnitude and differ only by
this bond-dependent phase.}
Recent comparisons of the $0$-flux and $\pi/2$-flux Ans\"atze show that the
$\pi/2$-flux Ansatz suppresses longer-distance spin correlations more strongly
and yields a real-space correlation pattern closer to the short-ranged form
expected in the Kitaev limit~\cite{Sasamoto-Ralko-Merino-Nasu-2026}.
We therefore use it as the reference spin-liquid saddle for the static and
dynamical response calculations.
In the pure Kitaev model for arbitrary $S$, local conserved quantities
impose an exact constraint on spin correlations~\cite{Baskaran-Sen-Shankar-2008}.
\red{Equivalently, apart from onsite terms,
$\langle S_i^\alpha S_j^\beta\rangle$ vanishes unless $i$ and $j$
are connected by a $\gamma$ bond and $\alpha=\beta=\gamma$.}

For the pure-Kitaev reference state, we follow the short-range-dynamics
prescription of Ref.~\cite{Sasamoto-Ralko-Merino-Nasu-2026}, retaining only the
onsite terms and these bond-resolved nearest-neighbor contributions.  The
perturbative RPA calculation around this reference state is described separately
in Appendix~\ref{app:rpa_calculation}.
\red{Figure~\ref{fig:mean_field_ansatz}(a) specifies the resulting mean-field
pattern.  Within the eight-site cell, the independent directed bonds with
$\eta_{ij}^{\gamma}=i$ are}
\begin{align}
\eta_{\mathrm{A}\mathrm{B}}^{z}
=\eta_{\mathrm{C}\mathrm{F}}^{y}
=\eta_{\mathrm{E}\mathrm{H}}^{x}=i,
\label{eq:pi_over_two_flux_factors}
\end{align}
whereas \red{the remaining independent directed bonds have
$\eta_{ij}^{\gamma}=1$.}
The eight-site cell corresponds to $M=8$ in
Eq.~\eqref{eq:schwinger_boson_nambu_spinor}, so that the Nambu spinor
$\Psi_{\bm{k}}$ has 32 components.
For the dynamical response, we present the spectra obtained along the path
$\mathrm{K}\!\rightarrow\!\Gamma\!\rightarrow\!\mathrm{M}
\!\rightarrow\!\mathrm{K}^{\prime}\!\rightarrow\!\Gamma^{\prime}
\!\rightarrow\!\mathrm{M}^{\prime}\!\rightarrow\!\mathrm{K}$,
which consists of the six directed segments shown in
Fig.~\ref{fig:mean_field_ansatz}(b).

\subsection{Static instabilities and phase diagram}
\label{sec:static_instabilities_and_phase_diagram}

\begin{figure}[t]
\centering
\includegraphics[width=\columnwidth]{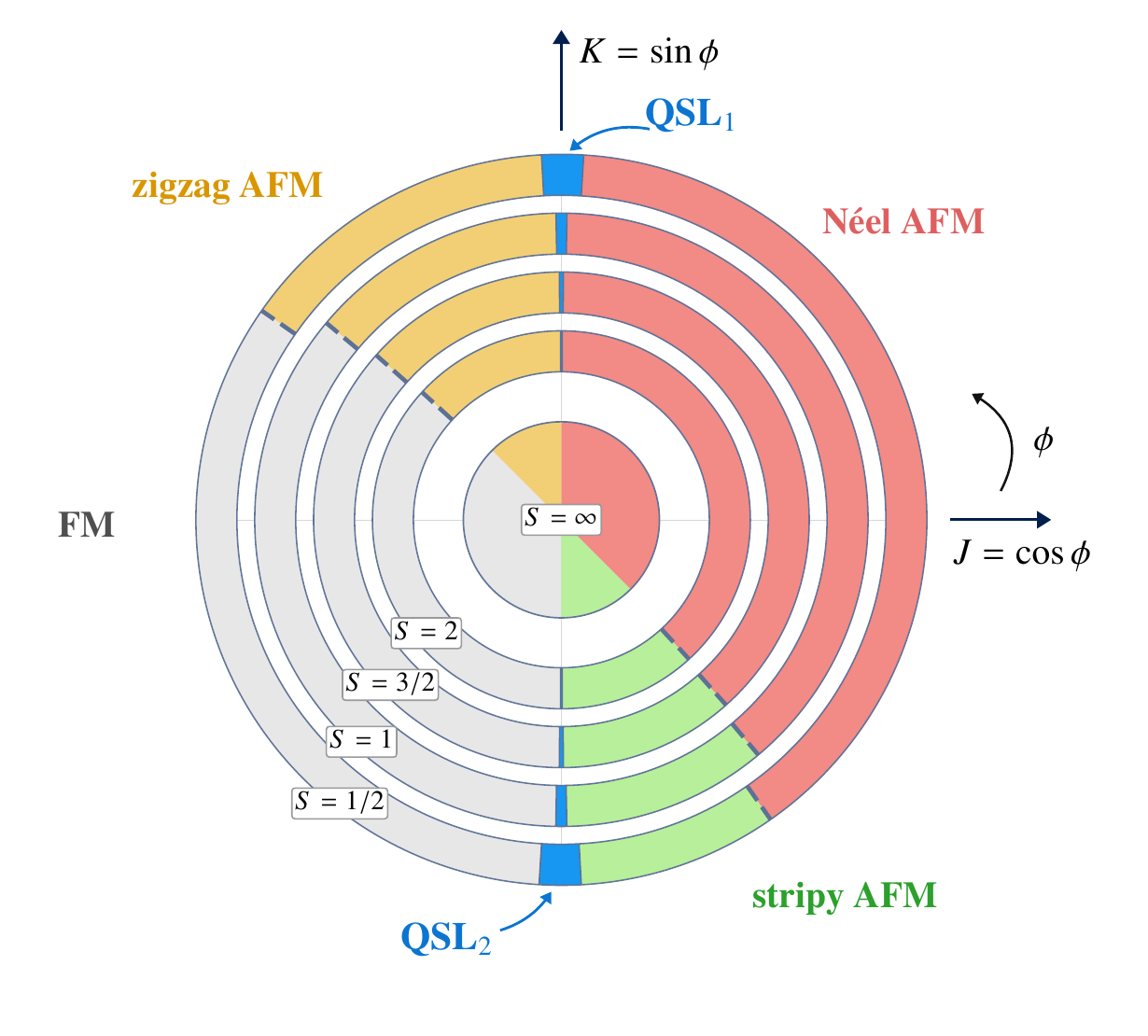}
\caption{
Ground-state phase diagram of the spin-$S$ Kitaev-Heisenberg model parametrized
by $J=\cos\phi$ and $K=\sin\phi$.  The central disk represents the classical
$S=\infty$ limit, while the concentric annuli correspond to $S=2$, $3/2$, $1$,
and $1/2$.  The blue QSL boundaries are obtained from the static RPA
instabilities at $\delta=0.01$.  The remaining finite-$S$ ordered-state
boundaries are supplied by a calculation based on linear spin-wave theory,
as described in Appendix~\ref{app:linear_spin_wave}, and are drawn as dashed radial lines.
}
\label{fig:phase_diagram}
\end{figure}

\begin{figure}[t]
\centering
\includegraphics[width=\columnwidth]{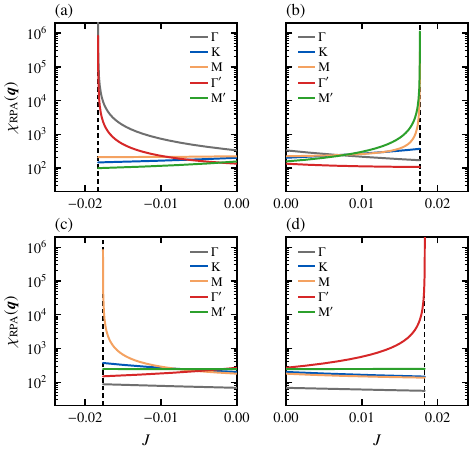}
\caption{
Static RPA \red{susceptibility} for $S=1$ and $\delta=0.01$ as a function of the
Heisenberg coupling $J$.  Panels show (a) FM Kitaev with FM Heisenberg, (b) FM Kitaev
with AFM Heisenberg, (c) AFM Kitaev with FM Heisenberg, and (d) AFM Kitaev
with AFM Heisenberg interactions.  The curves show
\red{$\chi_{\mathrm{RPA}}(\bm{q})$} defined in
Eq.~\eqref{eq:static_sublattice_response} at the five high-symmetry wave
vectors indicated in Fig.~\ref{fig:Kitaev_Heisenberg}(b).  Black dashed
vertical lines mark the critical couplings.
}
\label{fig:static_instabilities}
\end{figure}

Figure~\ref{fig:phase_diagram} shows the ground-state phase diagram of the
spin-$S$ Kitaev-Heisenberg model obtained in this work.  The phase diagram is
parametrized by the angle $\phi$ in
Eq.~\eqref{eq:kitaev_heisenberg_parametrization}.  The central disk represents
the classical $S=\infty$ limit, and the concentric annuli correspond to $S=2$,
$3/2$, $1$, and $1/2$.  The blue boundaries are the QSL boundaries determined
from the instabilities of the static RPA susceptibility.  QSL$_1$ and QSL$_2$ are
centered at the antiferromagnetic and ferromagnetic Kitaev points,
$\phi/\pi=1/2$ and $3/2$, respectively.  The remaining regions are the FM,
zigzag AFM, N\'eel AFM, and stripy AFM magnetically ordered phases, and the
finite-$S$ boundaries between these magnetically ordered phases are obtained
from the calculation based on linear spin-wave theory described in
Appendix~\ref{app:linear_spin_wave} (dashed radial lines).  In determining
the QSL boundaries, \red{we choose the analytic-continuation parameter
$\delta=0.01$.}  We also
checked the dependence on this numerical parameter using a substantially
smaller $\delta$ and confirmed that the phase boundaries are well converged
at $\delta=0.01$.  The RPA vertex is obtained directly from the
perturbing part of the Hamiltonian decomposition in
Eq.~\eqref{eq:rpa_perturbation_hamiltonian}.  The size of each QSL
region, measured by its extent in $\phi$, decreases
monotonically with increasing $S$.  A finite QSL region remains for every spin
examined here, up to and including
$S=2$, but the $S=2$ region is very narrow and its boundaries lie close to the
pure Kitaev points.  Accordingly, within the present analysis, a small
Heisenberg perturbation is sufficient to destabilize the $S=2$ QSL.

For $S=1$, the four static susceptibilities used to locate the QSL boundaries are
shown in Fig.~\ref{fig:static_instabilities}.  In panel (a), the $\Gamma$ and
$\Gamma^{\prime}$ susceptibilities increase sharply on approaching the boundary of
the FM Kitaev QSL on the FM-Heisenberg side.  In panel (b), the
$\mathrm{M}$ and $\mathrm{M}^{\prime}$ susceptibilities increase sharply at the
boundary on the AFM-Heisenberg side.  The corresponding enhancements for the
AFM Kitaev QSL occur at $\mathrm{M}$ on the FM-Heisenberg side [panel (c)] and
at $\Gamma^{\prime}$ on the AFM-Heisenberg side [panel (d)].  These wave
vectors coincide with those assigned to the FM, stripy AFM, zigzag AFM, and
N\'eel AFM ordering patterns in Fig.~\ref{fig:Kitaev_Heisenberg}(c).

\subsection{\texorpdfstring{Dynamical response for the spin-$1$ Kitaev spin liquid}{Dynamical response for the spin-1 Kitaev spin liquid}}
\label{sec:dynamical_response_spin_one}

An important advantage of SBMFT supplemented by perturbative RPA is that the
same quasiparticle framework used to diagnose static instabilities also gives
direct access to the momentum- and frequency-dependent dynamical spin
response.  In this
formulation, \red{the RPA susceptibility is constructed from the Schwinger
boson Green's function and analytically continued to real frequency within this
procedure}, allowing the spectral weight to be followed continuously along a
chosen momentum path.
This makes it possible to track the evolution of the excitation continuum as a
magnetic instability is approached, which is difficult to access from methods
designed primarily to determine ground-state energies or static observables.
This direct access is a major strength of the present approach, which
deals directly with the elementary spinon excitations.

Although the formalism developed above applies to general $S$, we focus here
on $S=1$ for the dynamical calculation.  The even--odd effect established for
the higher-spin Kitaev model assigns fermionic $\mathbb{Z}_2$ gauge charges to
half-integer spins and bosonic $\mathbb{Z}_2$ gauge charges to integer
spins~\cite{Ma-2023}.  While this statement concerns the statistics of the
gauge charge rather than the complete excitation spectrum, it provides a
microscopic motivation for expecting a bosonic-parton description to be
particularly appropriate in the integer-spin sector.  The spin-$1$ model is
the smallest such system and therefore provides the most direct setting for
examining the dynamical content of SBMFT.  Consistently, tensor-network
calculations of the spin-$1$ Kitaev model have identified bosonic excitation
structure and the lower edge of the two-particle continuum~\cite{Chen-2022},
while Schwinger boson studies of the spin-$1$ Kitaev model and the integer-spin
Kitaev-Heisenberg model have obtained bosonic QSL descriptions and calculated
their dynamical spin structure factors~\cite{Sasamoto-Nasu-2025,
Ralko-Merino-2024,Sasamoto-Ralko-Merino-Nasu-2026}.
We therefore use $S=1$ to illustrate the dynamical response of the present
general-$S$ construction.

\begin{figure*}[t]
\centering
\includegraphics[width=2\columnwidth]{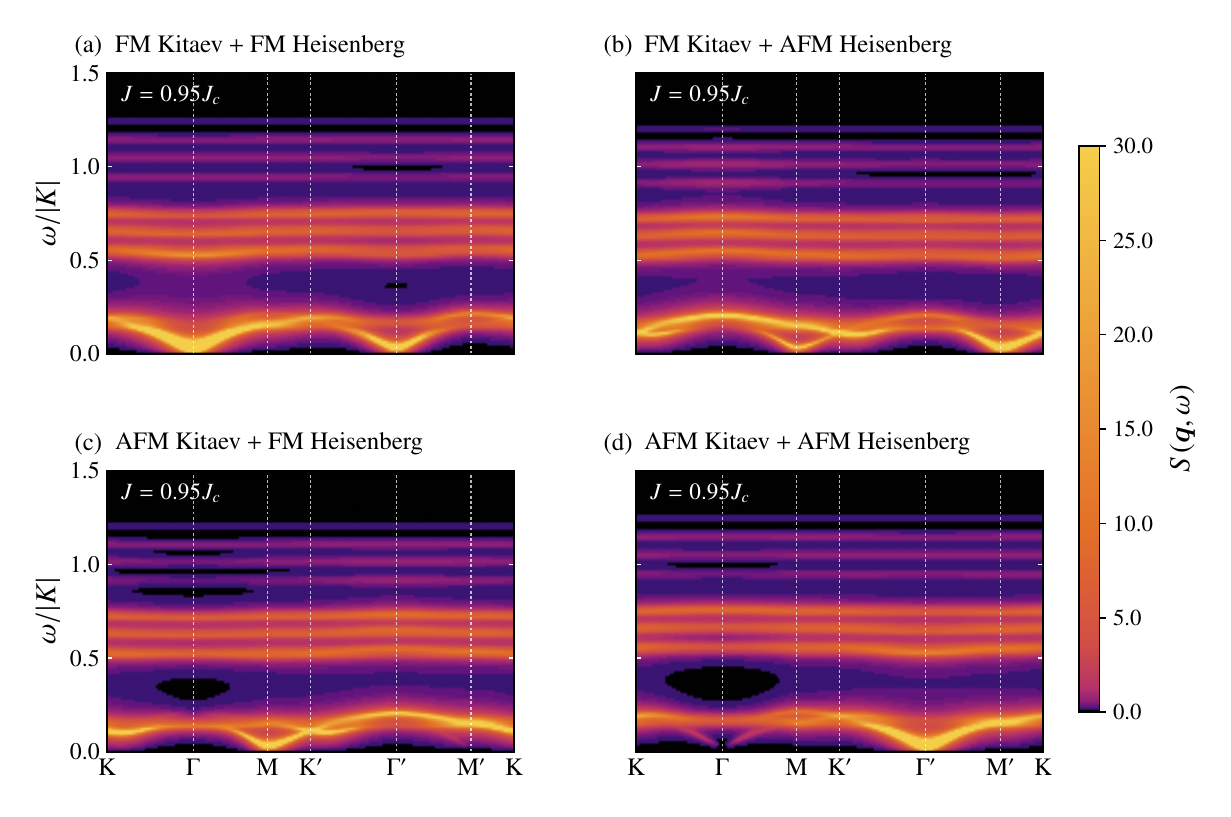}
\caption{
Dynamical spin structure factor \red{$S(\bm{q},\omega)$} in the ground state for $S=1$,
computed with $\delta=0.01$ and $J=0.95J_c$, where $J_c$ is the
corresponding static critical coupling.  Panels show (a) FM Kitaev with FM Heisenberg,
(b) FM Kitaev with AFM Heisenberg, (c) AFM Kitaev with FM Heisenberg,
and (d) AFM Kitaev with AFM Heisenberg interactions.  The momentum path is
    $\mathrm{K}$--$\Gamma$--$\mathrm{M}$--$\mathrm{K}^{\prime}$--$\Gamma^{\prime}$--
$\mathrm{M}^{\prime}$--$\mathrm{K}$.
}
\label{fig:dynamical_response}
\end{figure*}

Along the path
$\mathrm{K}\!\rightarrow\!\Gamma\!\rightarrow\!\mathrm{M}
\!\rightarrow\!\mathrm{K}^{\prime}\!\rightarrow\!\Gamma^{\prime}
\!\rightarrow\!\mathrm{M}^{\prime}\!\rightarrow\!\mathrm{K}$,
\red{we evaluate $S(\bm{q},\omega)$ by tracing over the diagonal spin
components as}
\begin{align}
S(\bm{q},\omega)
&=
\frac{1}{M}
\sum_{\alpha=x,y,z}
\sum_{a,b=1}^{M}
\left[
S^{\alpha\alpha}(\bm{q},\omega)
\right]_{ab}
\nonumber\\
&=
\frac{1}{M}
\frac{1}{\pi}
\frac{1}{1-e^{-\beta\omega}}
\sum_{\alpha=x,y,z}
\sum_{a,b=1}^{M}
\operatorname{Im}
\left[
\chi_{\mathrm{RPA}}(\bm{q},\omega+i\delta)
\right]_{a\alpha,b\alpha}.
\label{eq:plotted_dynamical_spin_structure_factor}
\end{align}
\red{The prefactor $1/M$ is introduced to normalize the dynamical spin structure factor per site.}

Figure~\ref{fig:dynamical_response} presents this ground-state quantity for
$S=1$, with $\delta=0.01$ and $J=0.95J_c$.  The same perturbing
Hamiltonian in Eq.~\eqref{eq:rpa_perturbation_hamiltonian} is used, and $J_c$ is the
corresponding static critical coupling for each panel.
Panels (a)--(d) correspond,
respectively, to FM Kitaev with FM Heisenberg, FM Kitaev with AFM Heisenberg,
AFM Kitaev with FM Heisenberg, and AFM Kitaev with AFM Heisenberg
interactions.  All four spectra exhibit a broad continuum over a substantial
part of the displayed energy range, which extends over the entire momentum
path with little dispersion.  In
the present Schwinger boson description, this signal is the two-spinon
continuum formed by pairs of fractionalized bosonic excitations.  The
continuum remains visible throughout the momentum path after the RPA
dressing.

In the low-energy region, pronounced softening is observed at $\Gamma$ and
$\Gamma^{\prime}$ in panel (a), at $\mathrm{M}$ and
$\mathrm{M}^{\prime}$ in panel (b), at $\mathrm{M}$ in panel (c), and at
$\Gamma^{\prime}$ in panel (d).  These are the same wave vectors at which the
corresponding zero-frequency susceptibilities increase on approaching the static
instabilities in Fig.~\ref{fig:static_instabilities}.  No comparably sharp
low-energy enhancement is present at $\mathrm{K}$ or
$\mathrm{K}^{\prime}$ in any of the four panels.  Among the two softened
wave vectors in panel (a), the enhancement at $\Gamma$ is stronger than that
at $\Gamma^{\prime}$, whereas in panel (b) the enhancement at
$\mathrm{M}^{\prime}$ is stronger than that at $\mathrm{M}$.  Thus, the
low-energy softening remains localized at specific wave vectors, while the
finite-energy continuum persists over the entire momentum path in all four
panels.

\section{Discussion}
\label{sec:discussion}

\subsection{Phase boundaries and comparison with previous studies}
\label{sec:comparison_with_previous_studies}

The principal static trend is the rapid shrinkage of both QSL regions with
increasing spin length.  Within the Schwinger boson formulation, $S$ enters
through the local boson-number constraint and the self-consistent saddle-point
parameters, whereas the form of the RPA instability criterion is unchanged.
The same construction can therefore be applied throughout the spin sequence
without introducing a different diagnostic at each $S$.  The resulting loss
of stability is consistent with the approach toward the semiclassical limit,
in which magnetic order becomes increasingly
competitive~\cite{Fukui-Kato-Nasu-Motome-2022,Georgiou-2024}.  Whether
such a narrow QSL region survives at larger $S$ can depend on the choice of
SBMFT saddle point.

\begin{table}[t]
\caption{
Sizes of the QSL$_1$ and QSL$_2$ regions, measured in degrees by their full
extent in the parameter angle $\phi$ defined by $K=\sin\phi$ and
$J=\cos\phi$, and grouped by spin length.
Only the absolute region sizes are tabulated.  The
abbreviations denote the coupled cluster method (CCM), exact diagonalization
(ED), cluster mean-field theory (CMFT), pseudofermion functional
renormalization group (PFFRG), and infinite-density-matrix renormalization
group (iDMRG).  References for previous results are given in the first column.
}
\label{tab:qsl_region_size_comparison}
\begin{ruledtabular}
\begin{tabular}{lcc}
Method & QSL$_1$ & QSL$_2$ \\
\colrule
\multicolumn{3}{c}{$S=1/2$} \\
Present work & $6.63^\circ$ & $6.63^\circ$ \\
CCM~\cite{Georgiou-2024}
             & $4.83^\circ$  & $4.46^\circ$ \\
ED~\cite{Gotfryd-2017}
             & $2.16^\circ$  & $16.38^\circ$ \\
CMFT~\cite{Gotfryd-2017}
             & $1.62^\circ$  & $7.38^\circ$ \\
PFFRG~\cite{Fukui-Kato-Nasu-Motome-2022}
             & $6.32^\circ$  & $37.55^\circ$ \\
\colrule
\multicolumn{3}{c}{$S=1$} \\
Present work & $2.06^\circ$ & $2.06^\circ$ \\
CCM~\cite{Georgiou-2024}
             & $1.80^\circ$  & $1.80^\circ$ \\
iDMRG~\cite{Dong-Sheng-2020}
             & $2.16^\circ$  & $5.22^\circ$ \\
PFFRG~\cite{Fukui-Kato-Nasu-Motome-2022}
             & $4.50^\circ$  & $11.37^\circ$ \\
\colrule
\multicolumn{3}{c}{$S=3/2$} \\
Present work & $0.96^\circ$ & $0.96^\circ$ \\
CCM~\cite{Georgiou-2024}
             & $1.08^\circ$  & $1.08^\circ$ \\
PFFRG~\cite{Fukui-Kato-Nasu-Motome-2022}
             & $2.70^\circ$  & $6.32^\circ$ \\
\colrule
\multicolumn{3}{c}{$S=2$} \\
Present work & $0.53^\circ$ & $0.53^\circ$ \\
\end{tabular}
\end{ruledtabular}
\end{table}

Table~\ref{tab:qsl_region_size_comparison} compares the sizes of the two QSL
regions with previous numerical estimates, where each size is measured by the
full extent in the parameter angle $\phi$ introduced in
Eq.~\eqref{eq:kitaev_heisenberg_parametrization}.  We use this table
to compare the absolute scale of the QSL regions and its spin
dependence.  The equal sizes of the present QSL$_1$ and QSL$_2$ regions follow
from the Hamiltonian decomposition in
Eq.~\eqref{eq:klein_even_odd_decomposition}.  The comparison below therefore
focuses on the spin-length dependence of the QSL-region size.
We note that the CCM study of
Ref.~\cite{Georgiou-2024} treated only $S=1/2$, $1$, and $3/2$, so that the
$S=2$ value in Table~\ref{tab:qsl_region_size_comparison} has no CCM counterpart.
The present $S=2$ result constitutes an extension of this trend to larger
spin.

The comparison is notable because the two calculations approach the boundary
from opposite sides.  CCM is formulated directly in the thermodynamic limit
using magnetically ordered reference states, and identifies the QSL boundaries
from termination points of the truncated CCM equations before the
relevant ordered-state energy crossings~\cite{Georgiou-2024}.  By contrast,
the present calculation starts from an SBMFT spin-liquid saddle point, rather
than from an ordered reference state, and identifies the boundary from the
softening of the RPA \red{susceptibility}.  The common decrease in the sizes of the
QSL regions with $S$ is therefore obtained from complementary instability
criteria.

Several other methods, including ED and CMFT for $S=1/2$, iDMRG for
$S=1$, and PFFRG for the reported spin values, find a substantially broader
QSL$_2$ region around the ferromagnetic Kitaev
point~\cite{Gotfryd-2017,Fukui-Kato-Nasu-Motome-2022,Dong-Sheng-2020}.

A useful point of comparison with CCM is that both calculations work
directly in the thermodynamic limit and locate the boundary from an
instability of a reference state, rather than from finite-size level
crossings.  In CCM, the
QSL boundaries are extracted from termination points of the CCM equations
obtained from expansions around the neighboring quasiclassical ordered
states~\cite{Georgiou-2024}.  In the present SBMFT+RPA calculation, the
RPA \red{susceptibility} on the spin-liquid saddle diverges at zero frequency, equivalently
signaled
by the vanishing of
\red{$\det[\bm{1}_{24}+X^{(0)}V]$}.  Both criteria are
thus controlled by the soft collective mode at the boundary.
The Klein-even/Klein-odd decomposition in
Eq.~\eqref{eq:klein_even_odd_decomposition} \red{keeps the RPA construction
consistent with Klein duality}, while the actual boundaries are set by
\red{zeros of the denominator of the RPA susceptibility}.
Thus the common trend across methods is the monotonic shrinkage of the QSL
regions with increasing spin, while the relative sizes of QSL$_1$ and QSL$_2$
remain method dependent.

\subsection{Dynamical signatures of the magnetic instability}
\label{sec:discussion_dynamical_signatures}

The static and dynamical calculations are linked by the same RPA-denominator
structure.
\red{The phase boundary is detected when the zero-frequency RPA denominator
$\det[\bm{1}_{24}+X^{(0)}V]$ approaches zero, while its finite-frequency
continuation determines the spectral evolution.}  The agreement
between the soft wave vectors in Figs.~\ref{fig:static_instabilities} and
\ref{fig:dynamical_response} is therefore an internal consistency check of the
construction, since the low-energy spectral weight accumulates in precisely the
channels associated with the adjacent FM, stripy AFM, zigzag AFM, and N\'eel
AFM states.  The absence of analogous softening at $\mathrm{K}$ and
$\mathrm{K}^{\prime}$ further shows that the enhancement is selective in
momentum space rather than a uniform increase of the low-energy response.

At the same time, the finite-energy continuum survives along the full momentum
path as the instability is approached.  Because a spin operator is bilinear in
Schwinger bosons, the bare response already contains a two-spinon continuum.
\red{Through the RPA procedure, this spectral weight is redistributed and a
strongly enhanced low-energy channel appears without removing the broad
background.}  The
resulting coexistence provides a useful dynamical picture of the boundary, in
which incipient magnetic order is concentrated in a specific soft channel,
while the higher-energy response retains the continuum inherited from the
spin-liquid saddle.  This simultaneous access to the ordering tendency and
the surrounding continuum is a central advantage of treating static and
dynamical correlations in one parton framework.

\red{This continuum can be regarded as a two-spinon continuum because, in the
Schwinger boson mean-field description used here, the spin operator creates
pairs of bosonic spinons.}  Gauge fluctuations, spinon self-energies, and the feedback of
the growing collective mode on the single-particle propagator are not included
at the present RPA level.  The spectra therefore establish how the chosen SBMFT
spin-liquid saddle loses stability, but they do not determine the critical
exponents or the excitation spectrum inside the ordered phase.

Further improvements would require allowing the saddle point to evolve with
the perturbation and incorporating self-energy and vertex corrections beyond
RPA.  Whether a different spin-liquid saddle intervenes before the magnetic
instability is reached would also need to be examined by comparing multiple
projective-symmetry-allowed Ans\"atze~\cite{Wang-2010}.

\section{Summary}
\label{sec:summary}

We have developed a Schwinger boson perturbative framework for the spin-$S$
Kitaev-Heisenberg model in which the SBMFT spin-liquid saddle of the pure
Kitaev model serves as the unperturbed state.  We used the $\pi/2$-flux
QSL Ansatz as the reference saddle throughout the present calculation.
We separated the
Kitaev-Heisenberg Hamiltonian based on Klein duality into a pure-Kitaev part
and a remaining Klein-odd perturbation, and used \red{the latter} as the RPA
vertex.  We determined the
QSL boundaries for $S=1/2$, $1$, $3/2$, and $2$.  The remaining ordered-state
boundaries were completed using linear spin-wave theory.  We found that the QSL
\red{regions shrink rapidly with increasing $S$ and become very narrow at
$S=2$.}  For $S=1$, the calculated dynamical spin structure factor further
shows that momentum-selective low-energy softening develops in the ordering
channels identified by the static susceptibility, while a broad two-spinon
continuum remains at finite energies throughout the momentum path.  These
results demonstrate that the present general-$S$ construction can describe
both \red{static instabilities of the QSL saddle and their dynamical
signatures} within one thermodynamic-limit framework.

The present approach can be extended by allowing the mean-field saddle to
evolve with the perturbation and by incorporating spinon self-energies, gauge
fluctuations, and vertex corrections beyond RPA.  A systematic
comparison among competing symmetry-allowed spin-liquid Ans\"atze would test
whether an alternative QSL intervenes before the magnetic instability, while
a complementary calculation formulated around the ordered state would be
needed to determine how this low-energy softening is related to the magnon
spectrum on the ordered side.  Applying the same framework to additional
non-Kitaev interactions and to realistic higher-spin models may also provide
momentum- and frequency-resolved predictions directly comparable with
inelastic neutron-scattering experiments.  Such developments would turn the
present instability analysis into a broader tool for tracking the evolution
from fractionalized spin-liquid dynamics to conventional magnetic order.

\begin{acknowledgments}
  The authors thank A.~Ono for fruitful discussions.
  D.S. thanks K.~Fukui for valuable advice and feedback in discussions.
  Parts of the numerical calculations were performed in the supercomputing systems in ISSP, the University of Tokyo.
  This work was supported by Grant-in-Aid for Scientific Research from
  JSPS, KAKENHI Grant Nos.~JP23H01129, JP23H04865, JP24K00563, JP26H00624, and
  JP26H02230.
  D.S. acknowledges support from GP-Spin at Tohoku University.
\end{acknowledgments}

\appendix

\section{\texorpdfstring{RPA calculation for the Klein-odd perturbation}{RPA calculation for the Klein-odd perturbation}}
\label{app:rpa_calculation}

In this Appendix, the pure-Kitaev SBMFT BdG Hamiltonian
$\mathcal{H}_{\mathrm{MF}}$, constructed with coupling $K+J$, is the
unperturbed Hamiltonian.  The perturbation is the second term in
Eq.~\eqref{eq:klein_even_odd_decomposition}, and it enters only through the RPA
vertex.  All
imaginary-time expectation values are therefore evaluated with
$\mathcal{H}_{\mathrm{MF}}$.  This is left implicit below.  The two-point
correlation functions entering the RPA expansion are organized in terms of the
following coefficient-free normal and anomalous boson bilinears
\begin{align}
\mathcal{X}_{ij;\mu\nu}^{\mathrm{N}}(\tau)
&=
\bar b_{i\mu}(\tau)b_{j\nu}(\tau),
&
\overline{\mathcal{X}}_{ij;\mu\nu}^{\mathrm{N}}(\tau)
&=
b_{i\mu}(\tau)\bar b_{j\nu}(\tau),
\nonumber\\
\mathcal{X}_{ij;\mu\nu}^{\mathrm{A}}(\tau)
&=
b_{i\mu}(\tau)b_{j\nu}(\tau),
&
\overline{\mathcal{X}}_{ij;\mu\nu}^{\mathrm{A}}(\tau)
&=
\bar b_{i\mu}(\tau)\bar b_{j\nu}(\tau).
\label{eq:appendix_rpa_bilinear_operator}
\end{align}
The normal sector contains the
$\mathcal{B}_{ij},\mathcal{C}_{ij}^{\alpha}$ channels, while the anomalous
sector contains $\mathcal{A}_{ij},\mathcal{D}_{ij}^{\alpha}$.

Following the generalized bilinear response formulation and related RPA-type
dynamical response calculations in
Refs.~\cite{Rao-Moessner-Knolle-2025,Sasamoto-Nasu-2026,
Willsher-Knolle-2025}, we collect the operators in
Eq.~\eqref{eq:appendix_rpa_bilinear_operator} into $\mathcal{X}_r$ and write an
external observable as
\begin{align}
\mathcal{O}_{\lambda}
=
\sum_r \Gamma_r^{(\lambda)}\mathcal{X}_r.
\label{eq:appendix_rpa_external_vertex}
\end{align}
All coefficients belong to the external vertex $\Gamma^{(\lambda)}$.  The
bare bubble to which the Feynman rules are applied is
\begin{align}
\left[X^{(0)}(\tau)\right]_{rs}
&=
\left\langle
\mathcal{T}_{\tau}
\mathcal{X}_r(\tau)\mathcal{X}_s(0)
\right\rangle_{c}.
\label{eq:appendix_rpa_time_ordered_correlator}
\end{align}
Equation~\eqref{eq:appendix_rpa_time_ordered_correlator} is a
coefficient-free four-boson correlator evaluated as products of two-point
correlation functions.  The labels $r,s$ enumerate these internal bilinear
channels, analogous to the one-loop tensor indices in generalized RPA
formulations.  \red{The physical susceptibility is obtained by contracting the external
spin vertices within the RPA procedure.}  For the diagonal spin components used in
the numerical calculation, the relevant internal channels are collected into
the 24-component representation $r=(\alpha,a)$, with $\alpha=x,y,z$ and
$a=\text{A},\ldots,\text{H}$.
\begin{align}
X^{(0)}(\bm{q},i\omega_n)
=
\begin{pmatrix}
X^{(0)xx}(\bm{q},i\omega_n) & 0 & 0 \\
0 & X^{(0)yy}(\bm{q},i\omega_n) & 0 \\
0 & 0 & X^{(0)zz}(\bm{q},i\omega_n)
\end{pmatrix}.
\label{eq:appendix_rpa_bare_matrix}
\end{align}
After analytic continuation, $i\omega_n$ is replaced by $\omega+i\delta$.
The physical susceptibility $\chi^{(0)}$ used in the main text is the
corresponding external-vertex projection of $X^{(0)}$.

The RPA-channel perturbation is
\begin{align}
\mathcal{H}_{\mathrm{pert}}
=
\sum_{\gamma=x,y,z}
\sum_{\langle i,j\rangle_\gamma}
J\left[
\bm{\mathit{S}}_i\cdot\bm{\mathit{S}}_j
-S_i^\gamma S_j^\gamma
\right].
\label{eq:appendix_rpa_perturbation_hamiltonian}
\end{align}
Thus, on a $\gamma$ bond, the $\gamma$-component of the Heisenberg
exchange cancels against the second term in
Eq.~\eqref{eq:appendix_rpa_perturbation_hamiltonian}, leaving only the two spin
components transverse to $\gamma$.
We use the directed bond convention
$\bm{r}_b-\bm{r}_a=\bm{\delta}_{\gamma}$.  The three bond-vector
orientations are illustrated in Fig.~\ref{fig:mean_field_ansatz}(a).
Explicitly,
\begin{align}
\bm{\delta}_x
&=
\left(-\frac{1}{2},-\frac{1}{2\sqrt{3}}\right),
&
(a,b)_x
&=
(\text{A},\text{D}),(\text{C},\text{B}),
(\text{E},\text{H}),(\text{G},\text{F}),
\nonumber\\
\bm{\delta}_y
&=
\left(\frac{1}{2},-\frac{1}{2\sqrt{3}}\right),
&
(a,b)_y
&=
(\text{A},\text{H}),(\text{C},\text{F}),
(\text{E},\text{D}),(\text{G},\text{B}),
\nonumber\\
\bm{\delta}_z
&=
\left(0,\frac{1}{\sqrt{3}}\right),
&
(a,b)_z
&=
(\text{A},\text{B}),(\text{C},\text{D}),
(\text{E},\text{F}),(\text{G},\text{H}).
\label{eq:appendix_rpa_directed_bonds}
\end{align}
For a $\gamma$ bond directed from sublattice $a$ to $b$, the nonzero vertex
element is written directly as
\begin{align}
V_{(\alpha a),(\beta b)}(\bm{q})
&=
\frac{1}{4}\delta_{\alpha\beta}
J\left(1-\delta_{\alpha\gamma}\right)
e^{i\bm{q}\cdot\bm{\delta}_{\gamma}},
\nonumber\\
V_{(\alpha b),(\beta a)}(\bm{q})
&=
V_{(\alpha a),(\beta b)}(\bm{q})^{*}.
\label{eq:appendix_rpa_vertex}
\end{align}
Here $\bm{\delta}_{\gamma}$ is the displacement along the directed bond and
$\delta_{\alpha\gamma}$ is a Kronecker delta.  The factor
$1-\delta_{\alpha\gamma}$ implements the cancellation of the spin component
parallel to the Kitaev axis on each $\gamma$ bond.
The positive phase follows directly from
$e^{-i\bm{q}\cdot(\bm{r}_a-\bm{r}_b)}
=e^{+i\bm{q}\cdot\bm{\delta}_{\gamma}}$.
All spin-off-diagonal elements vanish.

With the positive retarded-response convention used in this work, the Dyson
equation in the 24-component basis is
\begin{align}
X_{\mathrm{RPA}}(\bm{q},i\omega_n)
&=
X^{(0)}(\bm{q},i\omega_n)
-X^{(0)}(\bm{q},i\omega_n)V(\bm{q})
X_{\mathrm{RPA}}(\bm{q},i\omega_n).
\label{eq:appendix_rpa_dyson}
\end{align}
Solving this equation for $X_{\mathrm{RPA}}$ gives
\begin{align}
X_{\mathrm{RPA}}(\bm{q},i\omega_n)
&=
\left[
\bm{1}_{24}+X^{(0)}(\bm{q},i\omega_n)V(\bm{q})
\right]^{-1}
X^{(0)}(\bm{q},i\omega_n).
\label{eq:appendix_rpa_solution}
\end{align}
\red{The physical susceptibility is obtained by contracting the}
external vertices,
\begin{align}
\chi_{\lambda\lambda^{\prime}}(\bm{q},i\omega_n)
&=
\sum_{r,s}
\Gamma_r^{(\lambda)}
\left[X_{\mathrm{RPA}}(\bm{q},i\omega_n)\right]_{rs}
\Gamma_s^{(\lambda^{\prime})}.
\label{eq:appendix_rpa_physical_projection}
\end{align}
Numerically, an instability is located by the vanishing of the smallest
singular value of
$\bm{1}_{24}+X^{(0)}(\bm{q},0)V(\bm{q})$, accompanied by the enhancement of the
static spin susceptibility.  The dynamical
structure factor follows from the analytic continuation of
$\chi$ in Eq.~\eqref{eq:appendix_rpa_physical_projection} through
Eq.~\eqref{eq:plotted_dynamical_spin_structure_factor}.  As checks, $V$ is
Hermitian for real couplings, and
$J=0$ gives
$X_{\mathrm{RPA}}=X^{(0)}$.

\section{Linear spin-wave correction to the ordered-state boundaries}
\label{app:linear_spin_wave}

The static RPA analysis determines where a Schwinger boson spin-liquid saddle
point becomes unstable.  To complete the ordered sectors of
Fig.~\ref{fig:phase_diagram}, we determine the boundaries between conventional
magnetic phases by an independent calculation based on linear spin-wave
theory~\cite{Gotfryd-2017,Koyama-Nasu-2021,Consoli-2020}.  This
auxiliary calculation is used only for those ordered-state boundaries and does
not enter the RPA calculation near the Kitaev spin liquids.  For the present
pure spin model, the harmonic expansion keeps the usual transverse
$\Delta m=1$ spin fluctuation of the spin-$S$ irreducible representation on
each sublattice, so the calculation is the conventional linear spin-wave
theory for general $S$.  We retain below only the
ingredients needed to determine the phase boundaries.

For each of the FM, N\'eel, zigzag, and stripy states, we rotate the local spin
axes so that every reference moment points along the local negative $z$
direction.  Choosing one of the symmetry-related cubic axes for the ordered
moment, the ground-state energy per site has the expansion
\begin{align}
\frac{E}{N}
=S^2 e^{(0)}(\phi)+S e^{(1)}(\phi)+O(S^0),
\label{eq:lsw_energy_expansion}
\end{align}
where the four classical coefficients are
\begin{align}
e_{\mathrm{FM}}^{(0)}&=\frac{3J+K}{2}, &
e_{\text{N\'eel}}^{(0)}&=-\frac{3J+K}{2},
\nonumber\\
e_{\text{zigzag}}^{(0)}&=\frac{J-K}{2}, &
e_{\text{stripy}}^{(0)}&=\frac{K-J}{2}.
\label{eq:lsw_classical_energies}
\end{align}
These classical ordered states and their symmetry relations are standard for
the nearest-neighbor Kitaev-Heisenberg model~\cite{Chaloupka-2010,
Gotfryd-2017,Georgiou-2024}.
Consequently,
\begin{align}
e_{\text{zigzag}}^{(0)}-e_{\mathrm{FM}}^{(0)}&=-(J+K),
\nonumber\\
e_{\text{stripy}}^{(0)}-e_{\text{N\'eel}}^{(0)}&=J+K,
\label{eq:lsw_classical_differences}
\end{align}
and the corresponding classical crossings are at $\phi_0=3\pi/4$ and
$7\pi/4$, respectively.

In the rotated frame, the harmonic expansion is generated by
\begin{align}
\widetilde S_i^z&=-S+a_i^\dagger a_i,
&
\widetilde S_i^x&=\sqrt{\frac{S}{2}}(a_i+a_i^\dagger),
\nonumber\\
\widetilde S_i^y&=-i\sqrt{\frac{S}{2}}(a_i^\dagger-a_i).
\label{eq:lsw_hp_expansion}
\end{align}
Equation~\eqref{eq:lsw_hp_expansion} is the Holstein--Primakoff
representation~\cite{Holstein-Primakoff-1940}.  The local-axis construction and
the subsequent bosonic Bogoliubov treatment follow the linear spin-wave
formulations for Kitaev-Heisenberg magnets~\cite{Gotfryd-2017,
Koyama-Nasu-2021,Consoli-2020}.
The linear terms vanish because each reference state is a stationary classical
configuration.  After Fourier transformation, we introduce the two-sublattice
Nambu spinor
\begin{align}
\Psi_{\bm{k}}
=
\begin{pmatrix}
a_{\bm{k},\text{A}} \\
a_{\bm{k},\text{B}} \\
a_{-\bm{k},\text{A}}^\dagger \\
a_{-\bm{k},\text{B}}^\dagger
\end{pmatrix},
\notag
\end{align}
with which the quadratic Hamiltonian is written as
\begin{align}
\mathcal{H}_2
=\frac{S}{2}\sum_{\bm{k}}
\Psi_{\bm{k}}^\dagger\mathcal{M}_{\bm{k}}\Psi_{\bm{k}}
-\frac{S}{2}\sum_{\bm{k}}\operatorname{Tr}A_{\bm{k}}.
\label{eq:lsw_bdg_hamiltonian}
\end{align}
Here $\mathcal{M}_{\bm{k}}$ is the $4\times4$ bosonic Bogoliubov--de Gennes
matrix and $A_{\bm{k}}$ is its particle-conserving $2\times2$ block.  The two
positive eigenvalues of
$\Sigma_3\mathcal{M}_{\bm{k}}$, with
$\Sigma_3=\operatorname{diag}(1,1,-1,-1)$, are denoted by
$\epsilon_{1\bm{k}}$ and $\epsilon_{2\bm{k}}$~\cite{Colpa-1978}.  The harmonic coefficient in
Eq.~\eqref{eq:lsw_energy_expansion}, evaluated separately for each ordered
state on an $N_k\times N_k$ midpoint mesh, is
\begin{align}
e^{(1)}(N_k)
=\frac{1}{4N_k^2}\sum_{\bm{k}\in\mathrm{BZ}}
\left[
\epsilon_{1\bm{k}}+\epsilon_{2\bm{k}}
-\operatorname{Tr}A_{\bm{k}}
\right].
\label{eq:lsw_zero_point_energy}
\end{align}
The trace term is the normal-ordering constant and must be retained when
comparing different phases.

The shift of a first-order boundary is obtained as a strict next-to-leading
order expansion in $1/S$~\cite{Consoli-2020}.  Let $\Delta e^{(n)}$ denote the
difference between the two competing phases at order $S^{2-n}$.  Expanding the
energy-crossing condition about the classical angle gives
\begin{align}
\phi_c
=\phi_0
-\frac{\Delta e^{(1)}(\phi_0)}
{S\,\partial_\phi\Delta e^{(0)}(\phi_0)}
+O(S^{-2}).
\label{eq:lsw_boundary_shift}
\end{align}
Importantly, the zero-point energies in Eq.~\eqref{eq:lsw_boundary_shift} are
evaluated at the classical crossing.  Continuing one of the harmonic vacua
beyond that point generally produces imaginary magnon frequencies and does not
define a controlled energy comparison.

Using midpoint Brillouin-zone meshes with
$N_k=80,100,120,160,$ and $200$, followed by an extrapolation in
$N_k^{-2}$ and $N_k^{-3}$, we obtain
\begin{align}
e_{\text{zigzag}}^{(1)}-e_{\mathrm{FM}}^{(1)}
&\simeq -0.12454,
\nonumber\\
e_{\text{stripy}}^{(1)}-e_{\text{N\'eel}}^{(1)}
&\simeq +0.12454.
\label{eq:lsw_zero_point_differences}
\end{align}
Since $\partial_\phi\Delta e^{(0)}=\sqrt{2}$ at both classical crossings,
the resulting boundaries are written in radians, with $a\simeq0.08807$, as
\begin{align}
\phi_c(\text{zigzag--FM})
&= \frac{3\pi}{4}+\frac{a}{S}+O(S^{-2}),
\nonumber\\
\phi_c(\text{stripy--N\'eel})
&= \frac{7\pi}{4}-\frac{a}{S}+O(S^{-2}).
\label{eq:lsw_final_boundaries}
\end{align}

\begin{table}[h]
\caption{
Ordered-state boundaries obtained from the harmonic $1/S$ correction and used
in Fig.~\ref{fig:phase_diagram}.  Angles are given in degrees.
}
\label{tab:lsw_boundaries}
\centering
\begin{ruledtabular}
\begin{tabular}{ccc}
$S$ & zigzag--FM & stripy--N\'eel \\
\hline
$1/2$ & $145.09$ & $304.91$ \\
$1$   & $140.05$ & $309.95$ \\
$3/2$ & $138.36$ & $311.64$ \\
$2$   & $137.52$ & $312.48$ \\
\end{tabular}
\end{ruledtabular}
\end{table}

{Equations~\eqref{eq:lsw_final_boundaries} are harmonic next-to-leading-order
results in which quartic magnon interactions and feedback of the magnon
covariance on the reference state are not included.}  The $S=1/2$ entries should therefore be
read as formal direct-crossing estimates within the two competing collinear
Ans\"atze.  The calculation cannot resolve an intervening nonmagnetic or
larger-unit-cell phase.  This limitation is relevant because high-order CCM
calculations find narrow intermediate regions between the corresponding
ordered phases for $S=1/2$~\cite{Georgiou-2024}.

\bibliography{./refs}

\begin{thebibliography}{100}%
\makeatletter
\providecommand \@ifxundefined [1]{%
 \@ifx{#1\undefined}
}%
\providecommand \@ifnum [1]{%
 \ifnum #1\expandafter \@firstoftwo
 \else \expandafter \@secondoftwo
 \fi
}%
\providecommand \@ifx [1]{%
 \ifx #1\expandafter \@firstoftwo
 \else \expandafter \@secondoftwo
 \fi
}%
\providecommand \natexlab [1]{#1}%
\providecommand \enquote  [1]{``#1''}%
\providecommand \bibnamefont  [1]{#1}%
\providecommand \bibfnamefont [1]{#1}%
\providecommand \citenamefont [1]{#1}%
\providecommand \href@noop [0]{\@secondoftwo}%
\providecommand \href [0]{\begingroup \@sanitize@url \@href}%
\providecommand \@href[1]{\@@startlink{#1}\@@href}%
\providecommand \@@href[1]{\endgroup#1\@@endlink}%
\providecommand \@sanitize@url [0]{\catcode `\\12\catcode `\$12\catcode `\&12\catcode `\#12\catcode `\^12\catcode `\_12\catcode `\%12\relax}%
\providecommand \@@startlink[1]{}%
\providecommand \@@endlink[0]{}%
\providecommand \url  [0]{\begingroup\@sanitize@url \@url }%
\providecommand \@url [1]{\endgroup\@href {#1}{\urlprefix }}%
\providecommand \urlprefix  [0]{URL }%
\providecommand \Eprint [0]{\href }%
\providecommand \doibase [0]{https://doi.org/}%
\providecommand \selectlanguage [0]{\@gobble}%
\providecommand \bibinfo  [0]{\@secondoftwo}%
\providecommand \bibfield  [0]{\@secondoftwo}%
\providecommand \translation [1]{[#1]}%
\providecommand \BibitemOpen [0]{}%
\providecommand \bibitemStop [0]{}%
\providecommand \bibitemNoStop [0]{.\EOS\space}%
\providecommand \EOS [0]{\spacefactor3000\relax}%
\providecommand \BibitemShut  [1]{\csname bibitem#1\endcsname}%
\let\auto@bib@innerbib\@empty
\bibitem [{\citenamefont {Anderson}(1973)}]{Anderson-1973}%
  \BibitemOpen
  \bibfield  {author} {\bibinfo {author} {\bibfnamefont {P.}~\bibnamefont {Anderson}},\ }\bibfield  {title} {\bibinfo {title} {{R}esonating valence bonds: {A} new kind of insulator?},\ }\href {https://doi.org/10.1016/0025-5408(73)90167-0} {\bibfield  {journal} {\bibinfo  {journal} {Mater. Res. Bull.}\ }\textbf {\bibinfo {volume} {8}},\ \bibinfo {pages} {153} (\bibinfo {year} {1973})}\BibitemShut {NoStop}%
\bibitem [{\citenamefont {Kalmeyer}\ and\ \citenamefont {Laughlin}(1987)}]{Kalmeyer-Laughlin-1987}%
  \BibitemOpen
  \bibfield  {author} {\bibinfo {author} {\bibfnamefont {V.}~\bibnamefont {Kalmeyer}}\ and\ \bibinfo {author} {\bibfnamefont {R.~B.}\ \bibnamefont {Laughlin}},\ }\bibfield  {title} {\bibinfo {title} {{E}quivalence of the resonating-valence-bond and fractional quantum {H}all states},\ }\href {https://doi.org/10.1103/PhysRevLett.59.2095} {\bibfield  {journal} {\bibinfo  {journal} {Phys. Rev. Lett.}\ }\textbf {\bibinfo {volume} {59}},\ \bibinfo {pages} {2095} (\bibinfo {year} {1987})}\BibitemShut {NoStop}%
\bibitem [{\citenamefont {Wen}(1989)}]{Wen-1989}%
  \BibitemOpen
  \bibfield  {author} {\bibinfo {author} {\bibfnamefont {X.~G.}\ \bibnamefont {Wen}},\ }\bibfield  {title} {\bibinfo {title} {{V}acuum degeneracy of chiral spin states in compactified space},\ }\href {https://doi.org/10.1103/PhysRevB.40.7387} {\bibfield  {journal} {\bibinfo  {journal} {Phys. Rev. B}\ }\textbf {\bibinfo {volume} {40}},\ \bibinfo {pages} {7387} (\bibinfo {year} {1989})}\BibitemShut {NoStop}%
\bibitem [{\citenamefont {Wen}(2002)}]{Wen-2002}%
  \BibitemOpen
  \bibfield  {author} {\bibinfo {author} {\bibfnamefont {X.-G.}\ \bibnamefont {Wen}},\ }\bibfield  {title} {\bibinfo {title} {{Q}uantum orders and symmetric spin liquids},\ }\href {https://doi.org/10.1103/PhysRevB.65.165113} {\bibfield  {journal} {\bibinfo  {journal} {Phys. Rev. B}\ }\textbf {\bibinfo {volume} {65}},\ \bibinfo {pages} {165113} (\bibinfo {year} {2002})}\BibitemShut {NoStop}%
\bibitem [{\citenamefont {Balents}(2010)}]{Balents-2010}%
  \BibitemOpen
  \bibfield  {author} {\bibinfo {author} {\bibfnamefont {L.}~\bibnamefont {Balents}},\ }\bibfield  {title} {\bibinfo {title} {{S}pin liquids in frustrated magnets},\ }\href {https://doi.org/10.1038/nature08917} {\bibfield  {journal} {\bibinfo  {journal} {Nature (London)}\ }\textbf {\bibinfo {volume} {464}},\ \bibinfo {pages} {199} (\bibinfo {year} {2010})}\BibitemShut {NoStop}%
\bibitem [{\citenamefont {Savary}\ and\ \citenamefont {Balents}(2016)}]{Savary-Balents-2016}%
  \BibitemOpen
  \bibfield  {author} {\bibinfo {author} {\bibfnamefont {L.}~\bibnamefont {Savary}}\ and\ \bibinfo {author} {\bibfnamefont {L.}~\bibnamefont {Balents}},\ }\bibfield  {title} {\bibinfo {title} {{Q}uantum spin liquids: a review},\ }\href {https://doi.org/10.1088/0034-4885/80/1/016502} {\bibfield  {journal} {\bibinfo  {journal} {Rep. Prog. Phys.}\ }\textbf {\bibinfo {volume} {80}},\ \bibinfo {pages} {016502} (\bibinfo {year} {2016})}\BibitemShut {NoStop}%
\bibitem [{\citenamefont {Zhou}\ \emph {et~al.}(2017)\citenamefont {Zhou}, \citenamefont {Kanoda},\ and\ \citenamefont {Ng}}]{Zhou-Kanoda-2017}%
  \BibitemOpen
  \bibfield  {author} {\bibinfo {author} {\bibfnamefont {Y.}~\bibnamefont {Zhou}}, \bibinfo {author} {\bibfnamefont {K.}~\bibnamefont {Kanoda}},\ and\ \bibinfo {author} {\bibfnamefont {T.-K.}\ \bibnamefont {Ng}},\ }\bibfield  {title} {\bibinfo {title} {{Q}uantum spin liquid states},\ }\href {https://doi.org/10.1103/RevModPhys.89.025003} {\bibfield  {journal} {\bibinfo  {journal} {Rev. Mod. Phys.}\ }\textbf {\bibinfo {volume} {89}},\ \bibinfo {pages} {025003} (\bibinfo {year} {2017})}\BibitemShut {NoStop}%
\bibitem [{\citenamefont {Broholm}\ \emph {et~al.}(2020)\citenamefont {Broholm}, \citenamefont {Cava}, \citenamefont {Kivelson}, \citenamefont {Nocera}, \citenamefont {Norman},\ and\ \citenamefont {Senthil}}]{Broholm-Cava-2020}%
  \BibitemOpen
  \bibfield  {author} {\bibinfo {author} {\bibfnamefont {C.}~\bibnamefont {Broholm}}, \bibinfo {author} {\bibfnamefont {R.~J.}\ \bibnamefont {Cava}}, \bibinfo {author} {\bibfnamefont {S.~A.}\ \bibnamefont {Kivelson}}, \bibinfo {author} {\bibfnamefont {D.~G.}\ \bibnamefont {Nocera}}, \bibinfo {author} {\bibfnamefont {M.~R.}\ \bibnamefont {Norman}},\ and\ \bibinfo {author} {\bibfnamefont {T.}~\bibnamefont {Senthil}},\ }\bibfield  {title} {\bibinfo {title} {{Q}uantum spin liquids},\ }\href {https://doi.org/10.1126/science.aay0668} {\bibfield  {journal} {\bibinfo  {journal} {Science}\ }\textbf {\bibinfo {volume} {367}},\ \bibinfo {pages} {eaay0668} (\bibinfo {year} {2020})}\BibitemShut {NoStop}%
\bibitem [{\citenamefont {Knolle}\ and\ \citenamefont {Moessner}(2019)}]{Knolle-Moessner-2019}%
  \BibitemOpen
  \bibfield  {author} {\bibinfo {author} {\bibfnamefont {J.}~\bibnamefont {Knolle}}\ and\ \bibinfo {author} {\bibfnamefont {R.}~\bibnamefont {Moessner}},\ }\bibfield  {title} {\bibinfo {title} {{A} {F}ield {G}uide to {S}pin {L}iquids},\ }\href {https://doi.org/10.1146/annurev-conmatphys-031218-013401} {\bibfield  {journal} {\bibinfo  {journal} {Annu. Rev. Condens. Matter Phys.}\ }\textbf {\bibinfo {volume} {10}},\ \bibinfo {pages} {451} (\bibinfo {year} {2019})}\BibitemShut {NoStop}%
\bibitem [{\citenamefont {Kitaev}(2006)}]{Kitaev-2006}%
  \BibitemOpen
  \bibfield  {author} {\bibinfo {author} {\bibfnamefont {A.}~\bibnamefont {Kitaev}},\ }\bibfield  {title} {\bibinfo {title} {{A}nyons in an exactly solved model and beyond},\ }\href {https://doi.org/10.1016/j.aop.2005.10.005} {\bibfield  {journal} {\bibinfo  {journal} {Ann. Phys. (N.Y.)}\ }\textbf {\bibinfo {volume} {321}},\ \bibinfo {pages} {2} (\bibinfo {year} {2006})}\BibitemShut {NoStop}%
\bibitem [{\citenamefont {Knolle}\ \emph {et~al.}(2014)\citenamefont {Knolle}, \citenamefont {Kovrizhin}, \citenamefont {Chalker},\ and\ \citenamefont {Moessner}}]{Knolle-2014}%
  \BibitemOpen
  \bibfield  {author} {\bibinfo {author} {\bibfnamefont {J.}~\bibnamefont {Knolle}}, \bibinfo {author} {\bibfnamefont {D.~L.}\ \bibnamefont {Kovrizhin}}, \bibinfo {author} {\bibfnamefont {J.~T.}\ \bibnamefont {Chalker}},\ and\ \bibinfo {author} {\bibfnamefont {R.}~\bibnamefont {Moessner}},\ }\bibfield  {title} {\bibinfo {title} {{D}ynamics of a {T}wo-{D}imensional {Q}uantum {S}pin {L}iquid: {S}ignatures of {E}mergent {M}ajorana {F}ermions and {F}luxes},\ }\href {https://doi.org/10.1103/PhysRevLett.112.207203} {\bibfield  {journal} {\bibinfo  {journal} {Phys. Rev. Lett.}\ }\textbf {\bibinfo {volume} {112}},\ \bibinfo {pages} {207203} (\bibinfo {year} {2014})}\BibitemShut {NoStop}%
\bibitem [{\citenamefont {Knolle}\ \emph {et~al.}(2015)\citenamefont {Knolle}, \citenamefont {Kovrizhin}, \citenamefont {Chalker},\ and\ \citenamefont {Moessner}}]{Knolle-2015}%
  \BibitemOpen
  \bibfield  {author} {\bibinfo {author} {\bibfnamefont {J.}~\bibnamefont {Knolle}}, \bibinfo {author} {\bibfnamefont {D.~L.}\ \bibnamefont {Kovrizhin}}, \bibinfo {author} {\bibfnamefont {J.~T.}\ \bibnamefont {Chalker}},\ and\ \bibinfo {author} {\bibfnamefont {R.}~\bibnamefont {Moessner}},\ }\bibfield  {title} {\bibinfo {title} {{D}ynamics of fractionalization in quantum spin liquids},\ }\href {https://doi.org/10.1103/PhysRevB.92.115127} {\bibfield  {journal} {\bibinfo  {journal} {Phys. Rev. B}\ }\textbf {\bibinfo {volume} {92}},\ \bibinfo {pages} {115127} (\bibinfo {year} {2015})}\BibitemShut {NoStop}%
\bibitem [{\citenamefont {Hermanns}\ \emph {et~al.}(2018)\citenamefont {Hermanns}, \citenamefont {Kimchi},\ and\ \citenamefont {Knolle}}]{Hermanns-Kimchi-2018}%
  \BibitemOpen
  \bibfield  {author} {\bibinfo {author} {\bibfnamefont {M.}~\bibnamefont {Hermanns}}, \bibinfo {author} {\bibfnamefont {I.}~\bibnamefont {Kimchi}},\ and\ \bibinfo {author} {\bibfnamefont {J.}~\bibnamefont {Knolle}},\ }\bibfield  {title} {\bibinfo {title} {{P}hysics of the {K}itaev {M}odel: {F}ractionalization, {D}ynamic {C}orrelations, and {M}aterial {C}onnections},\ }\href {https://doi.org/10.1146/annurev-conmatphys-033117-053934} {\bibfield  {journal} {\bibinfo  {journal} {Annu. Rev. Condens. Matter Phys.}\ }\textbf {\bibinfo {volume} {9}},\ \bibinfo {pages} {17} (\bibinfo {year} {2018})}\BibitemShut {NoStop}%
\bibitem [{\citenamefont {Motome}\ and\ \citenamefont {Nasu}(2020)}]{Motome-Nasu-2020}%
  \BibitemOpen
  \bibfield  {author} {\bibinfo {author} {\bibfnamefont {Y.}~\bibnamefont {Motome}}\ and\ \bibinfo {author} {\bibfnamefont {J.}~\bibnamefont {Nasu}},\ }\bibfield  {title} {\bibinfo {title} {{H}unting {M}ajorana {F}ermions in {K}itaev {M}agnets},\ }\href {https://doi.org/10.7566/JPSJ.89.012002} {\bibfield  {journal} {\bibinfo  {journal} {J. Phys. Soc. Jpn.}\ }\textbf {\bibinfo {volume} {89}},\ \bibinfo {pages} {012002} (\bibinfo {year} {2020})}\BibitemShut {NoStop}%
\bibitem [{\citenamefont {Nasu}\ and\ \citenamefont {Motome}(2021)}]{Nasu-Motome-2021}%
  \BibitemOpen
  \bibfield  {author} {\bibinfo {author} {\bibfnamefont {J.}~\bibnamefont {Nasu}}\ and\ \bibinfo {author} {\bibfnamefont {Y.}~\bibnamefont {Motome}},\ }\bibfield  {title} {\bibinfo {title} {{S}pin dynamics in the {K}itaev model with disorder: {Q}uantum {M}onte {C}arlo study of dynamical spin structure factor, magnetic susceptibility, and {N}{M}{R} relaxation rate},\ }\href {https://doi.org/10.1103/PhysRevB.104.035116} {\bibfield  {journal} {\bibinfo  {journal} {Phys. Rev. B}\ }\textbf {\bibinfo {volume} {104}},\ \bibinfo {pages} {035116} (\bibinfo {year} {2021})}\BibitemShut {NoStop}%
\bibitem [{\citenamefont {Jackeli}\ and\ \citenamefont {Khaliullin}(2009)}]{Jackeli-Khaliullin-2009}%
  \BibitemOpen
  \bibfield  {author} {\bibinfo {author} {\bibfnamefont {G.}~\bibnamefont {Jackeli}}\ and\ \bibinfo {author} {\bibfnamefont {G.}~\bibnamefont {Khaliullin}},\ }\bibfield  {title} {\bibinfo {title} {{M}ott {I}nsulators in the {S}trong {S}pin-{O}rbit {C}oupling {L}imit: {F}rom {H}eisenberg to a {Q}uantum {C}ompass and {K}itaev {M}odels},\ }\href {https://doi.org/10.1103/PhysRevLett.102.017205} {\bibfield  {journal} {\bibinfo  {journal} {Phys. Rev. Lett.}\ }\textbf {\bibinfo {volume} {102}},\ \bibinfo {pages} {017205} (\bibinfo {year} {2009})}\BibitemShut {NoStop}%
\bibitem [{\citenamefont {Foyevtsova}\ \emph {et~al.}(2013)\citenamefont {Foyevtsova}, \citenamefont {Jeschke}, \citenamefont {Mazin}, \citenamefont {Khomskii},\ and\ \citenamefont {Valent\'{\i}}}]{Foyevtsova-2013}%
  \BibitemOpen
  \bibfield  {author} {\bibinfo {author} {\bibfnamefont {K.}~\bibnamefont {Foyevtsova}}, \bibinfo {author} {\bibfnamefont {H.~O.}\ \bibnamefont {Jeschke}}, \bibinfo {author} {\bibfnamefont {I.~I.}\ \bibnamefont {Mazin}}, \bibinfo {author} {\bibfnamefont {D.~I.}\ \bibnamefont {Khomskii}},\ and\ \bibinfo {author} {\bibfnamefont {R.}~\bibnamefont {Valent\'{\i}}},\ }\bibfield  {title} {\bibinfo {title} {{A}b initio analysis of the tight-binding parameters and magnetic interactions in {N}a$_{2}${I}r{O}$_{3}$},\ }\href {https://doi.org/10.1103/PhysRevB.88.035107} {\bibfield  {journal} {\bibinfo  {journal} {Phys. Rev. B}\ }\textbf {\bibinfo {volume} {88}},\ \bibinfo {pages} {035107} (\bibinfo {year} {2013})}\BibitemShut {NoStop}%
\bibitem [{\citenamefont {Katukuri}\ \emph {et~al.}(2014)\citenamefont {Katukuri}, \citenamefont {Nishimoto}, \citenamefont {Yushankhai}, \citenamefont {Stoyanova}, \citenamefont {Kandpal}, \citenamefont {Choi}, \citenamefont {Coldea}, \citenamefont {Rousochatzakis}, \citenamefont {Hozoi},\ and\ \citenamefont {van~den Brink}}]{Katukuri-2014}%
  \BibitemOpen
  \bibfield  {author} {\bibinfo {author} {\bibfnamefont {V.~M.}\ \bibnamefont {Katukuri}}, \bibinfo {author} {\bibfnamefont {S.}~\bibnamefont {Nishimoto}}, \bibinfo {author} {\bibfnamefont {V.}~\bibnamefont {Yushankhai}}, \bibinfo {author} {\bibfnamefont {A.}~\bibnamefont {Stoyanova}}, \bibinfo {author} {\bibfnamefont {H.}~\bibnamefont {Kandpal}}, \bibinfo {author} {\bibfnamefont {S.}~\bibnamefont {Choi}}, \bibinfo {author} {\bibfnamefont {R.}~\bibnamefont {Coldea}}, \bibinfo {author} {\bibfnamefont {I.}~\bibnamefont {Rousochatzakis}}, \bibinfo {author} {\bibfnamefont {L.}~\bibnamefont {Hozoi}},\ and\ \bibinfo {author} {\bibfnamefont {J.}~\bibnamefont {van~den Brink}},\ }\bibfield  {title} {\bibinfo {title} {{K}itaev interactions between $j$ = 1/2 moments in honeycomb {N}a$_{2}${I}r{O}$_{3}$ are large and ferromagnetic: insights from ab initio quantum chemistry calculations},\ }\href {https://doi.org/10.1088/1367-2630/16/1/013056} {\bibfield  {journal} {\bibinfo  {journal} {New J. Phys.}\ }\textbf {\bibinfo
  {volume} {16}},\ \bibinfo {pages} {013056} (\bibinfo {year} {2014})}\BibitemShut {NoStop}%
\bibitem [{\citenamefont {Yamaji}\ \emph {et~al.}(2014)\citenamefont {Yamaji}, \citenamefont {Nomura}, \citenamefont {Kurita}, \citenamefont {Arita},\ and\ \citenamefont {Imada}}]{Yamaji-2014}%
  \BibitemOpen
  \bibfield  {author} {\bibinfo {author} {\bibfnamefont {Y.}~\bibnamefont {Yamaji}}, \bibinfo {author} {\bibfnamefont {Y.}~\bibnamefont {Nomura}}, \bibinfo {author} {\bibfnamefont {M.}~\bibnamefont {Kurita}}, \bibinfo {author} {\bibfnamefont {R.}~\bibnamefont {Arita}},\ and\ \bibinfo {author} {\bibfnamefont {M.}~\bibnamefont {Imada}},\ }\bibfield  {title} {\bibinfo {title} {{F}irst-{P}rinciples {S}tudy of the {H}oneycomb-{L}attice {I}ridates {N}a$_{2}${I}r{O}$_{3}$ in the presence of strong spin-orbit interaction and electron correlations},\ }\href {https://doi.org/10.1103/PhysRevLett.113.107201} {\bibfield  {journal} {\bibinfo  {journal} {Phys. Rev. Lett.}\ }\textbf {\bibinfo {volume} {113}},\ \bibinfo {pages} {107201} (\bibinfo {year} {2014})}\BibitemShut {NoStop}%
\bibitem [{\citenamefont {Hwan~Chun}\ \emph {et~al.}(2015)\citenamefont {Hwan~Chun}, \citenamefont {Kim}, \citenamefont {Kim}, \citenamefont {Zheng}, \citenamefont {Stoumpos}, \citenamefont {Malliakas}, \citenamefont {Mitchell}, \citenamefont {Mehlawat}, \citenamefont {Singh}, \citenamefont {Choi}, \citenamefont {Gog}, \citenamefont {Al-Zein}, \citenamefont {Sala}, \citenamefont {Krisch}, \citenamefont {Chaloupka}, \citenamefont {Jackeli}, \citenamefont {Khaliullin},\ and\ \citenamefont {Kim}}]{Chun-2015}%
  \BibitemOpen
  \bibfield  {author} {\bibinfo {author} {\bibfnamefont {S.}~\bibnamefont {Hwan~Chun}}, \bibinfo {author} {\bibfnamefont {J.-W.}\ \bibnamefont {Kim}}, \bibinfo {author} {\bibfnamefont {J.}~\bibnamefont {Kim}}, \bibinfo {author} {\bibfnamefont {H.}~\bibnamefont {Zheng}}, \bibinfo {author} {\bibfnamefont {C.~C.}\ \bibnamefont {Stoumpos}}, \bibinfo {author} {\bibfnamefont {C.~D.}\ \bibnamefont {Malliakas}}, \bibinfo {author} {\bibfnamefont {J.~F.}\ \bibnamefont {Mitchell}}, \bibinfo {author} {\bibfnamefont {K.}~\bibnamefont {Mehlawat}}, \bibinfo {author} {\bibfnamefont {Y.}~\bibnamefont {Singh}}, \bibinfo {author} {\bibfnamefont {Y.}~\bibnamefont {Choi}}, \bibinfo {author} {\bibfnamefont {T.}~\bibnamefont {Gog}}, \bibinfo {author} {\bibfnamefont {A.}~\bibnamefont {Al-Zein}}, \bibinfo {author} {\bibfnamefont {M.~M.}\ \bibnamefont {Sala}}, \bibinfo {author} {\bibfnamefont {M.}~\bibnamefont {Krisch}}, \bibinfo {author} {\bibfnamefont {J.}~\bibnamefont {Chaloupka}}, \bibinfo {author} {\bibfnamefont {G.}~\bibnamefont
  {Jackeli}}, \bibinfo {author} {\bibfnamefont {G.}~\bibnamefont {Khaliullin}},\ and\ \bibinfo {author} {\bibfnamefont {B.~J.}\ \bibnamefont {Kim}},\ }\bibfield  {title} {\bibinfo {title} {{D}irect evidence for dominant bond-directional interactions in a honeycomb lattice iridate {N}a$_{2}${I}r{O}$_{3}$},\ }\href {https://doi.org/10.1038/nphys3322} {\bibfield  {journal} {\bibinfo  {journal} {Nat. Phys.}\ }\textbf {\bibinfo {volume} {11}},\ \bibinfo {pages} {462} (\bibinfo {year} {2015})}\BibitemShut {NoStop}%
\bibitem [{\citenamefont {Yadav}\ \emph {et~al.}(2016)\citenamefont {Yadav}, \citenamefont {Bogdanov}, \citenamefont {Katukuri}, \citenamefont {Nishimoto}, \citenamefont {van~den Brink},\ and\ \citenamefont {Hozoi}}]{Yadav-2016}%
  \BibitemOpen
  \bibfield  {author} {\bibinfo {author} {\bibfnamefont {R.}~\bibnamefont {Yadav}}, \bibinfo {author} {\bibfnamefont {N.~A.}\ \bibnamefont {Bogdanov}}, \bibinfo {author} {\bibfnamefont {V.~M.}\ \bibnamefont {Katukuri}}, \bibinfo {author} {\bibfnamefont {S.}~\bibnamefont {Nishimoto}}, \bibinfo {author} {\bibfnamefont {J.}~\bibnamefont {van~den Brink}},\ and\ \bibinfo {author} {\bibfnamefont {L.}~\bibnamefont {Hozoi}},\ }\bibfield  {title} {\bibinfo {title} {{K}itaev exchange and field-induced quantum spin-liquid states in honeycomb $\alpha$-{R}u{C}l$_{3}$},\ }\href {https://doi.org/10.1038/srep37925} {\bibfield  {journal} {\bibinfo  {journal} {Sci. Rep.}\ }\textbf {\bibinfo {volume} {6}},\ \bibinfo {pages} {37925} (\bibinfo {year} {2016})}\BibitemShut {NoStop}%
\bibitem [{\citenamefont {Rau}\ \emph {et~al.}(2014)\citenamefont {Rau}, \citenamefont {Lee},\ and\ \citenamefont {Kee}}]{Rau-Lee-Kee-2014}%
  \BibitemOpen
  \bibfield  {author} {\bibinfo {author} {\bibfnamefont {J.~G.}\ \bibnamefont {Rau}}, \bibinfo {author} {\bibfnamefont {E.~K.-H.}\ \bibnamefont {Lee}},\ and\ \bibinfo {author} {\bibfnamefont {H.-Y.}\ \bibnamefont {Kee}},\ }\bibfield  {title} {\bibinfo {title} {{G}eneric {S}pin {M}odel for the {H}oneycomb {I}ridates beyond the {K}itaev {L}imit},\ }\href {https://doi.org/10.1103/PhysRevLett.112.077204} {\bibfield  {journal} {\bibinfo  {journal} {Phys. Rev. Lett.}\ }\textbf {\bibinfo {volume} {112}},\ \bibinfo {pages} {077204} (\bibinfo {year} {2014})}\BibitemShut {NoStop}%
\bibitem [{\citenamefont {Rau}\ \emph {et~al.}(2016)\citenamefont {Rau}, \citenamefont {Lee},\ and\ \citenamefont {Kee}}]{Rau-Lee-Kee-2016}%
  \BibitemOpen
  \bibfield  {author} {\bibinfo {author} {\bibfnamefont {J.~G.}\ \bibnamefont {Rau}}, \bibinfo {author} {\bibfnamefont {E.~K.-H.}\ \bibnamefont {Lee}},\ and\ \bibinfo {author} {\bibfnamefont {H.-Y.}\ \bibnamefont {Kee}},\ }\bibfield  {title} {\bibinfo {title} {{S}pin-{O}rbit {P}hysics {G}iving {R}ise to {N}ovel {P}hases in {C}orrelated {S}ystems: {I}ridates and {R}elated {M}aterials},\ }\href {https://doi.org/10.1146/annurev-conmatphys-031115-011319} {\bibfield  {journal} {\bibinfo  {journal} {Annu. Rev. Condens. Matter Phys.}\ }\textbf {\bibinfo {volume} {7}},\ \bibinfo {pages} {195} (\bibinfo {year} {2016})}\BibitemShut {NoStop}%
\bibitem [{\citenamefont {Winter}\ \emph {et~al.}(2017)\citenamefont {Winter}, \citenamefont {Tsirlin}, \citenamefont {Daghofer}, \citenamefont {van~den Brink}, \citenamefont {Singh}, \citenamefont {Gegenwart},\ and\ \citenamefont {Valent{\'\i}}}]{Winter-2017}%
  \BibitemOpen
  \bibfield  {author} {\bibinfo {author} {\bibfnamefont {S.~M.}\ \bibnamefont {Winter}}, \bibinfo {author} {\bibfnamefont {A.~A.}\ \bibnamefont {Tsirlin}}, \bibinfo {author} {\bibfnamefont {M.}~\bibnamefont {Daghofer}}, \bibinfo {author} {\bibfnamefont {J.}~\bibnamefont {van~den Brink}}, \bibinfo {author} {\bibfnamefont {Y.}~\bibnamefont {Singh}}, \bibinfo {author} {\bibfnamefont {P.}~\bibnamefont {Gegenwart}},\ and\ \bibinfo {author} {\bibfnamefont {R.}~\bibnamefont {Valent{\'\i}}},\ }\bibfield  {title} {\bibinfo {title} {{M}odels and materials for generalized {K}itaev magnetism},\ }\href {https://doi.org/10.1088/1361-648X/aa8cf5} {\bibfield  {journal} {\bibinfo  {journal} {J. Phys.: Condens. Matter}\ }\textbf {\bibinfo {volume} {29}},\ \bibinfo {pages} {493002} (\bibinfo {year} {2017})}\BibitemShut {NoStop}%
\bibitem [{\citenamefont {Takagi}\ \emph {et~al.}(2019)\citenamefont {Takagi}, \citenamefont {Takayama}, \citenamefont {Jackeli}, \citenamefont {Khaliullin},\ and\ \citenamefont {Nagler}}]{Takagi-2019}%
  \BibitemOpen
  \bibfield  {author} {\bibinfo {author} {\bibfnamefont {H.}~\bibnamefont {Takagi}}, \bibinfo {author} {\bibfnamefont {T.}~\bibnamefont {Takayama}}, \bibinfo {author} {\bibfnamefont {G.}~\bibnamefont {Jackeli}}, \bibinfo {author} {\bibfnamefont {G.}~\bibnamefont {Khaliullin}},\ and\ \bibinfo {author} {\bibfnamefont {S.~E.}\ \bibnamefont {Nagler}},\ }\bibfield  {title} {\bibinfo {title} {{C}oncept and realization of {K}itaev quantum spin liquids},\ }\href {https://doi.org/10.1038/s42254-019-0038-2} {\bibfield  {journal} {\bibinfo  {journal} {Nat. Rev. Phys.}\ }\textbf {\bibinfo {volume} {1}},\ \bibinfo {pages} {264} (\bibinfo {year} {2019})}\BibitemShut {NoStop}%
\bibitem [{\citenamefont {Motome}\ \emph {et~al.}(2020)\citenamefont {Motome}, \citenamefont {Sano}, \citenamefont {Jang}, \citenamefont {Sugita},\ and\ \citenamefont {Kato}}]{Motome-2020}%
  \BibitemOpen
  \bibfield  {author} {\bibinfo {author} {\bibfnamefont {Y.}~\bibnamefont {Motome}}, \bibinfo {author} {\bibfnamefont {R.}~\bibnamefont {Sano}}, \bibinfo {author} {\bibfnamefont {S.}~\bibnamefont {Jang}}, \bibinfo {author} {\bibfnamefont {Y.}~\bibnamefont {Sugita}},\ and\ \bibinfo {author} {\bibfnamefont {Y.}~\bibnamefont {Kato}},\ }\bibfield  {title} {\bibinfo {title} {{M}aterials design of {K}itaev spin liquids beyond the {J}ackeli--{K}haliullin mechanism},\ }\href {https://doi.org/10.1088/1361-648X/ab8525} {\bibfield  {journal} {\bibinfo  {journal} {J. Phys.: Condens. Matter}\ }\textbf {\bibinfo {volume} {32}},\ \bibinfo {pages} {404001} (\bibinfo {year} {2020})}\BibitemShut {NoStop}%
\bibitem [{\citenamefont {Trebst}\ and\ \citenamefont {Hickey}(2022)}]{Trebst-Hickey-2022}%
  \BibitemOpen
  \bibfield  {author} {\bibinfo {author} {\bibfnamefont {S.}~\bibnamefont {Trebst}}\ and\ \bibinfo {author} {\bibfnamefont {C.}~\bibnamefont {Hickey}},\ }\bibfield  {title} {\bibinfo {title} {{K}itaev materials},\ }\href {https://doi.org/10.1016/j.physrep.2021.11.003} {\bibfield  {journal} {\bibinfo  {journal} {Phys. Rep.}\ }\textbf {\bibinfo {volume} {950}},\ \bibinfo {pages} {1} (\bibinfo {year} {2022})}\BibitemShut {NoStop}%
\bibitem [{\citenamefont {Rousochatzakis}\ \emph {et~al.}(2024)\citenamefont {Rousochatzakis}, \citenamefont {Perkins}, \citenamefont {Luo},\ and\ \citenamefont {Kee}}]{Rousochatzakis-Perkins-Luo-Kee-2024}%
  \BibitemOpen
  \bibfield  {author} {\bibinfo {author} {\bibfnamefont {I.}~\bibnamefont {Rousochatzakis}}, \bibinfo {author} {\bibfnamefont {N.~B.}\ \bibnamefont {Perkins}}, \bibinfo {author} {\bibfnamefont {Q.}~\bibnamefont {Luo}},\ and\ \bibinfo {author} {\bibfnamefont {H.-Y.}\ \bibnamefont {Kee}},\ }\bibfield  {title} {\bibinfo {title} {{B}eyond {K}itaev physics in strong spin-orbit coupled magnets},\ }\href {https://doi.org/10.1088/1361-6633/ad208d} {\bibfield  {journal} {\bibinfo  {journal} {Rep. Prog. Phys.}\ }\textbf {\bibinfo {volume} {87}},\ \bibinfo {pages} {026502} (\bibinfo {year} {2024})}\BibitemShut {NoStop}%
\bibitem [{\citenamefont {Chaloupka}\ \emph {et~al.}(2010)\citenamefont {Chaloupka}, \citenamefont {Jackeli},\ and\ \citenamefont {Khaliullin}}]{Chaloupka-2010}%
  \BibitemOpen
  \bibfield  {author} {\bibinfo {author} {\bibfnamefont {J.}~\bibnamefont {Chaloupka}}, \bibinfo {author} {\bibfnamefont {G.}~\bibnamefont {Jackeli}},\ and\ \bibinfo {author} {\bibfnamefont {G.}~\bibnamefont {Khaliullin}},\ }\bibfield  {title} {\bibinfo {title} {{K}itaev-{H}eisenberg {M}odel on a {H}oneycomb {L}attice: {P}ossible {E}xotic {P}hases in {I}ridium {O}xides ${A}_{2}${I}r{O}$_{3}$},\ }\href {https://doi.org/10.1103/PhysRevLett.105.027204} {\bibfield  {journal} {\bibinfo  {journal} {Phys. Rev. Lett.}\ }\textbf {\bibinfo {volume} {105}},\ \bibinfo {pages} {027204} (\bibinfo {year} {2010})}\BibitemShut {NoStop}%
\bibitem [{\citenamefont {Jiang}\ \emph {et~al.}(2011)\citenamefont {Jiang}, \citenamefont {Gu}, \citenamefont {Qi},\ and\ \citenamefont {Trebst}}]{Jiang-2011}%
  \BibitemOpen
  \bibfield  {author} {\bibinfo {author} {\bibfnamefont {H.-C.}\ \bibnamefont {Jiang}}, \bibinfo {author} {\bibfnamefont {Z.-C.}\ \bibnamefont {Gu}}, \bibinfo {author} {\bibfnamefont {X.-L.}\ \bibnamefont {Qi}},\ and\ \bibinfo {author} {\bibfnamefont {S.}~\bibnamefont {Trebst}},\ }\bibfield  {title} {\bibinfo {title} {Possible proximity of the {M}ott insulating iridate {N}a$_2${I}r{O}$_3$ to a topological phase: {P}hase diagram of the {H}eisenberg-{K}itaev model in a magnetic field},\ }\href {https://doi.org/10.1103/PhysRevB.83.245104} {\bibfield  {journal} {\bibinfo  {journal} {Phys. Rev. B}\ }\textbf {\bibinfo {volume} {83}},\ \bibinfo {pages} {245104} (\bibinfo {year} {2011})}\BibitemShut {NoStop}%
\bibitem [{\citenamefont {Reuther}\ \emph {et~al.}(2011)\citenamefont {Reuther}, \citenamefont {Thomale},\ and\ \citenamefont {Trebst}}]{Reuther-2011}%
  \BibitemOpen
  \bibfield  {author} {\bibinfo {author} {\bibfnamefont {J.}~\bibnamefont {Reuther}}, \bibinfo {author} {\bibfnamefont {R.}~\bibnamefont {Thomale}},\ and\ \bibinfo {author} {\bibfnamefont {S.}~\bibnamefont {Trebst}},\ }\bibfield  {title} {\bibinfo {title} {Finite-temperature phase diagram of the {H}eisenberg-{K}itaev model},\ }\href {https://doi.org/10.1103/PhysRevB.84.100406} {\bibfield  {journal} {\bibinfo  {journal} {Phys. Rev. B}\ }\textbf {\bibinfo {volume} {84}},\ \bibinfo {pages} {100406} (\bibinfo {year} {2011})}\BibitemShut {NoStop}%
\bibitem [{\citenamefont {Schaffer}\ \emph {et~al.}(2012)\citenamefont {Schaffer}, \citenamefont {Bhattacharjee},\ and\ \citenamefont {Kim}}]{Schaffer-2012}%
  \BibitemOpen
  \bibfield  {author} {\bibinfo {author} {\bibfnamefont {R.}~\bibnamefont {Schaffer}}, \bibinfo {author} {\bibfnamefont {S.}~\bibnamefont {Bhattacharjee}},\ and\ \bibinfo {author} {\bibfnamefont {Y.~B.}\ \bibnamefont {Kim}},\ }\bibfield  {title} {\bibinfo {title} {Quantum phase transition in {H}eisenberg-{K}itaev model},\ }\href {https://doi.org/10.1103/PhysRevB.86.224417} {\bibfield  {journal} {\bibinfo  {journal} {Phys. Rev. B}\ }\textbf {\bibinfo {volume} {86}},\ \bibinfo {pages} {224417} (\bibinfo {year} {2012})}\BibitemShut {NoStop}%
\bibitem [{\citenamefont {Price}\ and\ \citenamefont {Perkins}(2012)}]{Price-Perkins-2012}%
  \BibitemOpen
  \bibfield  {author} {\bibinfo {author} {\bibfnamefont {C.~C.}\ \bibnamefont {Price}}\ and\ \bibinfo {author} {\bibfnamefont {N.~B.}\ \bibnamefont {Perkins}},\ }\bibfield  {title} {\bibinfo {title} {Critical properties of the {K}itaev-{H}eisenberg model},\ }\href {https://doi.org/10.1103/PhysRevLett.109.187201} {\bibfield  {journal} {\bibinfo  {journal} {Phys. Rev. Lett.}\ }\textbf {\bibinfo {volume} {109}},\ \bibinfo {pages} {187201} (\bibinfo {year} {2012})}\BibitemShut {NoStop}%
\bibitem [{\citenamefont {Chaloupka}\ \emph {et~al.}(2013)\citenamefont {Chaloupka}, \citenamefont {Jackeli},\ and\ \citenamefont {Khaliullin}}]{Chaloupka-2013}%
  \BibitemOpen
  \bibfield  {author} {\bibinfo {author} {\bibfnamefont {J.}~\bibnamefont {Chaloupka}}, \bibinfo {author} {\bibfnamefont {G.}~\bibnamefont {Jackeli}},\ and\ \bibinfo {author} {\bibfnamefont {G.}~\bibnamefont {Khaliullin}},\ }\bibfield  {title} {\bibinfo {title} {{Z}igzag {M}agnetic {O}rder in the {I}ridium {O}xide {N}a$_{2}${I}r{O}$_{3}$},\ }\href {https://doi.org/10.1103/PhysRevLett.110.097204} {\bibfield  {journal} {\bibinfo  {journal} {Phys. Rev. Lett.}\ }\textbf {\bibinfo {volume} {110}},\ \bibinfo {pages} {097204} (\bibinfo {year} {2013})}\BibitemShut {NoStop}%
\bibitem [{\citenamefont {Osorio~Iregui}\ \emph {et~al.}(2014)\citenamefont {Osorio~Iregui}, \citenamefont {Corboz},\ and\ \citenamefont {Troyer}}]{Iregui-2014}%
  \BibitemOpen
  \bibfield  {author} {\bibinfo {author} {\bibfnamefont {J.}~\bibnamefont {Osorio~Iregui}}, \bibinfo {author} {\bibfnamefont {P.}~\bibnamefont {Corboz}},\ and\ \bibinfo {author} {\bibfnamefont {M.}~\bibnamefont {Troyer}},\ }\bibfield  {title} {\bibinfo {title} {Probing the stability of the spin-liquid phases in the {K}itaev-{H}eisenberg model using tensor network algorithms},\ }\href {https://doi.org/10.1103/PhysRevB.90.195102} {\bibfield  {journal} {\bibinfo  {journal} {Phys. Rev. B}\ }\textbf {\bibinfo {volume} {90}},\ \bibinfo {pages} {195102} (\bibinfo {year} {2014})}\BibitemShut {NoStop}%
\bibitem [{\citenamefont {Sela}\ \emph {et~al.}(2014)\citenamefont {Sela}, \citenamefont {Jiang}, \citenamefont {Gerlach},\ and\ \citenamefont {Trebst}}]{Sela-2014}%
  \BibitemOpen
  \bibfield  {author} {\bibinfo {author} {\bibfnamefont {E.}~\bibnamefont {Sela}}, \bibinfo {author} {\bibfnamefont {H.-C.}\ \bibnamefont {Jiang}}, \bibinfo {author} {\bibfnamefont {M.~H.}\ \bibnamefont {Gerlach}},\ and\ \bibinfo {author} {\bibfnamefont {S.}~\bibnamefont {Trebst}},\ }\bibfield  {title} {\bibinfo {title} {Order-by-disorder and spin-orbital liquids in a distorted {H}eisenberg-{K}itaev model},\ }\href {https://doi.org/10.1103/PhysRevB.90.035113} {\bibfield  {journal} {\bibinfo  {journal} {Phys. Rev. B}\ }\textbf {\bibinfo {volume} {90}},\ \bibinfo {pages} {035113} (\bibinfo {year} {2014})}\BibitemShut {NoStop}%
\bibitem [{\citenamefont {Chaloupka}\ and\ \citenamefont {Khaliullin}(2015)}]{Chaloupka-Khaliullin-2015}%
  \BibitemOpen
  \bibfield  {author} {\bibinfo {author} {\bibfnamefont {J.}~\bibnamefont {Chaloupka}}\ and\ \bibinfo {author} {\bibfnamefont {G.}~\bibnamefont {Khaliullin}},\ }\bibfield  {title} {\bibinfo {title} {{H}idden symmetries of the extended {K}itaev-{H}eisenberg model: {I}mplications for the honeycomb-lattice iridates {$A_{2}\mathrm{IrO}_{3}$}},\ }\href {https://doi.org/10.1103/PhysRevB.92.024413} {\bibfield  {journal} {\bibinfo  {journal} {Phys. Rev. B}\ }\textbf {\bibinfo {volume} {92}},\ \bibinfo {pages} {024413} (\bibinfo {year} {2015})}\BibitemShut {NoStop}%
\bibitem [{\citenamefont {Winter}\ \emph {et~al.}(2016)\citenamefont {Winter}, \citenamefont {Li}, \citenamefont {Jeschke},\ and\ \citenamefont {Valent\'{\i}}}]{Winter-2016}%
  \BibitemOpen
  \bibfield  {author} {\bibinfo {author} {\bibfnamefont {S.~M.}\ \bibnamefont {Winter}}, \bibinfo {author} {\bibfnamefont {Y.}~\bibnamefont {Li}}, \bibinfo {author} {\bibfnamefont {H.~O.}\ \bibnamefont {Jeschke}},\ and\ \bibinfo {author} {\bibfnamefont {R.}~\bibnamefont {Valent\'{\i}}},\ }\bibfield  {title} {\bibinfo {title} {{C}hallenges in design of {K}itaev materials: {M}agnetic interactions from competing energy scales},\ }\href {https://doi.org/10.1103/PhysRevB.93.214431} {\bibfield  {journal} {\bibinfo  {journal} {Phys. Rev. B}\ }\textbf {\bibinfo {volume} {93}},\ \bibinfo {pages} {214431} (\bibinfo {year} {2016})}\BibitemShut {NoStop}%
\bibitem [{\citenamefont {Gotfryd}\ \emph {et~al.}(2017)\citenamefont {Gotfryd}, \citenamefont {Rusna{\v c}ko}, \citenamefont {Wohlfeld}, \citenamefont {Jackeli}, \citenamefont {Chaloupka},\ and\ \citenamefont {Ole{\'s}}}]{Gotfryd-2017}%
  \BibitemOpen
  \bibfield  {author} {\bibinfo {author} {\bibfnamefont {D.}~\bibnamefont {Gotfryd}}, \bibinfo {author} {\bibfnamefont {J.}~\bibnamefont {Rusna{\v c}ko}}, \bibinfo {author} {\bibfnamefont {K.}~\bibnamefont {Wohlfeld}}, \bibinfo {author} {\bibfnamefont {G.}~\bibnamefont {Jackeli}}, \bibinfo {author} {\bibfnamefont {J.}~\bibnamefont {Chaloupka}},\ and\ \bibinfo {author} {\bibfnamefont {A.~M.}\ \bibnamefont {Ole{\'s}}},\ }\bibfield  {title} {\bibinfo {title} {{P}hase diagram and spin correlations of the {K}itaev-{H}eisenberg model: {I}mportance of quantum effects},\ }\href {https://doi.org/10.1103/PhysRevB.95.024426} {\bibfield  {journal} {\bibinfo  {journal} {Phys. Rev. B}\ }\textbf {\bibinfo {volume} {95}},\ \bibinfo {pages} {024426} (\bibinfo {year} {2017})}\BibitemShut {NoStop}%
\bibitem [{\citenamefont {Singh}\ and\ \citenamefont {Gegenwart}(2010)}]{Singh-Gegenwart-2010}%
  \BibitemOpen
  \bibfield  {author} {\bibinfo {author} {\bibfnamefont {Y.}~\bibnamefont {Singh}}\ and\ \bibinfo {author} {\bibfnamefont {P.}~\bibnamefont {Gegenwart}},\ }\bibfield  {title} {\bibinfo {title} {{A}ntiferromagnetic {M}ott insulating state in single crystals of the honeycomb lattice material {N}a$_{2}${I}r{O}$_{3}$},\ }\href {https://doi.org/10.1103/PhysRevB.82.064412} {\bibfield  {journal} {\bibinfo  {journal} {Phys. Rev. B}\ }\textbf {\bibinfo {volume} {82}},\ \bibinfo {pages} {064412} (\bibinfo {year} {2010})}\BibitemShut {NoStop}%
\bibitem [{\citenamefont {Singh}\ \emph {et~al.}(2012)\citenamefont {Singh}, \citenamefont {Manni}, \citenamefont {Reuther}, \citenamefont {Berlijn}, \citenamefont {Thomale}, \citenamefont {Ku}, \citenamefont {Trebst},\ and\ \citenamefont {Gegenwart}}]{Singh-Manni-2012}%
  \BibitemOpen
  \bibfield  {author} {\bibinfo {author} {\bibfnamefont {Y.}~\bibnamefont {Singh}}, \bibinfo {author} {\bibfnamefont {S.}~\bibnamefont {Manni}}, \bibinfo {author} {\bibfnamefont {J.}~\bibnamefont {Reuther}}, \bibinfo {author} {\bibfnamefont {T.}~\bibnamefont {Berlijn}}, \bibinfo {author} {\bibfnamefont {R.}~\bibnamefont {Thomale}}, \bibinfo {author} {\bibfnamefont {W.}~\bibnamefont {Ku}}, \bibinfo {author} {\bibfnamefont {S.}~\bibnamefont {Trebst}},\ and\ \bibinfo {author} {\bibfnamefont {P.}~\bibnamefont {Gegenwart}},\ }\bibfield  {title} {\bibinfo {title} {{R}elevance of the {H}eisenberg-{K}itaev {M}odel for the {H}oneycomb {L}attice {I}ridates ${A}_{2}{\mathrm{iro}}_{3}$},\ }\href {https://doi.org/10.1103/PhysRevLett.108.127203} {\bibfield  {journal} {\bibinfo  {journal} {Phys. Rev. Lett.}\ }\textbf {\bibinfo {volume} {108}},\ \bibinfo {pages} {127203} (\bibinfo {year} {2012})}\BibitemShut {NoStop}%
\bibitem [{\citenamefont {Plumb}\ \emph {et~al.}(2014)\citenamefont {Plumb}, \citenamefont {Clancy}, \citenamefont {Sandilands}, \citenamefont {Shankar}, \citenamefont {Hu}, \citenamefont {Burch}, \citenamefont {Kee},\ and\ \citenamefont {Kim}}]{Plumb-2014}%
  \BibitemOpen
  \bibfield  {author} {\bibinfo {author} {\bibfnamefont {K.~W.}\ \bibnamefont {Plumb}}, \bibinfo {author} {\bibfnamefont {J.~P.}\ \bibnamefont {Clancy}}, \bibinfo {author} {\bibfnamefont {L.~J.}\ \bibnamefont {Sandilands}}, \bibinfo {author} {\bibfnamefont {V.~V.}\ \bibnamefont {Shankar}}, \bibinfo {author} {\bibfnamefont {Y.~F.}\ \bibnamefont {Hu}}, \bibinfo {author} {\bibfnamefont {K.~S.}\ \bibnamefont {Burch}}, \bibinfo {author} {\bibfnamefont {H.-Y.}\ \bibnamefont {Kee}},\ and\ \bibinfo {author} {\bibfnamefont {Y.-J.}\ \bibnamefont {Kim}},\ }\bibfield  {title} {\bibinfo {title} {$\alpha$-{R}u{C}l$_{3}$: A spin-orbit assisted {M}ott insulator on a honeycomb lattice},\ }\href {https://doi.org/10.1103/PhysRevB.90.041112} {\bibfield  {journal} {\bibinfo  {journal} {Phys. Rev. B}\ }\textbf {\bibinfo {volume} {90}},\ \bibinfo {pages} {041112} (\bibinfo {year} {2014})}\BibitemShut {NoStop}%
\bibitem [{\citenamefont {Kubota}\ \emph {et~al.}(2015)\citenamefont {Kubota}, \citenamefont {Tanaka}, \citenamefont {Ono}, \citenamefont {Narumi},\ and\ \citenamefont {Kindo}}]{Kubota-2015}%
  \BibitemOpen
  \bibfield  {author} {\bibinfo {author} {\bibfnamefont {Y.}~\bibnamefont {Kubota}}, \bibinfo {author} {\bibfnamefont {H.}~\bibnamefont {Tanaka}}, \bibinfo {author} {\bibfnamefont {T.}~\bibnamefont {Ono}}, \bibinfo {author} {\bibfnamefont {Y.}~\bibnamefont {Narumi}},\ and\ \bibinfo {author} {\bibfnamefont {K.}~\bibnamefont {Kindo}},\ }\bibfield  {title} {\bibinfo {title} {{S}uccessive magnetic phase transitions in $\alpha$-{R}u{C}l$_{3}$: {X}{Y}-like frustrated magnet on the honeycomb lattice},\ }\href {https://doi.org/10.1103/PhysRevB.91.094422} {\bibfield  {journal} {\bibinfo  {journal} {Phys. Rev. B}\ }\textbf {\bibinfo {volume} {91}},\ \bibinfo {pages} {094422} (\bibinfo {year} {2015})}\BibitemShut {NoStop}%
\bibitem [{\citenamefont {Sinn}\ \emph {et~al.}(2016)\citenamefont {Sinn}, \citenamefont {Kim}, \citenamefont {Kim}, \citenamefont {Lee}, \citenamefont {Won}, \citenamefont {Oh}, \citenamefont {Han}, \citenamefont {Chang}, \citenamefont {Hur}, \citenamefont {Sato}, \citenamefont {Park}, \citenamefont {Kim}, \citenamefont {Kim},\ and\ \citenamefont {Noh}}]{Sinn-2016}%
  \BibitemOpen
  \bibfield  {author} {\bibinfo {author} {\bibfnamefont {S.}~\bibnamefont {Sinn}}, \bibinfo {author} {\bibfnamefont {C.~H.}\ \bibnamefont {Kim}}, \bibinfo {author} {\bibfnamefont {B.~H.}\ \bibnamefont {Kim}}, \bibinfo {author} {\bibfnamefont {K.~D.}\ \bibnamefont {Lee}}, \bibinfo {author} {\bibfnamefont {C.~J.}\ \bibnamefont {Won}}, \bibinfo {author} {\bibfnamefont {J.~S.}\ \bibnamefont {Oh}}, \bibinfo {author} {\bibfnamefont {M.}~\bibnamefont {Han}}, \bibinfo {author} {\bibfnamefont {Y.~J.}\ \bibnamefont {Chang}}, \bibinfo {author} {\bibfnamefont {N.}~\bibnamefont {Hur}}, \bibinfo {author} {\bibfnamefont {H.}~\bibnamefont {Sato}}, \bibinfo {author} {\bibfnamefont {B.-G.}\ \bibnamefont {Park}}, \bibinfo {author} {\bibfnamefont {C.}~\bibnamefont {Kim}}, \bibinfo {author} {\bibfnamefont {H.-D.}\ \bibnamefont {Kim}},\ and\ \bibinfo {author} {\bibfnamefont {T.~W.}\ \bibnamefont {Noh}},\ }\bibfield  {title} {\bibinfo {title} {{E}lectronic {S}tructure of the {K}itaev {M}aterial $\alpha$-{R}u{C}l$_{3}$ {P}robed by
  {P}hotoemission and {I}nverse {P}hotoemission {S}pectroscopies},\ }\href {https://doi.org/10.1038/srep39544} {\bibfield  {journal} {\bibinfo  {journal} {Sci. Rep.}\ }\textbf {\bibinfo {volume} {6}},\ \bibinfo {pages} {39544} (\bibinfo {year} {2016})}\BibitemShut {NoStop}%
\bibitem [{\citenamefont {Banerjee}\ \emph {et~al.}(2016)\citenamefont {Banerjee}, \citenamefont {Bridges}, \citenamefont {Yan}, \citenamefont {Aczel}, \citenamefont {Li}, \citenamefont {Stone}, \citenamefont {Granroth}, \citenamefont {Lumsden}, \citenamefont {Yiu}, \citenamefont {Knolle}, \citenamefont {Bhattacharjee}, \citenamefont {Kovrizhin}, \citenamefont {Moessner}, \citenamefont {Tennant}, \citenamefont {Mandrus},\ and\ \citenamefont {Nagler}}]{Banerjee-2016}%
  \BibitemOpen
  \bibfield  {author} {\bibinfo {author} {\bibfnamefont {A.}~\bibnamefont {Banerjee}}, \bibinfo {author} {\bibfnamefont {C.~A.}\ \bibnamefont {Bridges}}, \bibinfo {author} {\bibfnamefont {J.-Q.}\ \bibnamefont {Yan}}, \bibinfo {author} {\bibfnamefont {A.~A.}\ \bibnamefont {Aczel}}, \bibinfo {author} {\bibfnamefont {L.}~\bibnamefont {Li}}, \bibinfo {author} {\bibfnamefont {M.~B.}\ \bibnamefont {Stone}}, \bibinfo {author} {\bibfnamefont {G.~E.}\ \bibnamefont {Granroth}}, \bibinfo {author} {\bibfnamefont {M.~D.}\ \bibnamefont {Lumsden}}, \bibinfo {author} {\bibfnamefont {Y.}~\bibnamefont {Yiu}}, \bibinfo {author} {\bibfnamefont {J.}~\bibnamefont {Knolle}}, \bibinfo {author} {\bibfnamefont {S.}~\bibnamefont {Bhattacharjee}}, \bibinfo {author} {\bibfnamefont {D.~L.}\ \bibnamefont {Kovrizhin}}, \bibinfo {author} {\bibfnamefont {R.}~\bibnamefont {Moessner}}, \bibinfo {author} {\bibfnamefont {D.~A.}\ \bibnamefont {Tennant}}, \bibinfo {author} {\bibfnamefont {D.~G.}\ \bibnamefont {Mandrus}},\ and\ \bibinfo {author}
  {\bibfnamefont {S.~E.}\ \bibnamefont {Nagler}},\ }\bibfield  {title} {\bibinfo {title} {{P}roximate {K}itaev quantum spin liquid behaviour in a honeycomb magnet},\ }\href {https://doi.org/10.1038/nmat4604} {\bibfield  {journal} {\bibinfo  {journal} {Nat. Mater.}\ }\textbf {\bibinfo {volume} {15}},\ \bibinfo {pages} {733} (\bibinfo {year} {2016})}\BibitemShut {NoStop}%
\bibitem [{\citenamefont {Banerjee}\ \emph {et~al.}(2017)\citenamefont {Banerjee}, \citenamefont {Yan}, \citenamefont {Knolle}, \citenamefont {Bridges}, \citenamefont {Stone}, \citenamefont {Lumsden}, \citenamefont {Mandrus}, \citenamefont {Tennant}, \citenamefont {Moessner},\ and\ \citenamefont {Nagler}}]{Banerjee-2017}%
  \BibitemOpen
  \bibfield  {author} {\bibinfo {author} {\bibfnamefont {A.}~\bibnamefont {Banerjee}}, \bibinfo {author} {\bibfnamefont {J.}~\bibnamefont {Yan}}, \bibinfo {author} {\bibfnamefont {J.}~\bibnamefont {Knolle}}, \bibinfo {author} {\bibfnamefont {C.~A.}\ \bibnamefont {Bridges}}, \bibinfo {author} {\bibfnamefont {M.~B.}\ \bibnamefont {Stone}}, \bibinfo {author} {\bibfnamefont {M.~D.}\ \bibnamefont {Lumsden}}, \bibinfo {author} {\bibfnamefont {D.~G.}\ \bibnamefont {Mandrus}}, \bibinfo {author} {\bibfnamefont {D.~A.}\ \bibnamefont {Tennant}}, \bibinfo {author} {\bibfnamefont {R.}~\bibnamefont {Moessner}},\ and\ \bibinfo {author} {\bibfnamefont {S.~E.}\ \bibnamefont {Nagler}},\ }\bibfield  {title} {\bibinfo {title} {{N}eutron scattering in the proximate quantum spin liquid {$\alpha$}-{R}u{C}l$_3$},\ }\href {https://doi.org/10.1126/science.aah6015} {\bibfield  {journal} {\bibinfo  {journal} {Science}\ }\textbf {\bibinfo {volume} {356}},\ \bibinfo {pages} {1055} (\bibinfo {year} {2017})}\BibitemShut {NoStop}%
\bibitem [{\citenamefont {Yoshitake}\ \emph {et~al.}(2016)\citenamefont {Yoshitake}, \citenamefont {Nasu},\ and\ \citenamefont {Motome}}]{Yoshitake-Nasu-Motome-2016}%
  \BibitemOpen
  \bibfield  {author} {\bibinfo {author} {\bibfnamefont {J.}~\bibnamefont {Yoshitake}}, \bibinfo {author} {\bibfnamefont {J.}~\bibnamefont {Nasu}},\ and\ \bibinfo {author} {\bibfnamefont {Y.}~\bibnamefont {Motome}},\ }\bibfield  {title} {\bibinfo {title} {{F}ractional {S}pin {F}luctuations as a {P}recursor of {Q}uantum {S}pin {L}iquids: {M}ajorana {D}ynamical {M}ean-{F}ield {S}tudy for the {K}itaev {M}odel},\ }\href {https://doi.org/10.1103/PhysRevLett.117.157203} {\bibfield  {journal} {\bibinfo  {journal} {Phys. Rev. Lett.}\ }\textbf {\bibinfo {volume} {117}},\ \bibinfo {pages} {157203} (\bibinfo {year} {2016})}\BibitemShut {NoStop}%
\bibitem [{\citenamefont {Yoshitake}\ \emph {et~al.}(2017)\citenamefont {Yoshitake}, \citenamefont {Nasu}, \citenamefont {Kato},\ and\ \citenamefont {Motome}}]{Yoshitake-Nasu-2017}%
  \BibitemOpen
  \bibfield  {author} {\bibinfo {author} {\bibfnamefont {J.}~\bibnamefont {Yoshitake}}, \bibinfo {author} {\bibfnamefont {J.}~\bibnamefont {Nasu}}, \bibinfo {author} {\bibfnamefont {Y.}~\bibnamefont {Kato}},\ and\ \bibinfo {author} {\bibfnamefont {Y.}~\bibnamefont {Motome}},\ }\bibfield  {title} {\bibinfo {title} {{M}ajorana dynamical mean-field study of spin dynamics at finite temperatures in the honeycomb {K}itaev model},\ }\href {https://doi.org/10.1103/PhysRevB.96.024438} {\bibfield  {journal} {\bibinfo  {journal} {Phys. Rev. B}\ }\textbf {\bibinfo {volume} {96}},\ \bibinfo {pages} {024438} (\bibinfo {year} {2017})}\BibitemShut {NoStop}%
\bibitem [{\citenamefont {Xu}\ \emph {et~al.}(2018)\citenamefont {Xu}, \citenamefont {Feng}, \citenamefont {Xiang},\ and\ \citenamefont {Bellaiche}}]{Xu-2018}%
  \BibitemOpen
  \bibfield  {author} {\bibinfo {author} {\bibfnamefont {C.}~\bibnamefont {Xu}}, \bibinfo {author} {\bibfnamefont {J.}~\bibnamefont {Feng}}, \bibinfo {author} {\bibfnamefont {H.}~\bibnamefont {Xiang}},\ and\ \bibinfo {author} {\bibfnamefont {L.}~\bibnamefont {Bellaiche}},\ }\bibfield  {title} {\bibinfo {title} {{I}nterplay between {K}itaev interaction and single ion anisotropy in ferromagnetic {C}r{I}$_{3}$ and {C}r{G}e{T}e$_{3}$ monolayers},\ }\href {https://doi.org/10.1038/s41524-018-0115-6} {\bibfield  {journal} {\bibinfo  {journal} {npj Comput. Mater.}\ }\textbf {\bibinfo {volume} {4}},\ \bibinfo {pages} {57} (\bibinfo {year} {2018})}\BibitemShut {NoStop}%
\bibitem [{\citenamefont {Stavropoulos}\ \emph {et~al.}(2019)\citenamefont {Stavropoulos}, \citenamefont {Pereira},\ and\ \citenamefont {Kee}}]{Stavropoulos-2019}%
  \BibitemOpen
  \bibfield  {author} {\bibinfo {author} {\bibfnamefont {P.~P.}\ \bibnamefont {Stavropoulos}}, \bibinfo {author} {\bibfnamefont {D.}~\bibnamefont {Pereira}},\ and\ \bibinfo {author} {\bibfnamefont {H.-Y.}\ \bibnamefont {Kee}},\ }\bibfield  {title} {\bibinfo {title} {{M}icroscopic {M}echanism for a {H}igher-{S}pin {K}itaev {M}odel},\ }\href {https://doi.org/10.1103/PhysRevLett.123.037203} {\bibfield  {journal} {\bibinfo  {journal} {Phys. Rev. Lett.}\ }\textbf {\bibinfo {volume} {123}},\ \bibinfo {pages} {037203} (\bibinfo {year} {2019})}\BibitemShut {NoStop}%
\bibitem [{\citenamefont {Lee}\ \emph {et~al.}(2020{\natexlab{a}})\citenamefont {Lee}, \citenamefont {Utermohlen}, \citenamefont {Weber}, \citenamefont {Hwang}, \citenamefont {Zhang}, \citenamefont {van Tol}, \citenamefont {Goldberger}, \citenamefont {Trivedi},\ and\ \citenamefont {Hammel}}]{Lee-2020}%
  \BibitemOpen
  \bibfield  {author} {\bibinfo {author} {\bibfnamefont {I.}~\bibnamefont {Lee}}, \bibinfo {author} {\bibfnamefont {F.~G.}\ \bibnamefont {Utermohlen}}, \bibinfo {author} {\bibfnamefont {D.}~\bibnamefont {Weber}}, \bibinfo {author} {\bibfnamefont {K.}~\bibnamefont {Hwang}}, \bibinfo {author} {\bibfnamefont {C.}~\bibnamefont {Zhang}}, \bibinfo {author} {\bibfnamefont {J.}~\bibnamefont {van Tol}}, \bibinfo {author} {\bibfnamefont {J.~E.}\ \bibnamefont {Goldberger}}, \bibinfo {author} {\bibfnamefont {N.}~\bibnamefont {Trivedi}},\ and\ \bibinfo {author} {\bibfnamefont {P.~C.}\ \bibnamefont {Hammel}},\ }\bibfield  {title} {\bibinfo {title} {{F}undamental {S}pin {I}nteractions {U}nderlying the {M}agnetic {A}nisotropy in the {K}itaev {F}erromagnet ${\mathrm{cri}}_{3}$},\ }\href {https://doi.org/10.1103/PhysRevLett.124.017201} {\bibfield  {journal} {\bibinfo  {journal} {Phys. Rev. Lett.}\ }\textbf {\bibinfo {volume} {124}},\ \bibinfo {pages} {017201} (\bibinfo {year} {2020}{\natexlab{a}})}\BibitemShut {NoStop}%
\bibitem [{\citenamefont {Stavropoulos}\ \emph {et~al.}(2021)\citenamefont {Stavropoulos}, \citenamefont {Liu},\ and\ \citenamefont {Kee}}]{Stavropoulos-2021}%
  \BibitemOpen
  \bibfield  {author} {\bibinfo {author} {\bibfnamefont {P.~P.}\ \bibnamefont {Stavropoulos}}, \bibinfo {author} {\bibfnamefont {X.}~\bibnamefont {Liu}},\ and\ \bibinfo {author} {\bibfnamefont {H.-Y.}\ \bibnamefont {Kee}},\ }\bibfield  {title} {\bibinfo {title} {{M}agnetic anisotropy in spin-3/2 with heavy ligand in honeycomb {M}ott insulators: {A}pplication to {C}r{I}$_{3}$},\ }\href {https://doi.org/10.1103/PhysRevResearch.3.013216} {\bibfield  {journal} {\bibinfo  {journal} {Phys. Rev. Res.}\ }\textbf {\bibinfo {volume} {3}},\ \bibinfo {pages} {013216} (\bibinfo {year} {2021})}\BibitemShut {NoStop}%
\bibitem [{\citenamefont {Xu}\ \emph {et~al.}(2020)\citenamefont {Xu}, \citenamefont {Feng}, \citenamefont {Kawamura}, \citenamefont {Yamaji}, \citenamefont {Nahas}, \citenamefont {Prokhorenko}, \citenamefont {Qi}, \citenamefont {Xiang},\ and\ \citenamefont {Bellaiche}}]{Xu-2020}%
  \BibitemOpen
  \bibfield  {author} {\bibinfo {author} {\bibfnamefont {C.}~\bibnamefont {Xu}}, \bibinfo {author} {\bibfnamefont {J.}~\bibnamefont {Feng}}, \bibinfo {author} {\bibfnamefont {M.}~\bibnamefont {Kawamura}}, \bibinfo {author} {\bibfnamefont {Y.}~\bibnamefont {Yamaji}}, \bibinfo {author} {\bibfnamefont {Y.}~\bibnamefont {Nahas}}, \bibinfo {author} {\bibfnamefont {S.}~\bibnamefont {Prokhorenko}}, \bibinfo {author} {\bibfnamefont {Y.}~\bibnamefont {Qi}}, \bibinfo {author} {\bibfnamefont {H.}~\bibnamefont {Xiang}},\ and\ \bibinfo {author} {\bibfnamefont {L.}~\bibnamefont {Bellaiche}},\ }\bibfield  {title} {\bibinfo {title} {{P}ossible {K}itaev {Q}uantum {S}pin {L}iquid {S}tate in 2{D} {M}aterials with ${S}=3/2$},\ }\href {https://doi.org/10.1103/PhysRevLett.124.087205} {\bibfield  {journal} {\bibinfo  {journal} {Phys. Rev. Lett.}\ }\textbf {\bibinfo {volume} {124}},\ \bibinfo {pages} {087205} (\bibinfo {year} {2020})}\BibitemShut {NoStop}%
\bibitem [{\citenamefont {Baskaran}\ \emph {et~al.}(2007)\citenamefont {Baskaran}, \citenamefont {Mandal},\ and\ \citenamefont {Shankar}}]{Baskaran-Mandal-2007}%
  \BibitemOpen
  \bibfield  {author} {\bibinfo {author} {\bibfnamefont {G.}~\bibnamefont {Baskaran}}, \bibinfo {author} {\bibfnamefont {S.}~\bibnamefont {Mandal}},\ and\ \bibinfo {author} {\bibfnamefont {R.}~\bibnamefont {Shankar}},\ }\bibfield  {title} {\bibinfo {title} {{E}xact {R}esults for {S}pin {D}ynamics and {F}ractionalization in the {K}itaev {M}odel},\ }\href {https://doi.org/10.1103/PhysRevLett.98.247201} {\bibfield  {journal} {\bibinfo  {journal} {Phys. Rev. Lett.}\ }\textbf {\bibinfo {volume} {98}},\ \bibinfo {pages} {247201} (\bibinfo {year} {2007})}\BibitemShut {NoStop}%
\bibitem [{\citenamefont {Baskaran}\ \emph {et~al.}(2008)\citenamefont {Baskaran}, \citenamefont {Sen},\ and\ \citenamefont {Shankar}}]{Baskaran-Sen-Shankar-2008}%
  \BibitemOpen
  \bibfield  {author} {\bibinfo {author} {\bibfnamefont {G.}~\bibnamefont {Baskaran}}, \bibinfo {author} {\bibfnamefont {D.}~\bibnamefont {Sen}},\ and\ \bibinfo {author} {\bibfnamefont {R.}~\bibnamefont {Shankar}},\ }\bibfield  {title} {\bibinfo {title} {{S}pin-${S}$ {K}itaev model: {C}lassical ground states, order from disorder, and exact correlation functions},\ }\href {https://doi.org/10.1103/PhysRevB.78.115116} {\bibfield  {journal} {\bibinfo  {journal} {Phys. Rev. B}\ }\textbf {\bibinfo {volume} {78}},\ \bibinfo {pages} {115116} (\bibinfo {year} {2008})}\BibitemShut {NoStop}%
\bibitem [{\citenamefont {Koga}\ \emph {et~al.}(2018)\citenamefont {Koga}, \citenamefont {Tomishige},\ and\ \citenamefont {Nasu}}]{Koga-2018}%
  \BibitemOpen
  \bibfield  {author} {\bibinfo {author} {\bibfnamefont {A.}~\bibnamefont {Koga}}, \bibinfo {author} {\bibfnamefont {H.}~\bibnamefont {Tomishige}},\ and\ \bibinfo {author} {\bibfnamefont {J.}~\bibnamefont {Nasu}},\ }\bibfield  {title} {\bibinfo {title} {{G}round-state and {T}hermodynamic {P}roperties of an ${S} = 1$ {K}itaev {M}odel},\ }\href {https://doi.org/10.7566/JPSJ.87.063703} {\bibfield  {journal} {\bibinfo  {journal} {J. Phys. Soc. Jpn.}\ }\textbf {\bibinfo {volume} {87}},\ \bibinfo {pages} {063703} (\bibinfo {year} {2018})}\BibitemShut {NoStop}%
\bibitem [{\citenamefont {Suzuki}\ and\ \citenamefont {Yamaji}(2018)}]{Suzuki-2018}%
  \BibitemOpen
  \bibfield  {author} {\bibinfo {author} {\bibfnamefont {T.}~\bibnamefont {Suzuki}}\ and\ \bibinfo {author} {\bibfnamefont {Y.}~\bibnamefont {Yamaji}},\ }\bibfield  {title} {\bibinfo {title} {{T}hermal properties of spin-${S}$ {K}itaev-{H}eisenberg model on a honeycomb lattice},\ }\href {https://doi.org/10.1016/j.physb.2017.09.105} {\bibfield  {journal} {\bibinfo  {journal} {Physica B}\ }\textbf {\bibinfo {volume} {536}},\ \bibinfo {pages} {637} (\bibinfo {year} {2018})}\BibitemShut {NoStop}%
\bibitem [{\citenamefont {Oitmaa}\ \emph {et~al.}(2018)\citenamefont {Oitmaa}, \citenamefont {Koga},\ and\ \citenamefont {Singh}}]{Oitmaa-2018}%
  \BibitemOpen
  \bibfield  {author} {\bibinfo {author} {\bibfnamefont {J.}~\bibnamefont {Oitmaa}}, \bibinfo {author} {\bibfnamefont {A.}~\bibnamefont {Koga}},\ and\ \bibinfo {author} {\bibfnamefont {R.~R.~P.}\ \bibnamefont {Singh}},\ }\bibfield  {title} {\bibinfo {title} {{I}ncipient and well-developed entropy plateaus in spin-${S}$ {K}itaev models},\ }\href {https://doi.org/10.1103/PhysRevB.98.214404} {\bibfield  {journal} {\bibinfo  {journal} {Phys. Rev. B}\ }\textbf {\bibinfo {volume} {98}},\ \bibinfo {pages} {214404} (\bibinfo {year} {2018})}\BibitemShut {NoStop}%
\bibitem [{\citenamefont {Minakawa}\ \emph {et~al.}(2019)\citenamefont {Minakawa}, \citenamefont {Nasu},\ and\ \citenamefont {Koga}}]{Minakawa-2019}%
  \BibitemOpen
  \bibfield  {author} {\bibinfo {author} {\bibfnamefont {T.}~\bibnamefont {Minakawa}}, \bibinfo {author} {\bibfnamefont {J.}~\bibnamefont {Nasu}},\ and\ \bibinfo {author} {\bibfnamefont {A.}~\bibnamefont {Koga}},\ }\bibfield  {title} {\bibinfo {title} {{Q}uantum and classical behavior of spin-${S}$ {K}itaev models in the anisotropic limit},\ }\href {https://doi.org/10.1103/PhysRevB.99.104408} {\bibfield  {journal} {\bibinfo  {journal} {Phys. Rev. B}\ }\textbf {\bibinfo {volume} {99}},\ \bibinfo {pages} {104408} (\bibinfo {year} {2019})}\BibitemShut {NoStop}%
\bibitem [{\citenamefont {Koga}\ \emph {et~al.}(2020)\citenamefont {Koga}, \citenamefont {Minakawa}, \citenamefont {Murakami},\ and\ \citenamefont {Nasu}}]{Koga-2020}%
  \BibitemOpen
  \bibfield  {author} {\bibinfo {author} {\bibfnamefont {A.}~\bibnamefont {Koga}}, \bibinfo {author} {\bibfnamefont {T.}~\bibnamefont {Minakawa}}, \bibinfo {author} {\bibfnamefont {Y.}~\bibnamefont {Murakami}},\ and\ \bibinfo {author} {\bibfnamefont {J.}~\bibnamefont {Nasu}},\ }\bibfield  {title} {\bibinfo {title} {{S}pin {T}ransport in the {Q}uantum {S}pin {L}iquid {S}tate in the ${S} = 1$ {K}itaev {M}odel: {R}ole of the {F}ractionalized {Q}uasiparticles},\ }\href {https://doi.org/10.7566/JPSJ.89.033701} {\bibfield  {journal} {\bibinfo  {journal} {J. Phys. Soc. Jpn.}\ }\textbf {\bibinfo {volume} {89}},\ \bibinfo {pages} {033701} (\bibinfo {year} {2020})}\BibitemShut {NoStop}%
\bibitem [{\citenamefont {Khait}\ \emph {et~al.}(2021)\citenamefont {Khait}, \citenamefont {Stavropoulos}, \citenamefont {Kee},\ and\ \citenamefont {Kim}}]{Khait-Stavropoulos-2021}%
  \BibitemOpen
  \bibfield  {author} {\bibinfo {author} {\bibfnamefont {I.}~\bibnamefont {Khait}}, \bibinfo {author} {\bibfnamefont {P.~P.}\ \bibnamefont {Stavropoulos}}, \bibinfo {author} {\bibfnamefont {H.-Y.}\ \bibnamefont {Kee}},\ and\ \bibinfo {author} {\bibfnamefont {Y.~B.}\ \bibnamefont {Kim}},\ }\bibfield  {title} {\bibinfo {title} {{C}haracterizing spin-one {K}itaev quantum spin liquids},\ }\href {https://doi.org/10.1103/PhysRevResearch.3.013160} {\bibfield  {journal} {\bibinfo  {journal} {Phys. Rev. Res.}\ }\textbf {\bibinfo {volume} {3}},\ \bibinfo {pages} {013160} (\bibinfo {year} {2021})}\BibitemShut {NoStop}%
\bibitem [{\citenamefont {Lee}\ \emph {et~al.}(2020{\natexlab{b}})\citenamefont {Lee}, \citenamefont {Kawashima},\ and\ \citenamefont {Kim}}]{Lee-Kawashima-Kim-2020}%
  \BibitemOpen
  \bibfield  {author} {\bibinfo {author} {\bibfnamefont {H.-Y.}\ \bibnamefont {Lee}}, \bibinfo {author} {\bibfnamefont {N.}~\bibnamefont {Kawashima}},\ and\ \bibinfo {author} {\bibfnamefont {Y.~B.}\ \bibnamefont {Kim}},\ }\bibfield  {title} {\bibinfo {title} {{T}ensor network wave function of ${S}=1$ {K}itaev spin liquids},\ }\href {https://doi.org/10.1103/PhysRevResearch.2.033318} {\bibfield  {journal} {\bibinfo  {journal} {Phys. Rev. Res.}\ }\textbf {\bibinfo {volume} {2}},\ \bibinfo {pages} {033318} (\bibinfo {year} {2020}{\natexlab{b}})}\BibitemShut {NoStop}%
\bibitem [{\citenamefont {Hickey}\ \emph {et~al.}(2020)\citenamefont {Hickey}, \citenamefont {Berke}, \citenamefont {Stavropoulos}, \citenamefont {Kee},\ and\ \citenamefont {Trebst}}]{Hickey-2020}%
  \BibitemOpen
  \bibfield  {author} {\bibinfo {author} {\bibfnamefont {C.}~\bibnamefont {Hickey}}, \bibinfo {author} {\bibfnamefont {C.}~\bibnamefont {Berke}}, \bibinfo {author} {\bibfnamefont {P.~P.}\ \bibnamefont {Stavropoulos}}, \bibinfo {author} {\bibfnamefont {H.-Y.}\ \bibnamefont {Kee}},\ and\ \bibinfo {author} {\bibfnamefont {S.}~\bibnamefont {Trebst}},\ }\bibfield  {title} {\bibinfo {title} {{F}ield-driven gapless spin liquid in the spin-1 {K}itaev honeycomb model},\ }\href {https://doi.org/10.1103/PhysRevResearch.2.023361} {\bibfield  {journal} {\bibinfo  {journal} {Phys. Rev. Res.}\ }\textbf {\bibinfo {volume} {2}},\ \bibinfo {pages} {023361} (\bibinfo {year} {2020})}\BibitemShut {NoStop}%
\bibitem [{\citenamefont {Zhu}\ \emph {et~al.}(2020)\citenamefont {Zhu}, \citenamefont {Weng},\ and\ \citenamefont {Sheng}}]{Zhu-2020}%
  \BibitemOpen
  \bibfield  {author} {\bibinfo {author} {\bibfnamefont {Z.}~\bibnamefont {Zhu}}, \bibinfo {author} {\bibfnamefont {Z.-Y.}\ \bibnamefont {Weng}},\ and\ \bibinfo {author} {\bibfnamefont {D.~N.}\ \bibnamefont {Sheng}},\ }\bibfield  {title} {\bibinfo {title} {{M}agnetic field induced spin liquids in ${S}=1$ {K}itaev honeycomb model},\ }\href {https://doi.org/10.1103/PhysRevResearch.2.022047} {\bibfield  {journal} {\bibinfo  {journal} {Phys. Rev. Res.}\ }\textbf {\bibinfo {volume} {2}},\ \bibinfo {pages} {022047} (\bibinfo {year} {2020})}\BibitemShut {NoStop}%
\bibitem [{\citenamefont {Bradley}\ and\ \citenamefont {Singh}(2022)}]{Bradley-2022}%
  \BibitemOpen
  \bibfield  {author} {\bibinfo {author} {\bibfnamefont {O.}~\bibnamefont {Bradley}}\ and\ \bibinfo {author} {\bibfnamefont {R.~R.~P.}\ \bibnamefont {Singh}},\ }\bibfield  {title} {\bibinfo {title} {{I}nstabilities of spin-1 {K}itaev spin liquid phase in presence of single-ion anisotropies},\ }\href {https://doi.org/10.1103/PhysRevB.105.L060405} {\bibfield  {journal} {\bibinfo  {journal} {Phys. Rev. B}\ }\textbf {\bibinfo {volume} {105}},\ \bibinfo {pages} {L060405} (\bibinfo {year} {2022})}\BibitemShut {NoStop}%
\bibitem [{\citenamefont {Chen}\ \emph {et~al.}(2022)\citenamefont {Chen}, \citenamefont {Genzor}, \citenamefont {Kim},\ and\ \citenamefont {Kao}}]{Chen-2022}%
  \BibitemOpen
  \bibfield  {author} {\bibinfo {author} {\bibfnamefont {Y.-H.}\ \bibnamefont {Chen}}, \bibinfo {author} {\bibfnamefont {J.}~\bibnamefont {Genzor}}, \bibinfo {author} {\bibfnamefont {Y.~B.}\ \bibnamefont {Kim}},\ and\ \bibinfo {author} {\bibfnamefont {Y.-J.}\ \bibnamefont {Kao}},\ }\bibfield  {title} {\bibinfo {title} {{E}xcitation spectrum of spin-1 {K}itaev spin liquids},\ }\href {https://doi.org/10.1103/PhysRevB.105.L060403} {\bibfield  {journal} {\bibinfo  {journal} {Phys. Rev. B}\ }\textbf {\bibinfo {volume} {105}},\ \bibinfo {pages} {L060403} (\bibinfo {year} {2022})}\BibitemShut {NoStop}%
\bibitem [{\citenamefont {Jin}\ \emph {et~al.}(2022)\citenamefont {Jin}, \citenamefont {Natori}, \citenamefont {Pollmann},\ and\ \citenamefont {Knolle}}]{Jin-2022}%
  \BibitemOpen
  \bibfield  {author} {\bibinfo {author} {\bibfnamefont {H.-K.}\ \bibnamefont {Jin}}, \bibinfo {author} {\bibfnamefont {W.~M.~H.}\ \bibnamefont {Natori}}, \bibinfo {author} {\bibfnamefont {F.}~\bibnamefont {Pollmann}},\ and\ \bibinfo {author} {\bibfnamefont {J.}~\bibnamefont {Knolle}},\ }\bibfield  {title} {\bibinfo {title} {{U}nveiling the ${S}=3/2$ {K}itaev honeycomb spin liquids},\ }\href {https://doi.org/10.1038/s41467-022-31503-0} {\bibfield  {journal} {\bibinfo  {journal} {Nat. Commun.}\ }\textbf {\bibinfo {volume} {13}},\ \bibinfo {pages} {3813} (\bibinfo {year} {2022})}\BibitemShut {NoStop}%
\bibitem [{\citenamefont {Ralko}\ and\ \citenamefont {Merino}(2024)}]{Ralko-Merino-2024}%
  \BibitemOpen
  \bibfield  {author} {\bibinfo {author} {\bibfnamefont {A.}~\bibnamefont {Ralko}}\ and\ \bibinfo {author} {\bibfnamefont {J.}~\bibnamefont {Merino}},\ }\bibfield  {title} {\bibinfo {title} {{C}hiral bosonic quantum spin liquid in the integer-spin {H}eisenberg-{K}itaev model},\ }\href {https://doi.org/10.1103/PhysRevB.110.134402} {\bibfield  {journal} {\bibinfo  {journal} {Phys. Rev. B}\ }\textbf {\bibinfo {volume} {110}},\ \bibinfo {pages} {134402} (\bibinfo {year} {2024})}\BibitemShut {NoStop}%
\bibitem [{\citenamefont {Sasamoto}\ and\ \citenamefont {Nasu}(2026{\natexlab{a}})}]{Sasamoto-Nasu-2025}%
  \BibitemOpen
  \bibfield  {author} {\bibinfo {author} {\bibfnamefont {D.}~\bibnamefont {Sasamoto}}\ and\ \bibinfo {author} {\bibfnamefont {J.}~\bibnamefont {Nasu}},\ }\bibfield  {title} {\bibinfo {title} {{S}chwinger boson theory for ${S}=1$ {K}itaev quantum spin liquids},\ }\href {https://doi.org/10.1103/5r5d-nczv} {\bibfield  {journal} {\bibinfo  {journal} {Phys. Rev. Res.}\ }\textbf {\bibinfo {volume} {8}},\ \bibinfo {pages} {033138} (\bibinfo {year} {2026}{\natexlab{a}})}\BibitemShut {NoStop}%
\bibitem [{\citenamefont {Sasamoto}\ \emph {et~al.}(2026)\citenamefont {Sasamoto}, \citenamefont {Ralko}, \citenamefont {Merino},\ and\ \citenamefont {Nasu}}]{Sasamoto-Ralko-Merino-Nasu-2026}%
  \BibitemOpen
  \bibfield  {author} {\bibinfo {author} {\bibfnamefont {D.}~\bibnamefont {Sasamoto}}, \bibinfo {author} {\bibfnamefont {A.}~\bibnamefont {Ralko}}, \bibinfo {author} {\bibfnamefont {J.}~\bibnamefont {Merino}},\ and\ \bibinfo {author} {\bibfnamefont {J.}~\bibnamefont {Nasu}},\ }\href@noop {} {\bibinfo {title} {{C}hiral bosonic mean-field {A}nsatz and spin dynamics in spin-1 {K}itaev magnets}} (\bibinfo {year} {2026}),\ \Eprint {https://arxiv.org/abs/2608.19178} {arXiv:2608.19178} \BibitemShut {NoStop}%
\bibitem [{\citenamefont {Dong}\ and\ \citenamefont {Sheng}(2020)}]{Dong-Sheng-2020}%
  \BibitemOpen
  \bibfield  {author} {\bibinfo {author} {\bibfnamefont {X.-Y.}\ \bibnamefont {Dong}}\ and\ \bibinfo {author} {\bibfnamefont {D.~N.}\ \bibnamefont {Sheng}},\ }\bibfield  {title} {\bibinfo {title} {{S}pin-1 {K}itaev-{H}eisenberg model on a honeycomb lattice},\ }\href {https://doi.org/10.1103/PhysRevB.102.121102} {\bibfield  {journal} {\bibinfo  {journal} {Phys. Rev. B}\ }\textbf {\bibinfo {volume} {102}},\ \bibinfo {pages} {121102} (\bibinfo {year} {2020})}\BibitemShut {NoStop}%
\bibitem [{\citenamefont {Fukui}\ \emph {et~al.}(2022)\citenamefont {Fukui}, \citenamefont {Kato}, \citenamefont {Nasu},\ and\ \citenamefont {Motome}}]{Fukui-Kato-Nasu-Motome-2022}%
  \BibitemOpen
  \bibfield  {author} {\bibinfo {author} {\bibfnamefont {K.}~\bibnamefont {Fukui}}, \bibinfo {author} {\bibfnamefont {Y.}~\bibnamefont {Kato}}, \bibinfo {author} {\bibfnamefont {J.}~\bibnamefont {Nasu}},\ and\ \bibinfo {author} {\bibfnamefont {Y.}~\bibnamefont {Motome}},\ }\bibfield  {title} {\bibinfo {title} {{G}round-state phase diagram of spin-${S}$ {K}itaev-{H}eisenberg models},\ }\href {https://doi.org/10.1103/PhysRevB.106.174416} {\bibfield  {journal} {\bibinfo  {journal} {Phys. Rev. B}\ }\textbf {\bibinfo {volume} {106}},\ \bibinfo {pages} {174416} (\bibinfo {year} {2022})}\BibitemShut {NoStop}%
\bibitem [{\citenamefont {Georgiou}\ \emph {et~al.}(2024)\citenamefont {Georgiou}, \citenamefont {Rousochatzakis}, \citenamefont {Farnell}, \citenamefont {Richter},\ and\ \citenamefont {Bishop}}]{Georgiou-2024}%
  \BibitemOpen
  \bibfield  {author} {\bibinfo {author} {\bibfnamefont {M.}~\bibnamefont {Georgiou}}, \bibinfo {author} {\bibfnamefont {I.}~\bibnamefont {Rousochatzakis}}, \bibinfo {author} {\bibfnamefont {D.~J.~J.}\ \bibnamefont {Farnell}}, \bibinfo {author} {\bibfnamefont {J.}~\bibnamefont {Richter}},\ and\ \bibinfo {author} {\bibfnamefont {R.~F.}\ \bibnamefont {Bishop}},\ }\bibfield  {title} {\bibinfo {title} {{S}pin-${S}$ {K}itaev-{H}eisenberg model on the honeycomb lattice: {A} high-order treatment via the many-body coupled cluster method},\ }\href {https://doi.org/10.1103/PhysRevResearch.6.033168} {\bibfield  {journal} {\bibinfo  {journal} {Phys. Rev. Res.}\ }\textbf {\bibinfo {volume} {6}},\ \bibinfo {pages} {033168} (\bibinfo {year} {2024})}\BibitemShut {NoStop}%
\bibitem [{\citenamefont {Tomishige}\ \emph {et~al.}(2018)\citenamefont {Tomishige}, \citenamefont {Nasu},\ and\ \citenamefont {Koga}}]{Tomishige-Nasu-Koga-2018}%
  \BibitemOpen
  \bibfield  {author} {\bibinfo {author} {\bibfnamefont {H.}~\bibnamefont {Tomishige}}, \bibinfo {author} {\bibfnamefont {J.}~\bibnamefont {Nasu}},\ and\ \bibinfo {author} {\bibfnamefont {A.}~\bibnamefont {Koga}},\ }\bibfield  {title} {\bibinfo {title} {Interlayer coupling effect on a bilayer {K}itaev model},\ }\href {https://doi.org/10.1103/PhysRevB.97.094403} {\bibfield  {journal} {\bibinfo  {journal} {Phys. Rev. B}\ }\textbf {\bibinfo {volume} {97}},\ \bibinfo {pages} {094403} (\bibinfo {year} {2018})}\BibitemShut {NoStop}%
\bibitem [{\citenamefont {Seifert}\ \emph {et~al.}(2018)\citenamefont {Seifert}, \citenamefont {Gritsch}, \citenamefont {Wagner}, \citenamefont {Joshi}, \citenamefont {Brenig}, \citenamefont {Vojta},\ and\ \citenamefont {Schmidt}}]{Seifert-2018}%
  \BibitemOpen
  \bibfield  {author} {\bibinfo {author} {\bibfnamefont {U.~F.~P.}\ \bibnamefont {Seifert}}, \bibinfo {author} {\bibfnamefont {J.}~\bibnamefont {Gritsch}}, \bibinfo {author} {\bibfnamefont {E.}~\bibnamefont {Wagner}}, \bibinfo {author} {\bibfnamefont {D.~G.}\ \bibnamefont {Joshi}}, \bibinfo {author} {\bibfnamefont {W.}~\bibnamefont {Brenig}}, \bibinfo {author} {\bibfnamefont {M.}~\bibnamefont {Vojta}},\ and\ \bibinfo {author} {\bibfnamefont {K.~P.}\ \bibnamefont {Schmidt}},\ }\bibfield  {title} {\bibinfo {title} {Bilayer {K}itaev models: {P}hase diagrams and novel phases},\ }\href {https://doi.org/10.1103/PhysRevB.98.155101} {\bibfield  {journal} {\bibinfo  {journal} {Phys. Rev. B}\ }\textbf {\bibinfo {volume} {98}},\ \bibinfo {pages} {155101} (\bibinfo {year} {2018})}\BibitemShut {NoStop}%
\bibitem [{\citenamefont {Tomishige}\ \emph {et~al.}(2019)\citenamefont {Tomishige}, \citenamefont {Nasu},\ and\ \citenamefont {Koga}}]{Tomishige-Nasu-Koga-2019}%
  \BibitemOpen
  \bibfield  {author} {\bibinfo {author} {\bibfnamefont {H.}~\bibnamefont {Tomishige}}, \bibinfo {author} {\bibfnamefont {J.}~\bibnamefont {Nasu}},\ and\ \bibinfo {author} {\bibfnamefont {A.}~\bibnamefont {Koga}},\ }\bibfield  {title} {\bibinfo {title} {Low-temperature properties in the bilayer {K}itaev model},\ }\href {https://doi.org/10.1103/PhysRevB.99.174424} {\bibfield  {journal} {\bibinfo  {journal} {Phys. Rev. B}\ }\textbf {\bibinfo {volume} {99}},\ \bibinfo {pages} {174424} (\bibinfo {year} {2019})}\BibitemShut {NoStop}%
\bibitem [{\citenamefont {Merino}\ and\ \citenamefont {Ralko}(2025)}]{Merino-Ralko-2025}%
  \BibitemOpen
  \bibfield  {author} {\bibinfo {author} {\bibfnamefont {J.}~\bibnamefont {Merino}}\ and\ \bibinfo {author} {\bibfnamefont {A.}~\bibnamefont {Ralko}},\ }\bibfield  {title} {\bibinfo {title} {Even-odd effect in multilayer {K}itaev honeycomb magnets},\ }\href {https://doi.org/10.1103/PhysRevB.111.085152} {\bibfield  {journal} {\bibinfo  {journal} {Phys. Rev. B}\ }\textbf {\bibinfo {volume} {111}},\ \bibinfo {pages} {085152} (\bibinfo {year} {2025})}\BibitemShut {NoStop}%
\bibitem [{\citenamefont {Ma}(2023)}]{Ma-2023}%
  \BibitemOpen
  \bibfield  {author} {\bibinfo {author} {\bibfnamefont {H.}~\bibnamefont {Ma}},\ }\bibfield  {title} {\bibinfo {title} {${{\mathbb{Z}}_{2}}$ {S}pin {L}iquids in the {H}igher {S}pin-${S}$ {K}itaev {H}oneycomb {M}odel: {A}n {E}xact {D}econfined ${{\mathbb{Z}}_{2}}$ {G}auge {S}tructure in a {N}onintegrable {M}odel},\ }\href {https://doi.org/10.1103/PhysRevLett.130.156701} {\bibfield  {journal} {\bibinfo  {journal} {Phys. Rev. Lett.}\ }\textbf {\bibinfo {volume} {130}},\ \bibinfo {pages} {156701} (\bibinfo {year} {2023})}\BibitemShut {NoStop}%
\bibitem [{\citenamefont {Khaliullin}(2005)}]{Khaliullin-2005}%
  \BibitemOpen
  \bibfield  {author} {\bibinfo {author} {\bibfnamefont {G.}~\bibnamefont {Khaliullin}},\ }\bibfield  {title} {\bibinfo {title} {{O}rbital order and fluctuations in {M}ott insulators},\ }\href {https://doi.org/10.1143/PTPS.160.155} {\bibfield  {journal} {\bibinfo  {journal} {Prog. Theor. Phys. Suppl.}\ }\textbf {\bibinfo {volume} {160}},\ \bibinfo {pages} {155} (\bibinfo {year} {2005})}\BibitemShut {NoStop}%
\bibitem [{\citenamefont {Kimchi}\ and\ \citenamefont {Vishwanath}(2014)}]{Kimchi-Vishwanath-2014}%
  \BibitemOpen
  \bibfield  {author} {\bibinfo {author} {\bibfnamefont {I.}~\bibnamefont {Kimchi}}\ and\ \bibinfo {author} {\bibfnamefont {A.}~\bibnamefont {Vishwanath}},\ }\bibfield  {title} {\bibinfo {title} {{K}itaev-{H}eisenberg models for iridates on the triangular, hyperkagome, kagome, fcc, and pyrochlore lattices},\ }\href {https://doi.org/10.1103/PhysRevB.89.014414} {\bibfield  {journal} {\bibinfo  {journal} {Phys. Rev. B}\ }\textbf {\bibinfo {volume} {89}},\ \bibinfo {pages} {014414} (\bibinfo {year} {2014})}\BibitemShut {NoStop}%
\bibitem [{\citenamefont {Read}\ and\ \citenamefont {Sachdev}(1991)}]{Read-Sachdev-1991}%
  \BibitemOpen
  \bibfield  {author} {\bibinfo {author} {\bibfnamefont {N.}~\bibnamefont {Read}}\ and\ \bibinfo {author} {\bibfnamefont {S.}~\bibnamefont {Sachdev}},\ }\bibfield  {title} {\bibinfo {title} {{L}arge-${N}$ expansion for frustrated quantum antiferromagnets},\ }\href {https://doi.org/10.1103/PhysRevLett.66.1773} {\bibfield  {journal} {\bibinfo  {journal} {Phys. Rev. Lett.}\ }\textbf {\bibinfo {volume} {66}},\ \bibinfo {pages} {1773} (\bibinfo {year} {1991})}\BibitemShut {NoStop}%
\bibitem [{\citenamefont {Arovas}\ and\ \citenamefont {Auerbach}(1988)}]{Arovas-Auerbach-1998}%
  \BibitemOpen
  \bibfield  {author} {\bibinfo {author} {\bibfnamefont {D.~P.}\ \bibnamefont {Arovas}}\ and\ \bibinfo {author} {\bibfnamefont {A.}~\bibnamefont {Auerbach}},\ }\bibfield  {title} {\bibinfo {title} {{F}unctional integral theories of low-dimensional quantum {H}eisenberg models},\ }\href {https://doi.org/10.1103/PhysRevB.38.316} {\bibfield  {journal} {\bibinfo  {journal} {Phys. Rev. B}\ }\textbf {\bibinfo {volume} {38}},\ \bibinfo {pages} {316} (\bibinfo {year} {1988})}\BibitemShut {NoStop}%
\bibitem [{\citenamefont {Sachdev}\ and\ \citenamefont {Read}(1991)}]{Sachdev-Read-1991}%
  \BibitemOpen
  \bibfield  {author} {\bibinfo {author} {\bibfnamefont {S.}~\bibnamefont {Sachdev}}\ and\ \bibinfo {author} {\bibfnamefont {N.}~\bibnamefont {Read}},\ }\bibfield  {title} {\bibinfo {title} {Large-{N} expansion for frustrated and doped quantum antiferromagnets},\ }\href {https://doi.org/10.1142/S0217979291000158} {\bibfield  {journal} {\bibinfo  {journal} {Int. J. Mod. Phys. B}\ }\textbf {\bibinfo {volume} {05}},\ \bibinfo {pages} {219} (\bibinfo {year} {1991})}\BibitemShut {NoStop}%
\bibitem [{\citenamefont {Sachdev}(1992)}]{Sachdev-1992}%
  \BibitemOpen
  \bibfield  {author} {\bibinfo {author} {\bibfnamefont {S.}~\bibnamefont {Sachdev}},\ }\bibfield  {title} {\bibinfo {title} {{K}agom{\'e}- and triangular-lattice {H}eisenberg antiferromagnets: {O}rdering from quantum fluctuations and quantum-disordered ground states with unconfined bosonic spinons},\ }\href {https://doi.org/10.1103/PhysRevB.45.12377} {\bibfield  {journal} {\bibinfo  {journal} {Phys. Rev. B}\ }\textbf {\bibinfo {volume} {45}},\ \bibinfo {pages} {12377} (\bibinfo {year} {1992})}\BibitemShut {NoStop}%
\bibitem [{\citenamefont {Kargarian}\ \emph {et~al.}(2012)\citenamefont {Kargarian}, \citenamefont {Langari},\ and\ \citenamefont {Fiete}}]{Kargarian-Langari-Fiete-2012}%
  \BibitemOpen
  \bibfield  {author} {\bibinfo {author} {\bibfnamefont {M.}~\bibnamefont {Kargarian}}, \bibinfo {author} {\bibfnamefont {A.}~\bibnamefont {Langari}},\ and\ \bibinfo {author} {\bibfnamefont {G.~A.}\ \bibnamefont {Fiete}},\ }\bibfield  {title} {\bibinfo {title} {{U}nusual magnetic phases in the strong interaction limit of two-dimensional topological band insulators in transition metal oxides},\ }\href {https://doi.org/10.1103/PhysRevB.86.205124} {\bibfield  {journal} {\bibinfo  {journal} {Phys. Rev. B}\ }\textbf {\bibinfo {volume} {86}},\ \bibinfo {pages} {205124} (\bibinfo {year} {2012})}\BibitemShut {NoStop}%
\bibitem [{\citenamefont {Kos}\ and\ \citenamefont {Punk}(2017)}]{Kos-Punk-2017}%
  \BibitemOpen
  \bibfield  {author} {\bibinfo {author} {\bibfnamefont {P.}~\bibnamefont {Kos}}\ and\ \bibinfo {author} {\bibfnamefont {M.}~\bibnamefont {Punk}},\ }\bibfield  {title} {\bibinfo {title} {{Q}uantum spin liquid ground states of the {H}eisenberg-{K}itaev model on the triangular lattice},\ }\href {https://doi.org/10.1103/PhysRevB.95.024421} {\bibfield  {journal} {\bibinfo  {journal} {Phys. Rev. B}\ }\textbf {\bibinfo {volume} {95}},\ \bibinfo {pages} {024421} (\bibinfo {year} {2017})}\BibitemShut {NoStop}%
\bibitem [{\citenamefont {Samajdar}\ \emph {et~al.}(2019)\citenamefont {Samajdar}, \citenamefont {Scheurer}, \citenamefont {Chatterjee}, \citenamefont {Guo}, \citenamefont {Xu},\ and\ \citenamefont {Sachdev}}]{Samajdar-Scheurer-2019}%
  \BibitemOpen
  \bibfield  {author} {\bibinfo {author} {\bibfnamefont {R.}~\bibnamefont {Samajdar}}, \bibinfo {author} {\bibfnamefont {M.~S.}\ \bibnamefont {Scheurer}}, \bibinfo {author} {\bibfnamefont {S.}~\bibnamefont {Chatterjee}}, \bibinfo {author} {\bibfnamefont {H.}~\bibnamefont {Guo}}, \bibinfo {author} {\bibfnamefont {C.}~\bibnamefont {Xu}},\ and\ \bibinfo {author} {\bibfnamefont {S.}~\bibnamefont {Sachdev}},\ }\bibfield  {title} {\bibinfo {title} {{E}nhanced thermal {H}all effect in the square-lattice {N}{\'e}el state},\ }\href {https://doi.org/10.1038/s41567-019-0669-3} {\bibfield  {journal} {\bibinfo  {journal} {Nat. Phys.}\ }\textbf {\bibinfo {volume} {15}},\ \bibinfo {pages} {1290} (\bibinfo {year} {2019})}\BibitemShut {NoStop}%
\bibitem [{\citenamefont {Mondal}\ and\ \citenamefont {Kadolkar}(2021)}]{Mondal-2021}%
  \BibitemOpen
  \bibfield  {author} {\bibinfo {author} {\bibfnamefont {K.}~\bibnamefont {Mondal}}\ and\ \bibinfo {author} {\bibfnamefont {C.}~\bibnamefont {Kadolkar}},\ }\bibfield  {title} {\bibinfo {title} {${Q} = 0$ order in quantum kagome {H}eisenberg antiferromagnet},\ }\href {https://doi.org/10.1088/1361-648X/abdc8e} {\bibfield  {journal} {\bibinfo  {journal} {J. Phys.: Condens. Matter}\ }\textbf {\bibinfo {volume} {33}},\ \bibinfo {pages} {145802} (\bibinfo {year} {2021})}\BibitemShut {NoStop}%
\bibitem [{\citenamefont {Schneider}\ \emph {et~al.}(2022)\citenamefont {Schneider}, \citenamefont {Halimeh},\ and\ \citenamefont {Punk}}]{Schneider-2022}%
  \BibitemOpen
  \bibfield  {author} {\bibinfo {author} {\bibfnamefont {B.}~\bibnamefont {Schneider}}, \bibinfo {author} {\bibfnamefont {J.~C.}\ \bibnamefont {Halimeh}},\ and\ \bibinfo {author} {\bibfnamefont {M.}~\bibnamefont {Punk}},\ }\bibfield  {title} {\bibinfo {title} {{P}rojective symmetry group classification of chiral ${{\mathbb{Z}}}_{2}$ spin liquids on the pyrochlore lattice: Application to the spin-$\frac{1}{2}$ {X}{X}{Z} {H}eisenberg model},\ }\href {https://doi.org/10.1103/PhysRevB.105.125122} {\bibfield  {journal} {\bibinfo  {journal} {Phys. Rev. B}\ }\textbf {\bibinfo {volume} {105}},\ \bibinfo {pages} {125122} (\bibinfo {year} {2022})}\BibitemShut {NoStop}%
\bibitem [{\citenamefont {Messio}\ \emph {et~al.}(2010)\citenamefont {Messio}, \citenamefont {C\'epas},\ and\ \citenamefont {Lhuillier}}]{Messio-Cepas-2010}%
  \BibitemOpen
  \bibfield  {author} {\bibinfo {author} {\bibfnamefont {L.}~\bibnamefont {Messio}}, \bibinfo {author} {\bibfnamefont {O.}~\bibnamefont {C\'epas}},\ and\ \bibinfo {author} {\bibfnamefont {C.}~\bibnamefont {Lhuillier}},\ }\bibfield  {title} {\bibinfo {title} {{S}chwinger-boson approach to the kagome antiferromagnet with {D}zyaloshinskii-{M}oriya interactions: {P}hase diagram and dynamical structure factors},\ }\href {https://doi.org/10.1103/PhysRevB.81.064428} {\bibfield  {journal} {\bibinfo  {journal} {Phys. Rev. B}\ }\textbf {\bibinfo {volume} {81}},\ \bibinfo {pages} {064428} (\bibinfo {year} {2010})}\BibitemShut {NoStop}%
\bibitem [{\citenamefont {Mondal}\ and\ \citenamefont {Kadolkar}(2017)}]{Mondal-2017}%
  \BibitemOpen
  \bibfield  {author} {\bibinfo {author} {\bibfnamefont {K.}~\bibnamefont {Mondal}}\ and\ \bibinfo {author} {\bibfnamefont {C.}~\bibnamefont {Kadolkar}},\ }\bibfield  {title} {\bibinfo {title} {{S}chwinger boson mean-field theory of the kagome {H}eisenberg antiferromagnet with {D}zyaloshinskii-{M}oriya interactions},\ }\href {https://doi.org/10.1103/PhysRevB.95.134404} {\bibfield  {journal} {\bibinfo  {journal} {Phys. Rev. B}\ }\textbf {\bibinfo {volume} {95}},\ \bibinfo {pages} {134404} (\bibinfo {year} {2017})}\BibitemShut {NoStop}%
\bibitem [{\citenamefont {Rossi}\ \emph {et~al.}(2023)\citenamefont {Rossi}, \citenamefont {Motruk}, \citenamefont {Rademaker},\ and\ \citenamefont {Abanin}}]{Rossi-Motruk-2023}%
  \BibitemOpen
  \bibfield  {author} {\bibinfo {author} {\bibfnamefont {D.}~\bibnamefont {Rossi}}, \bibinfo {author} {\bibfnamefont {J.}~\bibnamefont {Motruk}}, \bibinfo {author} {\bibfnamefont {L.}~\bibnamefont {Rademaker}},\ and\ \bibinfo {author} {\bibfnamefont {D.~A.}\ \bibnamefont {Abanin}},\ }\bibfield  {title} {\bibinfo {title} {{S}chwinger boson study of the ${J}_{1}$-${J}_{2}$-${J}_{3}$ kagome {H}eisenberg antiferromagnet with {D}zyaloshinskii-{M}oriya interactions},\ }\href {https://doi.org/10.1103/PhysRevB.108.144406} {\bibfield  {journal} {\bibinfo  {journal} {Phys. Rev. B}\ }\textbf {\bibinfo {volume} {108}},\ \bibinfo {pages} {144406} (\bibinfo {year} {2023})}\BibitemShut {NoStop}%
\bibitem [{\citenamefont {Sasamoto}\ and\ \citenamefont {Nasu}(2026{\natexlab{b}})}]{Sasamoto-Nasu-2026}%
  \BibitemOpen
  \bibfield  {author} {\bibinfo {author} {\bibfnamefont {D.}~\bibnamefont {Sasamoto}}\ and\ \bibinfo {author} {\bibfnamefont {J.}~\bibnamefont {Nasu}},\ }\bibfield  {title} {\bibinfo {title} {{D}ynamical spin correlations in kagome antiferromagnets: {C}omparison of the {A}brikosov fermion and {S}chwinger boson approaches beyond mean field},\ }\href {https://doi.org/10.1103/h85g-v6zx} {\bibfield  {journal} {\bibinfo  {journal} {Phys. Rev. B}\ }\textbf {\bibinfo {volume} {114}},\ \bibinfo {pages} {074405} (\bibinfo {year} {2026}{\natexlab{b}})}\BibitemShut {NoStop}%
\bibitem [{\citenamefont {Colpa}(1978)}]{Colpa-1978}%
  \BibitemOpen
  \bibfield  {author} {\bibinfo {author} {\bibfnamefont {J.~H.~P.}\ \bibnamefont {Colpa}},\ }\bibfield  {title} {\bibinfo {title} {{D}iagonalization of the quadratic boson hamiltonian},\ }\href {https://doi.org/10.1016/0378-4371(78)90160-7} {\bibfield  {journal} {\bibinfo  {journal} {Physica A}\ }\textbf {\bibinfo {volume} {93}},\ \bibinfo {pages} {327} (\bibinfo {year} {1978})}\BibitemShut {NoStop}%
\bibitem [{\citenamefont {Rao}\ \emph {et~al.}(2025)\citenamefont {Rao}, \citenamefont {Moessner},\ and\ \citenamefont {Knolle}}]{Rao-Moessner-Knolle-2025}%
  \BibitemOpen
  \bibfield  {author} {\bibinfo {author} {\bibfnamefont {P.}~\bibnamefont {Rao}}, \bibinfo {author} {\bibfnamefont {R.}~\bibnamefont {Moessner}},\ and\ \bibinfo {author} {\bibfnamefont {J.}~\bibnamefont {Knolle}},\ }\bibfield  {title} {\bibinfo {title} {{D}ynamical response theory of interacting {M}ajorana fermions and its application to generic {K}itaev quantum spin liquids in a field},\ }\href {https://doi.org/10.1103/f4b4-h1yr} {\bibfield  {journal} {\bibinfo  {journal} {Phys. Rev. B}\ }\textbf {\bibinfo {volume} {112}},\ \bibinfo {pages} {024440} (\bibinfo {year} {2025})}\BibitemShut {NoStop}%
\bibitem [{\citenamefont {Willsher}\ and\ \citenamefont {Knolle}(2025)}]{Willsher-Knolle-2025}%
  \BibitemOpen
  \bibfield  {author} {\bibinfo {author} {\bibfnamefont {J.}~\bibnamefont {Willsher}}\ and\ \bibinfo {author} {\bibfnamefont {J.}~\bibnamefont {Knolle}},\ }\href@noop {} {\bibinfo {title} {{D}ynamics and stability of {U}(1) spin liquids beyond mean-field theory: {T}riangular-lattice ${J}_{1}$-${J}_{2}$ {H}eisenberg model}} (\bibinfo {year} {2025}),\ \Eprint {https://arxiv.org/abs/2503.13831} {arXiv:2503.13831} \BibitemShut {NoStop}%
\bibitem [{\citenamefont {Wang}(2010)}]{Wang-2010}%
  \BibitemOpen
  \bibfield  {author} {\bibinfo {author} {\bibfnamefont {F.}~\bibnamefont {Wang}},\ }\bibfield  {title} {\bibinfo {title} {{S}chwinger boson mean field theories of spin liquid states on a honeycomb lattice: {P}rojective symmetry group analysis and critical field theory},\ }\href {https://doi.org/10.1103/PhysRevB.82.024419} {\bibfield  {journal} {\bibinfo  {journal} {Phys. Rev. B}\ }\textbf {\bibinfo {volume} {82}},\ \bibinfo {pages} {024419} (\bibinfo {year} {2010})}\BibitemShut {NoStop}%
\bibitem [{\citenamefont {Koyama}\ and\ \citenamefont {Nasu}(2021)}]{Koyama-Nasu-2021}%
  \BibitemOpen
  \bibfield  {author} {\bibinfo {author} {\bibfnamefont {S.}~\bibnamefont {Koyama}}\ and\ \bibinfo {author} {\bibfnamefont {J.}~\bibnamefont {Nasu}},\ }\bibfield  {title} {\bibinfo {title} {{F}ield-angle dependence of thermal {H}all conductivity in a magnetically ordered {K}itaev-{H}eisenberg system},\ }\href {https://doi.org/10.1103/PhysRevB.104.075121} {\bibfield  {journal} {\bibinfo  {journal} {Phys. Rev. B}\ }\textbf {\bibinfo {volume} {104}},\ \bibinfo {pages} {075121} (\bibinfo {year} {2021})}\BibitemShut {NoStop}%
\bibitem [{\citenamefont {C{\^o}nsoli}\ \emph {et~al.}(2020)\citenamefont {C{\^o}nsoli}, \citenamefont {Janssen}, \citenamefont {Vojta},\ and\ \citenamefont {Andrade}}]{Consoli-2020}%
  \BibitemOpen
  \bibfield  {author} {\bibinfo {author} {\bibfnamefont {P.~M.}\ \bibnamefont {C{\^o}nsoli}}, \bibinfo {author} {\bibfnamefont {L.}~\bibnamefont {Janssen}}, \bibinfo {author} {\bibfnamefont {M.}~\bibnamefont {Vojta}},\ and\ \bibinfo {author} {\bibfnamefont {E.~C.}\ \bibnamefont {Andrade}},\ }\bibfield  {title} {\bibinfo {title} {{H}eisenberg-{K}itaev model in a magnetic field: {$1/S$} expansion},\ }\href {https://doi.org/10.1103/PhysRevB.102.155134} {\bibfield  {journal} {\bibinfo  {journal} {Phys. Rev. B}\ }\textbf {\bibinfo {volume} {102}},\ \bibinfo {pages} {155134} (\bibinfo {year} {2020})}\BibitemShut {NoStop}%
\bibitem [{\citenamefont {Holstein}\ and\ \citenamefont {Primakoff}(1940)}]{Holstein-Primakoff-1940}%
  \BibitemOpen
  \bibfield  {author} {\bibinfo {author} {\bibfnamefont {T.}~\bibnamefont {Holstein}}\ and\ \bibinfo {author} {\bibfnamefont {H.}~\bibnamefont {Primakoff}},\ }\bibfield  {title} {\bibinfo {title} {{F}ield dependence of the intrinsic domain magnetization of a ferromagnet},\ }\href {https://doi.org/10.1103/PhysRev.58.1098} {\bibfield  {journal} {\bibinfo  {journal} {Phys. Rev.}\ }\textbf {\bibinfo {volume} {58}},\ \bibinfo {pages} {1098} (\bibinfo {year} {1940})}\BibitemShut {NoStop}%
\end{thebibliography}%

\end{document}